\documentclass[12pt,openright]{report}

\usepackage{dbl12,rackham}
\usepackage{natbib}

\usepackage{amsmath}
\usepackage{ifthen,psfig,epsfig}
\usepackage{rotating}

\begin{document}                                    


\titlepage{ION HEATING IN COLLISIONLESS SHOCKS IN SUPERNOVAE AND THE HELIOSPHERE}                
{Kelly Elizabeth Korreck}                            
{Doctor of Philosophy}                             
{Space and Planetary Physics}                      
{2005}                                             
{Associate Professor Thomas H. Zurbuchen, Co-Chairperson \\        
 Dr. John C. Raymond, Co-Chairperson, Smithsonian Astrophysical Observatory\\
 Professor August Evrard \\
 Professor Lennard A. Fisk  \\
 Associate Professor Timothy A. McKay
}


\unnumberedpage                             

\copyrightpage{Kelly E. Korreck}

\initializefrontsections                    


\dedicationpage{For my past, present, and future.}    



\startacknowledgementspage

I would like to thank Thomas Zurbuchen, my advisor for his belief
in me and his willingness to take me on in my third year.  I
admire your drive and passion for all that you do.

I thank Dr. John C. Raymond who was my advisor for my year and a
half that I spent at the Harvard-Smithsonian Center for
Astrophysics.  He is patient.  His knowledge of the field is
incredible.  And he is a great mentor.  I especially thank him for
his enthusiasm for collaborating on and discussing the neutral
code.

Of course, I am grateful to my parents for their patience and {\em
love}.  I have a very large extended family and I would like to
thank them all for their support throughout all of my life.
Grandpa Korreck, you are an inspiration to us all, 92 and still
going strong! I think I get my natural curiosity and need to keep
going from you!

To my best friend, my sister.  Somehow we grew up and became
friends.  I know mom said it would happen but I remember times
when I thought it never could.  Thank you for your support and the
endless shopping trips!

And to my little brother, who isn't so little anymore.  I know you
love the Maize and Blue as much as I do.  Do whatever your heart
tells you and I know you will do it well!

And now my newest brother - well brother-in-law if you want to get
technical.  Mike, I am so happy you are part of our family.  I
hope you enjoy your copy of the thesis as you may be the only one
besides me or my future grad students to read it!

The only way to get through grad school is with support of friends
and family. I have been so lucky to have such wonderful friends
and collaborators. The list is long and I apologize for anyone who
I missed.  Margaret Reid, Jan Beltran, Sue Griffith and all the
AOSS staff - You have been there to listen to gripes and help
workout the hard problems.  Thank you so much!   As part of the
FUSE team,  Ravi Sankrit helped with the night extraction of the
FUSE data as well as fruitful conversations about shock physics
and supernova remnants. Thanks also goes to Parviz Ghavamian for
his advice, guidance, and ds9 skills.

Amy Reighard played sphere with me in the early years. She also
kept me going and believing in myself when I was in doubt. Amy is
an amazing physicist and I hope we can collaborate in the future.

The Empress Alysha, I will be the advisor for you anytime. Thank
you for your help getting me through classes and the qualifiers.
You are a great friend-just watch out for bears!

Pat Koehn for reading many chapters and the niffy template!  I am
so excited for you to start your teaching position.  You just seem
to be one of those who just has 'it' when it comes to teaching!

Kevin Kane thanks for the support especially though the last few
months!  Thank you for lending me your house to finish the last
two chapters of the thesis. And Lena Adams and all the women of
AOSS you make the department wonderful!

A special thanks to my kitty Orion, who warmed my lap many a
nights as I wrote this work.

Sue Lepri from my roommate to my officemate--I got though Jackson
because of you.  I enjoy working on everything from science
projects or social affairs or hair cutting with you!   And why
does everyone say, 'Here comes trouble' when they see us together?
And of course to keep with tradition I must mention your funny
faces that have entertained a generation of graduate students!

For Suni, my favorite! I am so happy that I get to come back to
Cambridge for a while so we can go to the cider mill and hopefully
learn how to sail! For Chelle Reno who always has a way with words
and a zest for adventure of which I can only sit in amazement.  I
can't thank you enough for coming into town to see my oral defense
of this thesis. Hopefully we can all stick together through the
years!

To My Cambridge Crew: Jim Carey, Raslyn Rendon, Dana Ozik(and Todd
too), Jeno Soloski, Paul Martini, Matt Rosen, Matt Povich, Travis
Metcalf, Maryam, Jenny Greene,  Cara Rakowski and probably many
more I have accidently forgotten.  Working with you all is an
amazing experience! I really enjoyed being a part of the CfA
community and can't wait to see what lies ahead!

Thank you to everyone who supported me in one way or another
throughout this learning process. I am a better person for knowing
you and going through the experience!

\bigskip
\noindent
Ann Arbor, Michigan \hfill Kelly E. Korreck\\
May 2005




\tableofcontents                     


\listoffigures                       


\listoftables                        


\startthechapters                         


\chapter{Introduction}

\def\baselinestretch{2.0}
\smallskip

Gazing into the night sky is a favorite pastime of the human race.
The earliest human records have symbols of astronomical events and
objects such as the moon and the Sun.  The ancient Greeks saw many
"pictures" in the stars that are now known as constellations that
rotate around the sky throughout the year.  Each constellation had
a story associated with it, mythical stories about heros,
princesses, and creatures.  Supernovae or "guest stars" have been
recorded by ancient Chinese astronomers \citep{str94}. The native
people of North America depended on the heavens to determine the
best time for planting crops as well as for spiritual and
religious rites. The Lakota, a nomadic people, used the equinoxes
and solstices to track seasons for hunting and to prepare for
their sacred spring ceremonies \citep{goo92}.

\begin{figure}[width=3cm]
  \def\baselinestretch{1.0}
  \centering
  \includegraphics{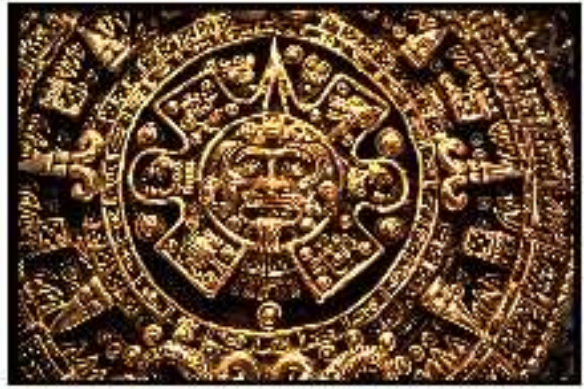}
  \caption{The photograph above is the Aztec Calendar Stone, also
called Sunstone. The carvings in the stone represent a rayed disk
with the four previous cycles of creation and destruction. The
skull at the center depicts the god Tonatiuh, the fifth Sun. Image
courtesy of Corel Corporation.\label{tonatiuh}}
  \def\baselinestretch{2.0}
\end{figure}

The bright lights of the Aurora Borealis, sometimes referred to as
the Northern Lights, have meant many things to ancient people. For
the Fox Indians of Wisconsin, they were an omen of war
\citep{spg98}. The lights were the ghosts of their slain enemies
who sought revenge. We know today these lights are caused by
cascading energetic particles from the Sun exciting emissions in
the atmosphere. But for the Salteaus Indians of eastern Canada and
the Kwakiutl and Tlingit of Southeastern Alaska the northern
lights were the dancing of human spirits or animal spirits,
especially those of deer, seals, salmon and beluga: the animals
that gave them sustenance \citep{ray58}.

The Aztecs, of present day Mexico, believed in the Sun as an
active living being, named Tonatiuh, pictured in Figure
\ref{tonatiuh}, that had a beginning and an end. A new Sun would
replace the old. In order to keep the Sun strong they needed to
make sacrifices to the Sun. Their temples and buildings were
aligned with the sunrise and sunsets of the solstices to track the
life of the current Sun\citep{mto04}.

The events unfolding in the heavens were viewed as powerful
symbols and signs and those chosen few that interpreted them were
considered high priests or magicians. Even today the popularity of
Star Trek, Star Wars, and other science fiction allows us to feed
our own curiosity about the cosmos.  Humans have always felt very
connected and as if they participated in what occurred in the
cosmos.

Modern science has proven through remote and in situ observations
that the cosmos is more mysterious and amazing than previously
thought.  But "ordinary" in the sense that terrestrial physics
applies to objects in space. Stars, including our Sun, and most of
the objects in "outer space" are in the fourth state of matter
called a plasma. A plasma can be defined as 'a quasi-neutral gas
of charged and neutral particles which exhibits collective
behavior'\citep{che84}. Quasi-neutral means that although the
atoms in the plasma are ionized, there are equal number of
electrons and positively charged ions.  Their collective behavior
allows the plasma to be described as a fluid instead of individual
atoms or ions.  The plasma can be characterized by fluid dynamics.
One parameter often used when discussing magentic plasmas is the
Alfvenic speed.  The Alfvenic speed is the speed that magnetic
information travels in the plasma.

\begin{equation}
\label{va}
  v_{A}=\frac{B}{\sqrt{\rho\mu_{o}}}
  \end{equation}

One interesting aspect of fluid dynamics is the steepening of
waves into shocks. These shocks create a difference in density,
velocity, and thermal energy between regions of the plasma and are
important in the study of astrophysical systems.

\section{Collisionless Shocks in Astrophysical Plasmas}

If a disturbance propagates through a plasma faster than the
characteristic or Alfvenic speed of the local plasma, a shock is
formed.  Qualitatively, this is similar to the formation of a
sonic boom or water waves breaking close to the beach.

In general, shocks transfer energy and momentum through the
collision of atoms or molecules. However, most space-based shocks
are collisionless. A collisionless shock occurs when the
characteristic shock length is much smaller than the collisional
mean free path of the particles in the plasma. Collisionless
shocks were confirmed by the discovery and study of the Earth,
Heliosphere, and solar wind in the late 1960's \citep{son63,
ken85}. These shocks transfer energy and momentum via
electromagnetic and wave interactions and not collisions.
Collisionless shocks appear in many different physical systems
such as those produced by Coronal Mass Ejections (CMEs) in the
Heliosphere, the termination shock, new stars (Herbig-Haro
objects), jets from Active Galactic Nuclei (AGN), and supernova
remnants.

Collisionless shocks have been studied for several decades but are
still not well understood.  There are several questions that
remain to be answered about collisionless shock physics
\citep{lem04}. They include the electron heating and dynamics at
the collisionless shock front, the particle diffusion via
turbulence, electric, and magnetic fields, general particle
acceleration by these shocks and pickup ions interaction with the
collisionless shocks.  Pickup ions are atoms from the Interstellar
Medium (ISM) which are ionized by solar radiation and then carried
along with the solar wind.

This thesis will address the heating and acceleration mechanisms
at collisionless shocks fronts and the differences in heating
observed for various ion species.  Observed ion heating thus far
have shown that the heating "fractionates" according to the mass
or charge of a particle.  Ion heating creates a thermal seed
population necessary to accelerate ions to higher cosmic ray
energies. Several outstanding questions are -

What is the heating mechanism responsible for the fractionation of
the ions?

Why does the heating process fractionate the ions according to
mass?

How does the mass fractionation of the heating process affect the
seed population?

To cover a wide range of parameter space in investigating these
heating questions, two very different shock systems were used: the
heliospheric shocks that occur in front of CMEs and the shocks of
a supernova remnant, SN1006. Heliospheric shocks, including the
bow shock, have been studied as the acceleration mechanisms for
ions. Supernova remnants are also known to be the most powerful
particle accelerators which produce high energy cosmic rays.  The
heliospheric shocks represent lower velocity shocks and lower Mach
number shocks, with v$_{shock}$ less than 1000 km s$^{-1}$ and
M$_{A}$$\sim$ 1-5.  The SN1006 shock represents the faster
velocity and higher Mach shocks at v$_{shock}$ $\sim$ 3000 km
s$^{-1}$ and M$_{A}$$\sim$100. However, shock properties are
fundamental in nature and thus can be related by scaling or a
specific physical parameter.

For a greater understanding of the ion heating that is implicit to
the acceleration method, a comparison study of the conditions of
shocks and the heating mechanisms are necessary. Bulk
thermalization of the kinetic energy of the shock would be the
simplest heat transfer method; however, work done in the
heliosphere indicates a different form of heating dominates.

Heliospheric shocks have been studied in much detail because of
the availability of in situ data. Initial analysis of data from
various satellites showed a mass proportional heating.
\cite{ogi80} used particle distribution data from the ISEE3
satellite to analyze helium (He$^{2+}$) and oxygen( O$^{7+}$) in
the solar wind. Although the accuracy of the data for heavier ions
(oxygen) data were poor, the average heating was approximately
mass proportional. Any heating was attributed to wave interaction
close to the Sun, rather than shock heating.

Studies using the Prognoz satellite at 1 AU by \cite{zer76} found
heating less than mass proportional for Helium. \citet{ber97}
studied the heating of ions in interplanetary shocks.  This work
contradicts the earlier work on shock heating and showed a greater
than mass proportional heating present in the shocks for helium
and oxygen ions. These shocks heated oxygen ions 19 - 48 times
more than the protons, such that the heating is 1.2 to 3.0 times
mass proportional.  These same shocks heated helium 4.6 - 10.8
times more than the protons.  The heating of the helium was 1.2 -
2.7 times more than mass proportional.

Supernovae interact with the ISM based on the nature of the
progenitor star and the make up of the medium surrounding the
specific supernova.  Study of these collisionless shocks must
explore the interaction and subsequent heating produced by the
shock with ions heavier than protons as well as the shock-neutral
ISM interaction. Past measurements of ion heating
\citep{kor04,gha02}, electron \citep{lam96}, proton and ion
temperature, and other emission features are studied to understand
both the supernova explosion and the interstellar medium into
which the shocks are expanding. Studies to date have shown a less
than mass proportional heating for supernova shocks
\citep{kor04,ray95}.  This directly contradicts what is found for
the heliospheric shocks and leads to questions about the injection
process necessary for cosmic ray acceleration, such as how the
mass of the ion species plays a role in the heating mechanisms.

In order to understand shock heating, a definition of several
plasma and shock characteristics is necessary. A shock most
generally is a transition layer which propagates through a plasma
causing discontinuous changes in the density, velocity, and
pressure of the plasma \citep{tid69}.  If a magnetic field is
present, the plasma can be described by the Magnetohydrodynamics
(MHD) equations for mass, momentum and energy conservation. From
\citet{gom99}, the following are the conservative form of the
ideal MHD equations in 3-D shown below, assuming no external
forces (i.e. gravity):
\medskip

Conservation of Mass:
\begin{equation}\label{mhdmass}
\frac{\partial\rho}{\partial t}+ \nabla\cdot (\rho \vec{u})=0
\end{equation}

Conservation of Momentum:
\begin{equation}\label{mhdp}
\frac{\partial(\rho \vec{u})}{\partial t}+ \nabla \cdot
(\rho\vec{u}\vec{u}+p\vec{I}+\frac{B^{2}}{2\mu_{0}}\vec{I}-\frac{\vec{B}\vec{B}}{\mu_{0}})=0
\end{equation}
Conservation of Energy:
\begin{equation}\label{mhdenergy}
\frac{\partial}{\partial t}(\frac{1}{2}\rho
u^{2}+\frac{1}{\gamma-1}p+\frac{B^{2}}{2\mu_{0}})+\nabla\cdot(\frac{1}{2}\rho
u^{2}\vec{u}+\frac{\gamma}{\gamma-1}p\vec{u}+\frac{(\vec{B}\cdot\vec{B}\vec{u}-\vec{B}(\vec{B}\cdot\vec{u}))}{\mu_{0}})=0
\end{equation}
Induction Equation:
\begin{equation}\label{btangent}
\frac{\partial\vec{B}}{\partial t}=\nabla\times(\vec{u} \times
\vec{B})
\end{equation}
Lack of Magnetic Monopoles:
\begin{equation}\label{bvector}
\nabla\cdot\vec{B}=0
\end{equation}
\medskip
where\\
$\rho$=mass density\\
$\vec{u}$= flow velocity\\
p=thermal pressure\\
$\vec{B}$=magnetic field
B= magnitude of the magnetic field\\
B$_{n}$=normal component of the magnetic field\\
B$_{t}$=tangential component of the magnetic field\\
$\vec{I}$=Identity matrix\\
$\mu_{0}$=permeability of free space\\
$\gamma$=adiabatic index\\

The MHD equations characterize the plasma as a fluid but do not
predict the shock conditions.  Three characteristic waves can
develop in the plasma described by the MHD equations and steepen
into shock waves or discontinuities. These waves are named
according to their speed: slow, intermediate, and fast waves. Each
wave is related to the Alfven speed and the angle between the
shock normal and the magnetic field.

If the discontinuity of the shock is considered infinitesimally
thin, the fluxes of the mass, momentum, and energy should be
conserved across the discontinuity. The Rankine-Hugoniot
relations, Equations \ref{rhcontinuity}-\ref{rhnomonopoles},
describe the relationship between pre-shock to post-shock physical
characteristics due to conservation of mass flux, momentum, and
energy across the shock front.

From the continuity equation:
\begin{equation}\label{rhcontinuity}
[\rho u_{n}]=0
\end{equation}

From the conservation of momentum equation:
\begin{equation}
[\rho u_{n} \vec{u_{t}} - \frac{B_{n}\vec{B_{t}}}{\mu_{0}} ]=0
\end{equation}
From the conservation of energy flux equations:
\begin{equation}
[\rho u_{n}^{2} + p + \frac{B_{t}^{2}-B_{n}^{2}}{2\mu_{0}} ]=0
\end{equation}

\begin{equation}
[\frac{1}{2}\rho(u_{n}^{2}+u_{t}^{2})u_{n} +
\frac{\gamma}{\gamma-1}pu_{n} +
\frac{B_{t}^{2}}{\mu_{0}}u_{n}-\frac{B_{n}}{\mu_{0}}(\vec{B_{t}}\cdot\vec{u_{t}})
]=0
\end{equation}

In addition from the induction and magnetic monopole equations we
have:
\begin{equation}
[u_{n} \vec{B_{t}} - B_{n}\vec{u_{t}} ]=0
\end{equation}
\begin{equation}\label{rhnomonopoles}
[B_{n}]=0
\end{equation}

where the subscript t indicates the tangential component and the
subscript n indicates the normal component with respect to the
shock front. The brackets indicate the difference from upstream to
downstream conditions. These equations allow for great insight
when observing shocks. With observations of atomic emission lines
or in situ plasma measurements, one can determine the density,
velocity, or temperature of the downstream side, and use the
Rankine-Hugoniot equations to infer the upstream characteristics
or vice versa.


Three plasma parameters are important in characterizing heating
and acceleration in a shock, plasma $\beta$, Mach number, and
$\theta$$_{Bn}$,the magnetic angle.  The Alfvenic Mach number,
Equation \ref{Mach}, is a measure of the speed of the shock versus
the Alfvenic speed.

\begin{equation}\label{Mach}
  M=\frac{v_{shock}}{v_{A}}
  \end{equation}
where
\begin{equation}\label{va}
  v_{A}=\frac{B}{\sqrt{\rho\mu_{o}}}
  \end{equation}
The Alfven speed \citep{alf45} is the speed at which magnetic
information can be transported through a plasma.  It is the
magnetic equivalent of the sound speed which is the speed at which
thermal pressure can be relayed.

The plasma $\beta$ is the magnetic pressure of the medium versus
the thermal pressure. $\beta$ is defined as

\begin{equation}\label{beta}
  \beta=\frac{\rho kT}{\frac{B^{2}}{2\mu_{0}}}
  \end{equation}

A plasma is defined as low $\beta$ plasma (magnetically dominated)
when $\beta$ is much less than 1 and defined as a high $\beta$
plasma (thermally dominated) when $\beta$ $\ge$ 1.

\begin{figure}
\def\baselinestretch{1.0}
  \centering{ \epsfig{file=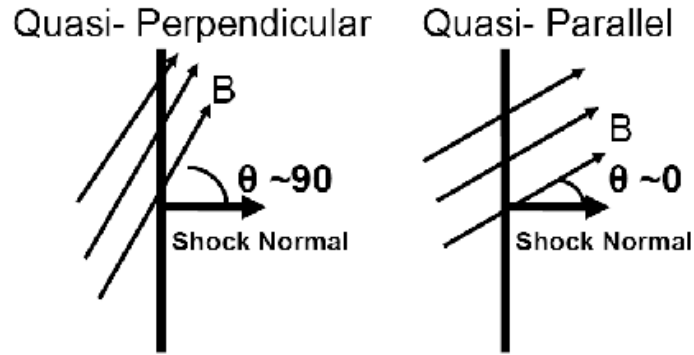,width=10cm}
  \caption{Orientation of the magnetic field versus the normal of
  the shock front. Perpendicular shocks, shown on the left, have
  a magnetic field oriented along the shock front and perpendicular
  to the shock normal. Parallel shocks, shown on the right, have a
  magnetic field oriented parallel to the shock normal.
  \label{shockstruc}} }
  \def\baselinestretch{2.0}
\end{figure}

 The geometry of the shock plays a critical role in heating mechanisms. Shocks can be classified by the geometry of
the magnetic field versus the shock normal.  Figure
\ref{shockstruc} shows an example of two types of shock geometry.

If the angle of the magnetic field to the shock normal is
approximately zero degrees the shock is quasi-parallel. This type
of shock allows the ions to easily cross the shock as the parallel
velocity of the ion is aligned with the bulk fluid flow,
v$_{shock}$.  If the angle of the magnetic field to the normal is
approximately 90 degrees, the shock is quasi-perpendicular.
Perpendicular shocks inhibit the ion movement with the fluid
across the shock front.

Parallel or quasi-parallel shocks are known to heat ions by a two
step process \citep{lee00}. At the shock front, a concentration of
ions occurs creating a density ramp. When the ions "see" this
density ramp the ions are scattered by whistler waves and by back
streaming ions that were reflected by the higher density material.
The backstreaming ions then heat the ions that are near the
density ramp as they flow upstream of the shock.

Perpendicular shocks are known to heat ions by processes based on
diffusion.  All shocks regardless of magnetic angle will dissipate
the ram energy of the plasma flow into thermal energy. If the
mechanism of dissipation is based on the resistivity and viscosity
due to waves excited by some instability due to departure from
equilibrium, the shock is considered subcritical. When a shock
cannot dissipate its energy by viscosity and resistivity alone, it
is classified as supercritical. Ions are heated more than
electrons creating a two fluid system which lends itself to many
instabilities such as the fire-hose or two-stream instabilities.
As the electrons and ions pass into the compressed magnetic field
downstream of the shock, their gyroradii are much different
setting up an effective potential.  This potential decelerates the
electrons and reflects a small amount of the protons upstream,
which can gyrate gaining energy \citep{bal02}. Once the reflected
ions are directed downstream through other scattering, this
effectively heats the ions that were directly transmitted
\citep{ler82}.  In a subcritical quasi-perpendicular shock,
heating is due to non-deflection of upstream ions at the shock's
ramp \citep{lee86,l87}. As they pass through the shock, the
direction of the magnetic field along which they are travelling
changes.  This causes the ions to start gyrating around the
magnetic field increasing their perpendicular velocity by the
proton gyro-velocity.

Although for simplicity the shock front is assumed to be planar
and laminar, in reality there is turbulence and a physical scale
over which the parameters change from the upstream to downstream
values.  The magnetic structure of the shock affects the density,
velocity and temperature.  For perpendicular shocks, the magnetic
field has a three part structure: a foot, a ramp and then an
overshoot of the downstream value for the magnetic field
\citep{bau97}. The shock foot is a gradual rise in the magnetic
field before the shock passes. Next a sharp increase called the
ramp occurs.  The ramp overshoots the downstream value before
coming to an average downstream value. For a laminar flow, these
transitions are rather abrupt.  However, as there is increasing
turbulence and non-linearity to the flow, the magnetic field is
characterized by waves and the region of the rise in magnetic
field is widened. Parallel shocks have a highly oscillatory
pre-shock magnetic field that is called a foreshock region. These
transition regions play a key role in heating of ions.

Another measure of the importance of the magnetic field to heating
is the relation of the jump conditions with respect to the
magnetic pressure. Shocks that have an increasing Alfvenic speed
or, in other words, an increasing magnetic pressure are classified
as fast shocks.  The fast shock bends the magnetic field toward
the shock surface and increases the magnetic field. The particles
crossing the shock increase their velocity due to the added
magnetic field that influences their gyration. If there is a
decreasing magnetic pressure across the shock, the shock is
classified as a slow shock.  The slow shock bends the magnetic
field towards the shock normal and decreases the field strength.
This thesis focuses on fast shocks because they are most prevelant
in the current data set.


Shocks in the heliosphere originate in some way from our Sun. A
brief introduction to the Sun and the processes in the solar wind
that creates shocks follows.
\goodbreak

\smallskip
\section{The Sun-Our Star}

The Sun is an ordinary dwarf variable star of spectral type G2V.
It is not the brightest, heaviest, or most unusual star known but
is our source of light, heat, energy, and our nearest stellar
laboratory.  The Sun is a gaseous sphere made up of an interior
and an atmosphere each with several layers of varying temperature,
density, and dynamics.  At a mass of 1.99 x 10$^{30}$ kg and a
radius of 7 x 10$^{5}$ km, the Sun dominates our solar system with
1000 times the mass of the rest of the solar system. The Sun is
made up of 74\% Hydrogen, 25\% Helium, and 1\% of other heavy
metals by mass \citep{kau91}.  The Sun is the greatest accelerator
of particles in the solar system.  In most cases, it does so by
forming shocks in its atmosphere.  We shall describe the Sun and
its atmosphere as a basis for our shock study involving Coronal
Mass Ejections.

\smallskip
\goodbreak

\subsection{Solar Structure}
The layer structure of the Sun is shown in Figure \ref{sunstruc}.
It consists of an interior and an atmosphere.  Three layers make
up the interior of the Sun: the core, the radiative zone, and the
convective zone. The solar atmosphere is also characterized by of
three temperature regimes: the photosphere (cooler=5800 K), the
chromosphere (warmer), and the corona, the hottest and part of the
atmosphere that is least understood.

\begin{figure}
\def\baselinestretch{1.0}
  \centering{
  \epsfig{file=layers,width=6cm}
  \caption{The structure of the Sun. In addition to the interior
  structure, coronal phenomena such as flares, holes, and
  prominences are shown.  Adapted from the SOHO website
  http://sohowww.nascom.nasa.gov\label{sunstruc}}
  }
  \def\baselinestretch{2.0}
\end{figure}

The inner most region of the Sun is the high temperature core.
Temperatures in the core reach 15 x 10$^{6}$ K \citep{car96}. The
core is approximately one quarter of the radius of the Sun yet
contains 50\% of its mass \citep{car96}. At these temperatures and
densities, atoms are stripped of all their electrons and protons
are readily available for fusion to occur. Nuclear fusion of
hydrogen to helium releases energy and neutrinos.  The fusion at
the core of the Sun creates immense heat which causes the
surrounding plasma to expand away from the core. However, gravity
counteracts this pressure and maintains the Sun's structure.  The
enormous pressure, density, and temperature produced by fusion is
a self fuelling process that results in further heating in order
to produce heavier fusion products. The Sun is 4.5 billion years
old and will continue to convert hydrogen into helium via nuclear
reactions for another 5 billion years \citep{car96}.

Energy produced by the Sun at its core then travels through the
five outer regions in order to reach interplanetary space.
Directly above the core is the radiative layer; photons carry
energy through this region hence the name.  It takes 1 million
years for a photon to diffuse through the radiative layers via
absorption and re-emission \citep{kau91}. As one moves out in
radius from the center of the Sun to the top of the radiative
layer, the temperature falls off to 2 $\times$ 10$^{6}$ K.

At the base of the chromosphere temperatures are approximately
4400 K however only 2000 km higher at the top of the chromosphere
the temperature rises to 25000 K. Then in the corona the
temperature rises from 25000K to 1-2 $\times$ 10$^{6}$ K. One of
the many remaining mysteries of the Sun is the heating that occurs
in the transition region. This region lies between the cool
chromosphere and the extremely hot corona. The rapid rise in
temperature indicates an explosive energy source in this region
\citep{moo99}.

\smallskip
The outermost layer of the atmosphere is the corona.  The corona
was identified in 968 A.D. by viewing an eclipse \citep{cha00}.
Further studies in the 1900's revealed that the corona is a highly
dynamic, complex, magnetically dominated region just beyond the
chromosphere.  From modern studies using coronagraphs, many
interesting features have been identified in the corona: flares,
prominences, arcades, and coronal mass ejections (CMEs) to name a
few. Flares, whose association with CMEs has been hotly debated,
are an explosive, rapid release of photons with a frequency range
from the X-rays to Radio \citep{kah92}. Prominences or filaments
consist of cool plasma on magnetic loops that extend above the
surface of the Sun \citep{van89}. Coronal mass ejections are a
release of large amounts of energetic particles (10$^{15}$ grams)
\citep{gom99}, magnetic energy, and lower energy charged particles
into the heliosphere; they will be discussed in the next section.

The number of sunspots, flares, CMEs, and streamers vary with an
11 year cycle. It takes 11 years to progress from minimum solar
activity through maximum solar activity and back to minimum
conditions.  This illustrates how these phenomena are closely tied
to the magnetic field of the Sun.

\goodbreak
\subsection{Solar Wind}
The Sun has a steady but highly variable supersonic outflow of
charged particles, magnetic field, and energy called the solar
wind. The solar wind is bimodal, fast or slow, with velocities
ranging from 400-900 km sec$^{-1}$.  At 1 AU, normal densities for
the slow wind are 8 cm$^{-3}$ \citep{gom99}, with a proton
temperature of 1$\times$ 10$^{5}$ K and a speed of v~400 km
sec$^{-1}$. In the fast solar wind the density drops to around 2.5
cm$^{-3}$ and with a mean speed of 770 km sec$^{-1}$\citep{gom99}.

The fast solar wind is associated with coronal holes located near
the poles of the Sun during solar minimum.  This fast wind is
relatively steady as well as relatively uniform in composition. In
contrast, the slow solar wind is highly variable and less
predictable.  The slow solar wind is associated with field lines
near closed magnetic regions that open up and allow an outflow of
material for a short time \citep{wan90,fis03}.

The solar wind varies with the solar cycle.  During solar minimum,
the fast wind originates mainly over the poles of the Sun but
expands in latitude to fill a large region of the heliosphere
\citep{hab97}. During solar minimum, the slow solar wind is
confined to the equatorial region.  During times of elevated solar
activity the solar magnetic field becomes highly disordered.
Coronal holes occur at all latitudes and are smaller.  Therefore,
the fast wind is not restricted to the polar area \citep{woo97}.
Similarly, slow solar wind sources extend to higher latitudes.

The Sun's magnetic field is carried radially outward by the solar
wind. However, the Sun differentially rotates.  The Sun's rate of
rotation from its equator to its poles varies but averages 27
days.  This rotation twists the magnetic field into a Parker
spiral as it is carried away from the Sun in the solar wind.
\begin{figure}[width=4cm]
  \def\baselinestretch{1.0}
  \centering
  \includegraphics{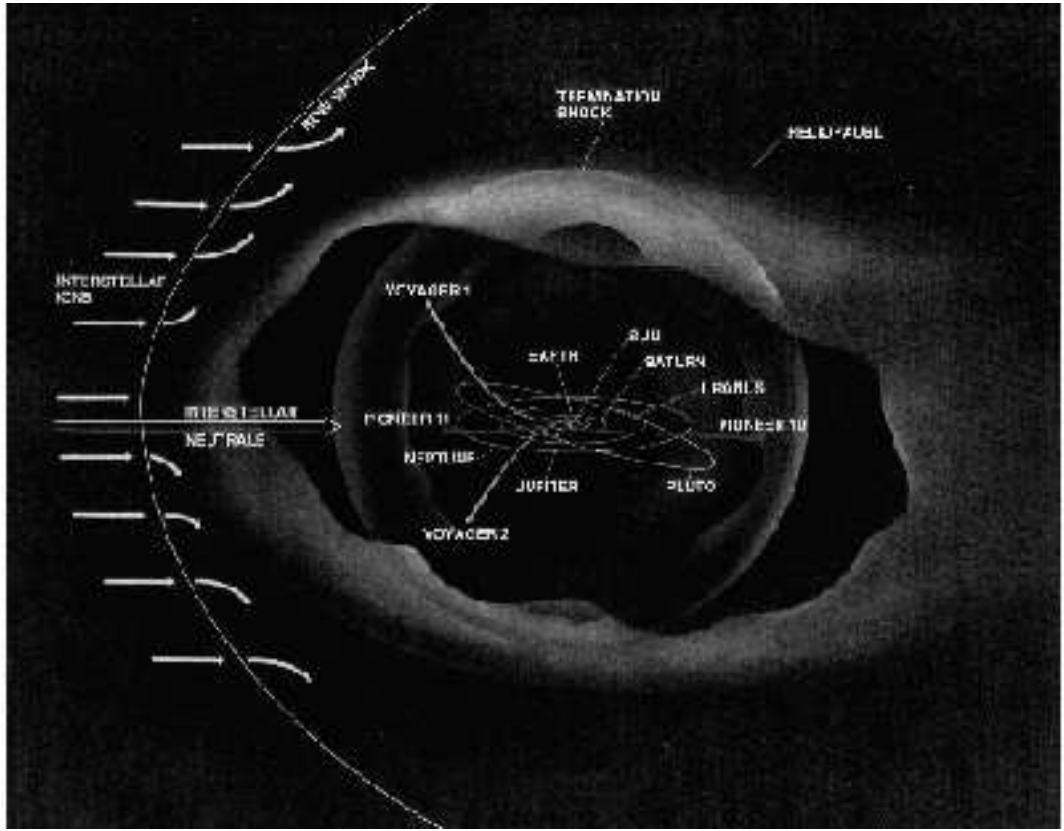}
  \caption{The heliosphere, where the solar wind ions dominate,
extends well beyond Earth.  This view of the heliosphere shows the
planets as well as the meeting of the heliosphere with the
interstellar space. From the ACE website
http://helios.gsfc.nasa.gov/ace/gallery.html\label{heliostruc}}
  \def\baselinestretch{2.0}
\end{figure}

\smallskip
The solar wind does not continue on indefinitely.  The heliosphere
is the region in space where the Sun's magnetic field and the
solar wind dominate, see Figure \ref{heliostruc}.  The solar wind
stretches well beyond the planets to a point where its pressure
eventually equals that of the interstellar medium.  A shock is
created where the solar wind meets the Interstellar Wind. At this
point the solar wind has slowed down and becomes subsonic forming
a termination shock.  The termination shock is thought to occur
between 80 and 100 AU \citep{bel93}. The heliopause marks the
boundary between the heliosphere and the Interstellar Medium
(ISM). Inside the heliopause, the Sun controls the environment,
whereas outside of the heliopause the environment is dominated by
the interstellar medium.
\goodbreak
\section{Coronal Mass Ejections}
\begin{figure}
  \def\baselinestretch{1.0}
  \centering
  \includegraphics{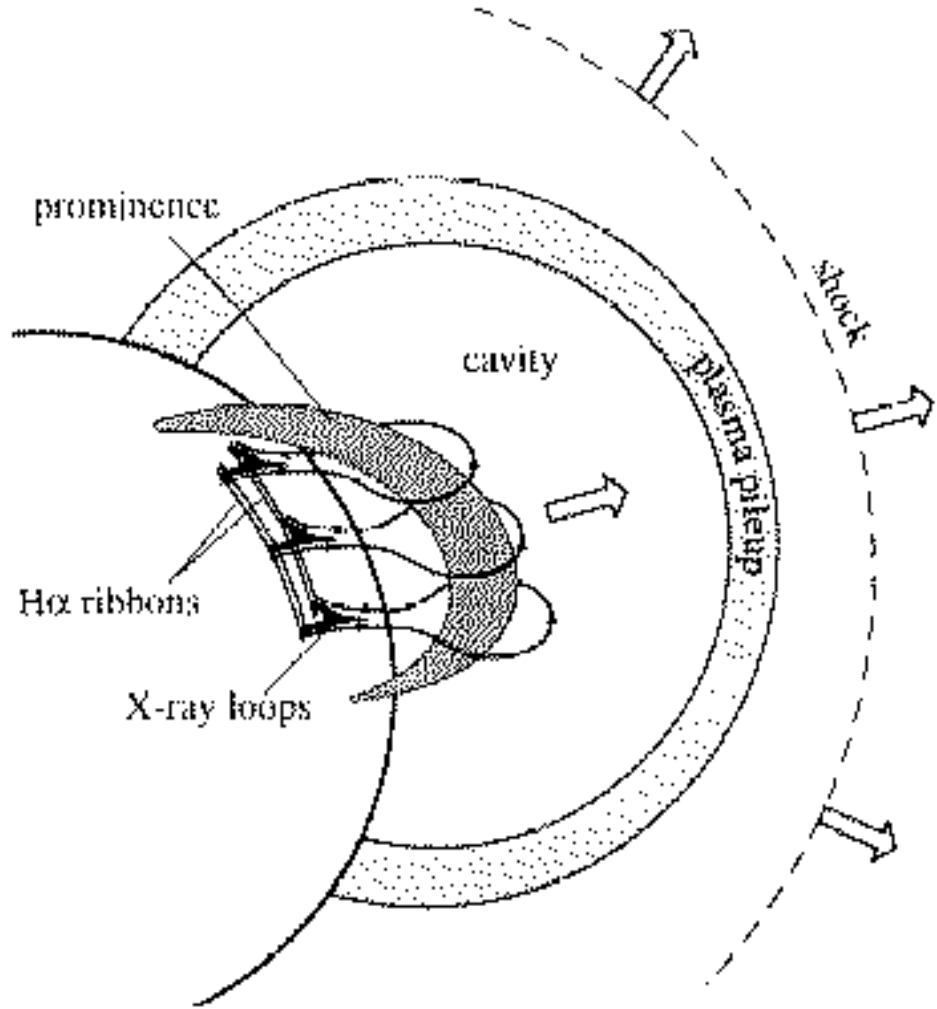}
  \caption{Schematic of the structure of CME as it leaves the
  corona. A three part structure is present, the plasma pile up,
  the low-density cavity, and the bright high density core formed
  by the prominence. Adapted from \citet{for00}\label{cmestruc}}
  \def\baselinestretch{2.0}
\end{figure}

Coronal mass ejections (CME) were first identified in the 1970's
\citep{mac80} using coronagraph images. CMEs are a transient
phenomena on the Sun that involve a catastrophic reorganization of
the magnetic field and release of mass.  Coronal magnetic loops
elongate and then pinch off, or reconnect, releasing vast amounts
of magnetic field and energy into the heliosphere. The material
released in CME, just like the solar wind, includes electrons,
protons and heavy ions. The mass released in CMEs is approximately
10$^{15}$ grams and carries along the embedded or frozen-in
magnetic field as it expands into interplanetary space. The
magnetic energy associated with this release is in the range of
10$^{31}$ - 10$^{32}$ erg \citep{gos74, gos97}.  CMEs generally
have a three part structure. First is the initial bright dense
front or plasma pile-up.  Behind the pileup is a dark, low density
cavity surrounding the inner most part of the CME, a bright high
density core. This structure is illustrated by Figure
\ref{cmestruc} adapted from Forbes (2000).

\begin{figure}[width=1cm]
  \def\baselinestretch{1.0}
  \centering
  \includegraphics{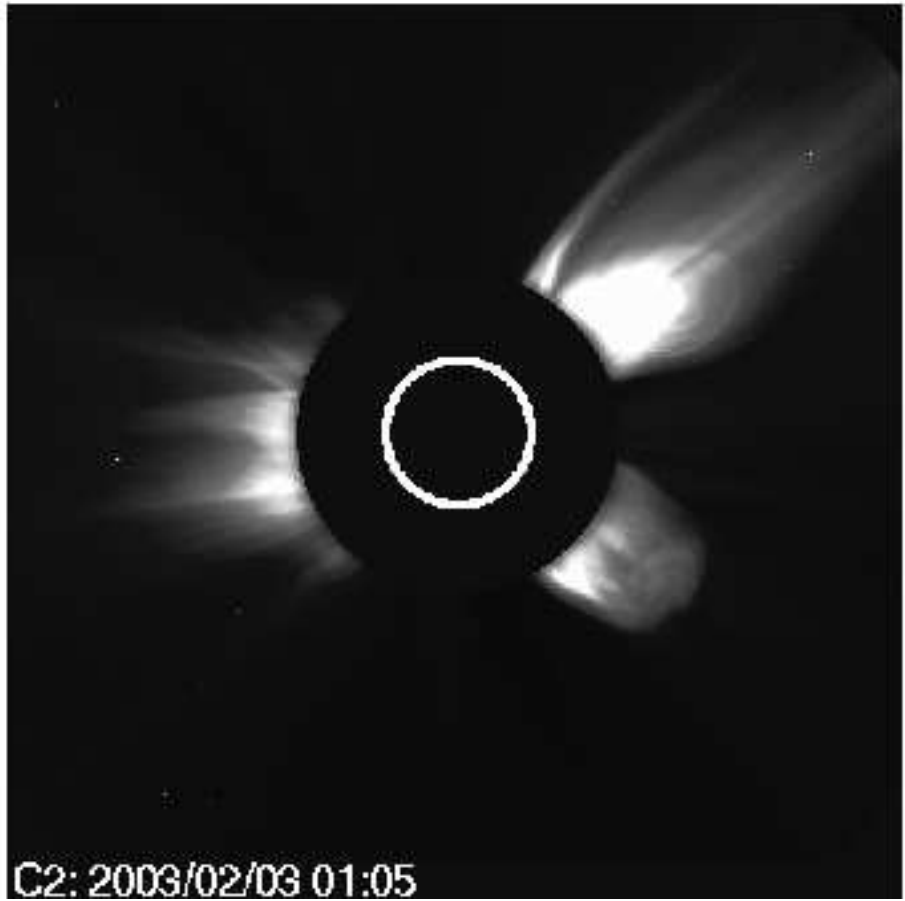}
  \caption{A LASCO C2 image of a CME expanding out of the corona in
  the top right corner of the picture.  Faint loop structures can
  be seen as the plasma is being expelled from the corona.  From sohowww.nascom.nasa.gov\label{cmesoho}}
  \def\baselinestretch{2.0}
\end{figure}

Once the CME leaves the corona, it expands into the heliosphere
with a velocity between 200 km s$^{-1}$ to 2000 km s$^{-1}$.  A
CME is seen leaving the corona in the SOHO-LASCO C2 image Figure
\ref{cmesoho}.

As the CME propagates through the heliosphere it is then termed an
Interplanetary CME or ICME.   ICMEs are of vast interest in the
space weather community as their effects pose a great danger to
our satellites and human activity in space. Due to their dynamic
interaction with the ambient solar wind, the ICMEs cause shocks to
form as they propagate \citep{ste00}. These shocks accelerate high
energy particles that pose a threat to astronauts.  The
collisionless shock ahead of the ICME heats ions and transfers
energy into the heliosphere.  These shocks are vital to
understanding the energetic particles that are detected.  This
thesis focuses on these shocks as well as other collisionless
shocks that are responsible for heating and acceleration of
particles such as those associated with supernova remnants.

\goodbreak
\section{Supernovae}

The most efficient accelerator in our galaxy and the source of the
highest energy particles are supernova.  These powerful explosions
expel the mass of several suns into the Interstellar Medium and a
shock precedes the ejecta.

A star with a mass $\ge$ 1.44 M$_{\odot}$, the Chandrasekhar limit
\citep{cha84}, ends its life in a spectacular explosion: a
supernova. Supernovae have been recorded as `guest stars' in the
sky by Chinese, Japanese, and Middle Eastern scholars as early as
386 A.D.\citep{str94}.  Their explosions can give off as much
light as that of their host galaxies and be hot enough to perform
nuclear synthesis during their explosion \citep{hor99}.

Supernovae are classified into two types based on the mechanism of
their detonation and their emission spectra. Type I supernovae
occur when a star runs out of its principle fuel, hydrogen, and a
gravitational collapse occurs.  The light curve of a Type I
supernova has a quick intensity rise to maximum luminosity of more
than 10$^{9}$ times the Sun's luminosity in two weeks. Type I
supernovae have a marked absence of hydrogen lines present in
their spectra \citep{cha95}. The dying star is in a constant
battle to balance the internal energy produced with the gravity
that is trying to collapse the core. The progenitor object, a
white dwarf with a companion accreting mass onto its surface,
burns hydrogen, then helium, up to carbon and oxygen. When the
mass of the white dwarf increases via accretion from its companion
to over the 1.44 M$_{\odot}$ Chandrasekhar limit, gravity is
greater than the electron degeneracy pressure in the stellar core
and sends a shock wave inward.  The shock wave heats the core so
that carbon and oxygen start to fuse. This causes an explosion or
deflagration, an explosion without an initial shock wave, that
rips apart the star while sustaining enough energy to fuse
elements up to radioactive $^{56}$Ni. The ejecta move outward with
an expansion velocity of up to 15000 km s$^{-1}$ \citep{cha95}.
There are supernova subtypes such as 1a and 1b that depend on the
mass of the progenitor star and brightness of the light curve.

\smallskip
A Type II supernova starts from a more massive star, M$\ge$10
M$_{\odot}$, which is a relatively young progenitor that still has
its hydrogen envelope, explaining the appearance of broad hydrogen
in its spectra.  Such a heavy star evolves through a series of
burning and contracting that uses increasingly heavier elements
for fuel.  This gives the star an onion like structure of
elements.  The last of the fusion products in the interior result
in an iron core.  There is no energy gain from fusing iron, thus
making this an endothermic reaction, requiring external heating to
continue. Iron itself cannot fuse, however the silicon in the
layer above the core is still fusing into iron increasing the mass
of the core and disturbing the delicate balance of electron
degeneracy pressure in the core and gravity. The core contracts
and iron fissions into lighter nuclei, adjusting the pressure
causing gravity to overcome the electron degeneracy and fuse the
center of the core into a nuclear density: a neutron star.  A
neutron star has the mass of the Sun within a radius of 10 km. The
light curve of a Type II supernova rises more slowly to maximum
and has a lower intensity than a Type I supernova. They are not
found in older stellar population but in gas rich young spiral
galaxies supporting the hypothesis that these are relatively young
stars.

\goodbreak
\subsection{Supernova Remnants}

Although the light from the initial explosion of a supernova can
be observed for many weeks, and in non-visible light for more than
a year, they leave a remnant behind that lasts for 10,000 years.
The ejecta expand spherically and supersonically into the local
ISM creating a shock around the supernova. There are also two
types of remnants: the shell-type remnant and the Crab-like
remnant. The shell-type remnant, such as that in the Cygnus Loop
or SN1006, blows out the center of the cavity creating a ring of
emitting material near the shock wave. The Crab-type remnants,
named for the Crab Nebula, have a central source, a neutron star,
with jets that fill the interior of the shock cavity.  There are
four stages of the supernova remnant expansion, regardless of the
type of supernovae. The four stages of evolution are free
expansion, adiabatic expansion or Sedov-Taylor phase, radiative,
and constant momentum phase.

During the free expansion phase \citep{che82}, the shock from the
explosion of the star is propagating through the interstellar
medium and sweeping up mass.  At this phase, the remnant is
expanding adiabatically into the ISM.  The temperature of the
remnant scales as
\begin{equation}
T=R^{-3(\gamma-1)}
\end{equation}
where $\gamma$ is the ratio of specific heats.

At this phase the kinetic energy of the shock is converted to
heating of the swept up ISM material.  During this phase, the
remnant can be observed in the x-ray and radio wavelengths.  The
end of this phase is reached when the amount of the material swept
up is equal to that of the mass of the ejecta from the supernova.
This occurs around 1000 years or $\sim$3 parsecs depending on the
density of the interstellar medium and the initial mass of the
ejecta.

The second phase, the Sedov-Taylor phase \citep{sed59}, slows the
bulk velocity of the shock because the mass that has been swept up
is greater than that of the ejecta.  The deceleration of the shock
causes a density build up near then leading edge of the shock. The
gas in this shell becomes supersonic and a reverse shock is formed
between the hot gas of the shell and the inner material of the
supernova remnant. This shock heats the outer portion of the
supernova remnant recycling the kinetic energy lost in adiabatic
expansion back to heat the ejecta. This heating leads to soft
X-Ray emission lines that can be used to study the shock
characteristics as well as the composition of the ejecta and the
interstellar medium. Specifically, the OVI ion is used to trace
temperatures greater than 3 $\times$ 10$^{5}$ Kelvin in the far UV
($\lambda$ = 1032,1038 \AA).

Due to further expansion, the remnant begins to cool. When it is
cooled to around 10$^{6}$K, the material radiates away most of its
internal energy and the remnant enters the radiative phase
\citep{mck77}. The emission of lines of heavy elements become a
key observable at this phase of the supernova remnant.  The
emission decreases as the remnant expands until it fades into
interstellar space or the constant momentum phase. At this time
the velocity becomes subsonic and no longer can continue
supporting a shock.  The constant momentum phase equilibrates the
ejecta with the surround interstellar medium spreading heavy
elements into the ISM.

\goodbreak
\subsection{Supernova 1006}
Supernova 1006 (SN1006), with a well known age, distance
\citep{wink03}, and shock speed \citep{gha02} was chosen to
perform shock studies in this thesis. Located in the southern
constellation Lupus, it was first recorded at its brilliant
optical peak April 30, 1006 A.D. It was recorded by astronomers in
present day China, Iran, Egypt, and Southern Europe.  The remnant
of SN1006, a Type 1a remnant, is large in the sky, about the size
of a full moon. It has been measured to be at a distance of 2.1
kpc \citep{wink03}. With a mean expansion rate of 8700 km
s$^{-1}$, it is $\sim$18 pc wide \citep{wink03}. This young
supernova remnant is entering the Sedov-Taylor phase of supernova
remnant evolution.

\begin{figure}[width=1cm]
  \def\baselinestretch{1.0}
  \centering
  \includegraphics{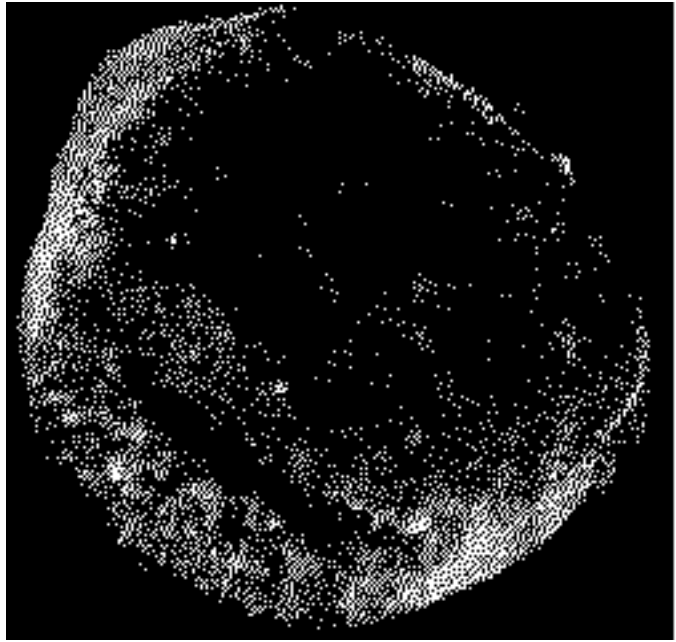}
  \caption{A composite ROSAT HRI image of SN1006 displayed in
false color derived from the ROSAT PSPC spectra. Blue represents
non-thermal emission, red represents thermal emission. Adapted
from \citet{wil96}\label{sn}}
  \def\baselinestretch{2.0}
\end{figure}

SN1006 has been observed at radio \citep{pye81}, optical
\citep{gha02,kwc87,smi91}, ultraviolet\\
\citep{ray95} and X-ray \citep{wink03,lon03,bam03} wavelengths.
Gamma ray observations \citep{tan98} were reported but not
confirmed.  Thin, pure Balmer line filaments were found in the
optical observations. In the radio and X-ray, the remnant has a
limb-brightened shell structure with cylindrical symmetry around a
SE to NW axis probably aligned with the ambient galactic magnetic
field \citep{rey86,jp88}.  The NE shock front of SN1006 shows
strong non-thermal X-ray and possible gamma ray emission while the
NW shock shows very little non-thermal emission at radio or X-ray
wavelengths.  Figure \ref{sn} is an image of SN1006 from the ROSAT
satellite.
\goodbreak

\section{Instrumentation}
\smallskip
Several satellites were used to collect the data for this thesis.
The Far Ultraviolet Spectroscopic Explore (FUSE) \citep{moo00} and
the Advanced Composition Explorer (ACE) \citep{sto98} were the
main satellites, although comparative data was used from Solar and
Heliospheric Observatory (SOHO), Cerro Tololo INTER-AMERICAN
OBSERVATORY (CTIO) 4-m Ground based Optical Telescope, ROSAT X-ray
Telescope, and CHANDRA X-ray Observatory.

The Advanced Composition Explorer (ACE) and the Solar and
Heliospheric Observatory (SOHO) take in-situ measurements of the
solar wind and the shocks that occur within the solar wind.
Observations are made of the atomic emission spectra with the FUSE
Satellite to find out physical processes from the shocks that
occur outside of the heliosphere allowing for the study of shock
parameters of distant astrophysical objects such as SN1006.

\medskip
\subsection{Far Ultraviolet Spectroscopic Explorer(FUSE)}
The UV observation of Supernova 1006 was performed with the Far
Ultraviolet Spectroscopic Explorer (FUSE) Satellite.  The
satellite was launched on June 24, 1999. FUSE is in orbit 760 km
(475 miles) above the Earth.  Its primary objective is to observe
in the ultraviolet from 900-1200 \AA\ .  The spectrograph is
optimized to observe the O VI line in the interstellar medium and
in stars.  A schematic of the spectrograph design is shown below
in Figure \ref{fuse}.

The FUSE spectrometer consists of four independent channels with
two segments each. When photons enter the instrument they are
directed onto one of four different mirrors of the spectrograph.
These photons are then reflected onto four different gratings.
These gratings then reflect the light into four distinct
wavelength regions on two detectors. Four of these eight segments
operate in the wavelength range for the O VI doublet,
$\lambda$=1031.91, 1037.61 \AA. However, the Silicon Carbon (SiC)
coated channels, because they are optimized for $\lambda$ $\le$
1020 \AA, add an unacceptable amount of noise to the faint signal,
so only the Lithium Fluoride (LiF) channels are used.  These two
segments are designated LiF1A and LiF2B.  The LiF1A channel covers
wavelengths 987.1 - 1082.3 \AA, while the LiF2B covers
979.2-1075.0 \AA.

\begin{figure}
\def\baselinestretch{1.0}
  \centering{
  \epsfig{file=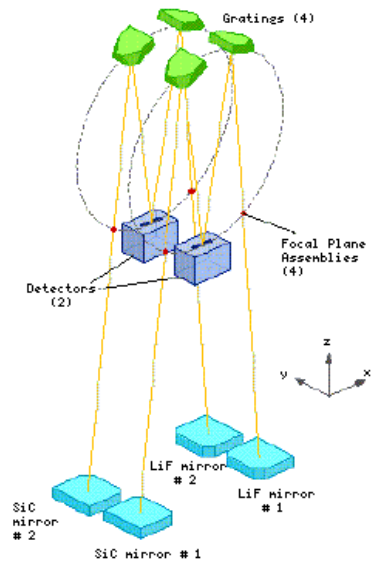,width=5cm}
  \caption{Schematic of the UV Spectrometer aboard the FUSE
  Satellite. From the FUSE website fuse.pha.jhu.edu/support/guide/guide.html\label{fuse}}}
  \def\baselinestretch{2.0}
\end{figure}

Data collected from the detector are processed through the FUSE
Pipeline in order to extract the photons per wavelength
information that can be analyzed for spectral emission
information.

\medskip
\subsection{Advanced Composition Explorer (ACE)}
The Advanced Composition Explorer satellite (ACE) provided the
data used for the Coronal Mass Ejection study.  ACE was launched
from a Delta II rocket in August 1997. ACE orbits the L1 point,
the point where the gravitational forces of the Earth and the Sun
are equal to the centripetal force required for the spacecraft to
rotate with them, keeping the position between the Sun and Earth
constant, about 1.5 million km from Earth and 148.5 million km
from the Sun. In its elliptical orbit, ACE can readily view the
Sun and the galactic region beyond the Sun.

Of ACE's suite of nine instruments, three were used in the coronal
mass ejection study, the Solar Wind Ion Composition Spectrometer
(SWICS), the Solar Wind Electron, Proton, and Alpha Monitor
(SWEPAM), and the Magnetometer instrument (MAG).

\begin{figure}[width=0.25cm]
  \def\baselinestretch{1.0}
  \centering
  \includegraphics{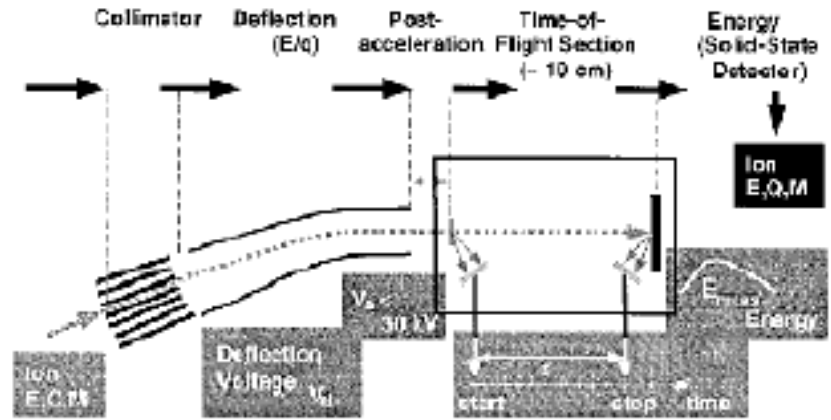}
  \caption{Schematic of time of flight setup for the SWICS instruments
  aboard the ACE Satellite. Adapted from \citet{glo98}\label{swics}}
  \def\baselinestretch{2.0}
\end{figure}

SWICS performs measurements of the chemical and ionic composition
of the solar wind \citep{glo98}.  This instrument uses
electrostatic analysis followed by a time-of-flight region and an
energy measurement, seen in Figure \ref{swics}, separating all
heavy components of the solar wind providing unique identification
of up to 40 ions.

SWEPAM measures the solar wind plasma electron and ion fluxes
(rates of particle flow) as functions of direction and energy
\citep{mcc98}. These data provide detailed knowledge of the solar
wind conditions and internal state every minute.  SWEPAM provided
temperatures, solar wind speed, and proton thermal speeds.

Electron and ion measurements are made with separate sensors. The
ion sensor measures particle energies between about 0.26 and 36
KeV, and the electron sensor's energy range is between 1 and 1350
eV. Both sensors use electrostatic analyzers with fan-shaped
fields-of-view. The electrostatic analyzers measure the energy per
charge of each particle by bending its flight path through the
system. The fields-of-view are swept across all solar wind
directions by the spin of the spacecraft.  This allows the
instrument to measure the mass, based on position on the detector,
and energy based on time of flight in the detector.

MAG is a magnetometer that is able to detect the magnitude as well
as the direction of the magnetic field \citep{smi98}. The basic
instrument is a twin triaxial fluxgate magnetometer system. The
two identical sensors are on booms that extend past the end of
diametrically opposite solar panels. The instrument measures small
fluctuations in the magnetic field. It is important to know the
magnetic field because the magnetic field direction and strength
are crucial to understanding shock geometry and the solar wind
flow properties.

\bigskip
\goodbreak
\section{Specific Topics in this Thesis}

This thesis examines the heating of particles as they pass through
a collisionless shock.  Specifically, the heating of heavy ions
and neutral particles at the shock front will be examined.  Three
parameters of the collisionless shocks, Mach number, M$_{A}$, the
orientation of the magnetic field to the shock normal,
$\theta$$_{Bn}$, and the plasma $\beta$ have been identified as
the important characteristics in heating at a shock front.  Using
these three parameters and other supplemental data, three
different systems are examined to explain the heating and
acceleration mechanisms of heavy ions in shocks.

\subsection{Heavy Ion Heating in Collisionless Shocks}
It was found by \citet{ber97} that the heating in shocks is not
proportional to mass, as would be found by bulk thermalization of
energy, but 1.2-3.0 times mass proportional for oxygen.  This
differential heating is of importance to understanding the
kinetics of the collisionless shock front as well as the
acceleration of particles.  The heated ion species are also needed
as a seed population for acceleration of particles to cosmic ray
energies. ACE satellite data provides plasma measurements of the
thermal speeds of several species of heavy ions. These ions were
shown to be heated preferentially to protons.  These results are
contrasted with those of supernovae shock studies
\citep{kor04,ray95}. Here heavy ions are heated less than mass
proportionally to the protons.  SNRs are known to be cosmic ray
accelerators.  However, the seed population is not well understood
and the less than mass proportional heating does not favor a
thermal seed population for cosmic ray acceleration.  The
differences in speed and density of upstream material which can be
represented by the plasma $\beta$, all play a role in the heating
mechanisms. By studying the parameter space afforded by CMEs and
SNR shocks a mechanism for the heating is sought.

\subsection{Neutrals at Collisionless Shock Fronts}
Neutrals at a collisionless shock front could act as a precursor
to the shock or as pick up ions.  Neutral atoms that can go
through the shock upstream from the downstream area can modify the
ramp structure of the shock front. Fast neutral atoms from
downstream that can avoid being affected by the shock's magnetic
field can flow upstream creating a precursor that would pre-heat
the shocked material.  The dynamics of the neutrals at the shock
front are of great interest in acceleration mechanisms as they
could have high energies creating a seed population for cosmic ray
acceleration. In \citet{cr78}, the authors describe the mechanism
for understanding and tracing the neutrals in the shock front. The
H$\alpha$ emission line is made up of two components when a
significant fraction of neutrals are present.  The two components
are a broad component made in two steps and a narrow component
that is made up of line emission from excited hydrogen atoms.  The
two steps to create the broad component are as follows:  first a
downstream proton must charge exchange with a neutral to become a
fast neutral.  Next, the fast neutral must be excited.  When the
fast neutral is excited it gives off the H$\alpha$ emission with a
shift according to the speed of the particle.  Since the
excitation is highly dependant on the proton and electron density
and energy, the H$\alpha$ intensity ratio is a tracer for the
plasma characteristics as well as the neutral fraction.

Several attempts to understand the effect of this sometimes minor
population of particles have been modeled by \citet{lim95} and
\citet{lim96}. Although the simulations did not match the observed
broad to narrow H$\alpha$ components, the distribution of neutrals
after the simulation was a ring distribution similar to that of a
pickup ion distribution. Since a neutral medium is rare in the
heliosphere, the neutral modeling is directed at understanding the
shocks such as those in SNRs that interact with the neutral ISM
material.

\section{Thesis Overview}

\subsection{Heating of Ions in the Shock of SN1006}
In Chapter Two, ultra-violet spectral observations from SN1006 are
discussed. The data from the FUSE satellite show strong OVI
spectral lines which are an indicator of temperatures in the shock
of over 100 million degrees. The non-radiative, thermal
collisionless shock of the Northwest region of the supernova
remnant and the non-radiative, non-thermal collisionless shocks in
the Northeast region of the supernova remnant will be contrasted.
A discussion of this specific collisionless shock will follow with
respect to the ion heating, neutral fraction of the pre-shock
medium and the role of turbulence in the shock front. This section
is based on \citet{kor04}.

\subsection{CME shock ion heating}
Chapter Three leads to analysis of Coronal Mass Ejections'
collisionless shocks through situ measurements. Using the ACE
satellite data from SWICS, MAG, and SWEPAM instruments, over 20
shocks were studied. Shocks were first classified as perpendicular
or parallel as this has been shown to be a parameter that greatly
changes the heating. The heating of the heavy ions, He$^{+2}$,
C$^{+5}$, C$^{+6}$, O$^{+6}$, O$^{+7}$, Fe$^{+10}$, were used as a
measure of heating versus Mach number and plasma $\beta$.  In
addition to the analysis of data, the parallel shock work was used
to test the theoretical model of the Rankine-Hugoniot conditions
for ions in parallel shocks laid out by \citet{bur91}.

\subsection{Neutral Atoms at the Shock Front: A Monte Carlo Model}
Chapter Four summarizes the Monte Carlo modelling done in order to
understand the neutrals present in collisionless shock fronts. The
main objective of the project was to simulate a broad to narrow
H-$\alpha$ component intensity ratios that would be affected by
the heating by neutrals. The effect of magnetic angle, initial
ionization fraction, shock speed, and equilibration between
electrons and protons is discussed.  Spectra were simulated from
the model and compared with observations made by \citet{smi91} of
other supernova shocks and simulations done by \citet{lim96}.

\subsection{Summary}
Chapter Five summarizes the contribution that this work makes to
the understanding of the heating processes in collisionless shocks
and outlines future work. The work in this thesis is the most
comprehensive study of heavy ion heating in shocks.  The knowledge
gained from this study impacts not only the system of the specific
study but also the remote sensing of shocks in the extreme
ultraviolet and X-rays wavelengths.  The study shows an ideal
example of the use of remote and in situ data to study a
fundamental physical phenomena.


\chapter{SN1006 Collisionless Shock Fronts}

\def\baselinestretch{2.0}

\goodbreak

\section{Introduction}
\smallskip

    SN1006 (G327.6+14.6) is a nearby Type $\textrm{I}$a supernova
remnant at a distance of 2.1 kpc \citep{wink03}.  With a mean
expansion rate of 8700 km s$^{-1}$ it is $\sim$18 pc wide
\citep{wink03}. The remnant has a high Galactic latitude and
modest foreground reddening, E(B-V)=0.11 $\pm$ 0.02 \citep{sch80}.
This young supernova remnant is entering the Sedov-Taylor phase of
supernova remnant evolution.
    SN1006 has been observed at radio \citep{pye81}, optical
\citep{gha02,kwc87,smi91}, ultraviolet\\ \citep{ray95} and X-ray
\citep{wink03,lon03,bam03} wavelengths. Gamma ray observations
\citep{tan98} were reported but not confirmed.  Thin, pure Balmer
line filaments were found in the optical. In the radio and X-ray,
the remnant has a limb-brightened shell structure with cylindrical
symmetry around a southeast (SE) to northwest (NW) axis probably
aligned with the ambient galactic magnetic field
\citep{rey86,jp88}.  The NE shock front of SN1006 shows strong
non-thermal X-ray and possible gamma ray emission while the NW
shock shows very little non-thermal emission at radio or X-ray
wavelengths.
\smallskip
    Ly-$\beta$, He II, C VI, and O VI lines were observed from the
faint optical Balmer line filament of the NW shock of the
supernova remnant, by the Hopkins Ultraviolet Telescope (HUT),
flown during the Astro-2 space shuttle mission. The observed FWHM
of the lines were 2230, 2558, 2641 km s$^{-1}$, respectively (the
O VI line width could not be measured).  A kinetic temperature
could be calculated from these line widths. The kinetic
temperatures of these species are not equal, because the line
widths do not scale inversely with the square root of their atomic
mass. Instead, the UV observations do suggest that $T_{ion} \sim
\frac{m_{ion}}{m_{p}} T_{proton}$ indicating lack of temperature
equilibration between species \citep{gha02}.

\smallskip
    SN1006 provides an opportunity to investigate parameters of
non-radiative collisionless shocks faster than 2000 km s$^{-1}$.
Collisionless shocks appear in many astrophysical phenomena, from
coronal mass ejections (CMEs) in the heliosphere to jets in
Herbig-Haro objects.  When a shock is non-radiative the detection
of emission from the shock front is possible, as all of the
optical and UV emission of a non-radiative shock comes from a
narrow zone directly behind the shock front. Interactions at the
collisionless shock front depend upon mechanisms such as plasma
waves to transfer heat, kinetic energy and momentum, and it is not
well understood how particles of different masses and charges are
affected by these processes. The temperature of the species and
the degree of temperature equilibration between electrons, protons
and other ions are central to the interpretation of X-ray spectra,
which effectively measure electron temperature. The energy
distribution of a particle species is important to cosmic ray
studies as only those particles at a high energy tail of a
particle distribution are available for cosmic ray acceleration.

\smallskip
    The method of using H$\alpha$ lines to determine collisionless
shock parameters was originated by Chevalier \& Raymond(1978) and
Chevalier, Kirshner, \& Raymond (1980). The H$\alpha$ line has a
two component profile. The width of the broad component of the
H$\alpha$ line is related to the post-shock proton temperature as
a result of charge exchange between neutrals and protons, which
produces a hot neutral population behind the shock. The narrow
component of the H$\alpha$ line is produced when cold ambient
neutrals pass through the shock and emit line radiation before
being ionized by a proton or electron.  The ratio of the broad to
narrow flux is sensitive to electron-ion equilibrium and the
pre-shock neutral fraction. The FWHM of H$\alpha$ line was
measured to be 2290 $\pm$ 80 km s$^{-1}$, with models implying the
speed of the shock is v$_{shock}= 2890 \pm 100$ km s$^{-1}$
\citep{gha02}. The H$\alpha$ broad to narrow intensity ratio
measured to be 0.84 implies an electron temperature much lower
than the ion temperature.

    This UV observation from the FUSE satellite focused on the shock
front in the NW observed by \citet{ray95} and \citet{gha02} and on
a region in the NE dominated by non-thermal emission.  From the
spectra, a broad Lyman $\beta$ line ($\lambda$$\sim$1025 \AA) and
the doublet of O VI ($\lambda$$\sim$1032, 1038 \AA) were analyzed
for spectral width, intensity, and flux.  We use the line widths
of the NW and the intensities of the O VI lines in the NE and NW
shock fronts to compare the electron-ion and ion-ion temperature
equilibration efficiencies as well as densities.  The heating of
different particle species by the shock front as well as
parameters of collisionless shocks that affect particle species
heating will be discussed.

\goodbreak
\section{Observations}
\smallskip

    The Far Ultraviolet Spectroscopic Explorer (FUSE) has a wavelength
range of approximately 900-1180 \AA.  The Large Square Aperture
(LWRS), with a field-of-view of 30" x 30", with a roll angle of
167$^{o}$, was chosen for this observation because models
predicted that the O VI emission behind the shock would be spread
over 35" \citep{ray95,lam96}.  The LWRS has a filled-aperture
resolution of about 100 km s$^{-1}$.

\begin{figure}[h]
  \def\baselinestretch{1.0}
\centering
\includegraphics[width=12 cm]{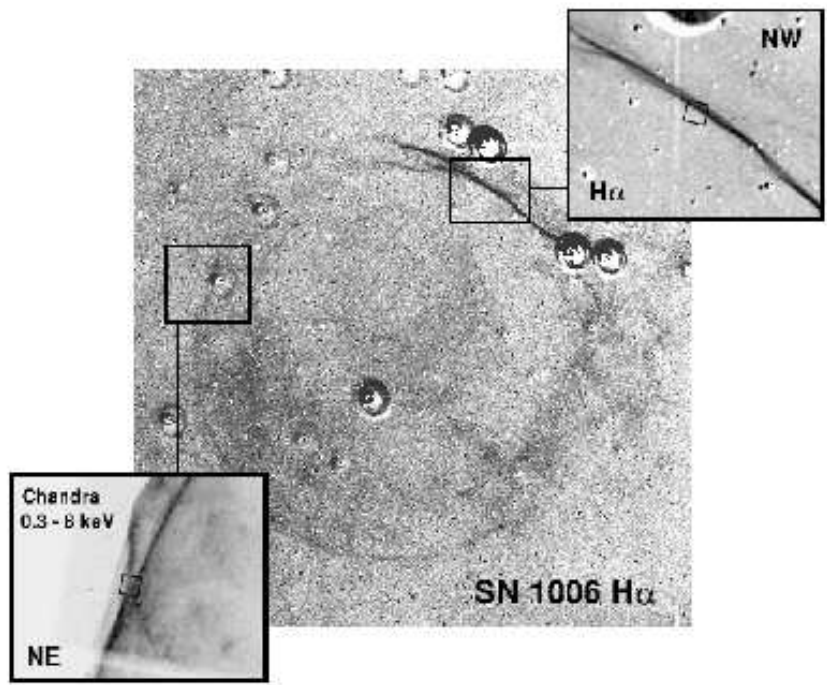}
\caption{H$\alpha$ image of SN1006 taken by the CTIO Schmidt
telescope from Winkler et al. (2003). Closeup images of the
observed filaments are shown in the insets. In each case the
interior box, drawn to scale, shows the location of the 30" x 30"
FUSE LWRS aperture.  The NW blowup is from the same H$\alpha$
image while the NE blowup is a 0.3-8 keV Chandra image (Long et
al. 2003).} \label{halpha}
  \def\baselinestretch{2.0}
\end{figure}

    Although the northwest region of the remnant has been observed
before in the UV \citep{ray95}, we have much better spectral
resolution and a more optimal aperture size to include the entire
ionization region given that it may be larger than 19"
\citep{lam96}. The apertures used for past
observations were 19" x 197" \citep{ray95} in HUT and 2" x 51" CTIO RC Spectrometer \\
\citep{wink03,gha02}. We positioned the aperture center to be
5"-10" behind the H$\alpha$ filament where the peak formation of O
VI occurs. The NE position was chosen based on the edge of the
X-ray filament from Long et al.(2003).
\smallskip
    FUSE observations of the northwest region, centered at
$\alpha_{2000}$ =$15^{{h}}$ $2^{{m}}$ $19.17^{{s}}$,
$\delta_{2000}$ =-41$^o$ 44' 50.4", were obtained on 23 June 2001
and 26 February 2002 with total exposure times of 35,627 s and
6,690 s. Observations of the northeast region, centered at
$\alpha_{2000}$=$15^{{h}}$ $4^{{m}}$ $5.0^{{s}}$,
$\delta_{2000}$=-41$^o$ 50' 40.5", were obtained on 25 June 2001
and 27 February 2002 with exposure times of 42,365 s and 9,666 s.
The locations of observations are shown superimposed on an
H$\alpha$ image of the remnant taken with the CTIO Schmidt
telescope in Figure \ref{halpha} \citep{wink03}.

Inserted in the figure is a close up from Chandra \citep{lon03} of
the NE region of observation to illustrate the x-ray morphology,
although no optical emission is obviously present.

    There are four components to the background of this observation;
detector background, geocoronal lines, the diffuse galactic UV
continuum and diffuse galactic O VI emission.   The background
count distribution on the FUSE detectors is composed of two
separate components \citep{fuse03}.  The `intrinsic' background
forms from the $\beta$-decay of potassium in the microchannel
plate (MCP) detector glass and the spacecraft radiation
environment. The effect of the spacecraft radiation environment on
the detector background varies from night to day and with solar
activity, but over a short observing time this variation is not
significant.  The second component is caused by scattered light,
primarily geocoronal Ly-$\alpha$. This line produces detector
averaged count rates as small as 20$\%$ of the intrinsic
background during the night and increasing to 1-3 times the
intrinsic rate during the day.  The other two components of the
background, galactic UV emission and diffuse O VI emission, will
be discussed later.

\smallskip
    The observations were calibrated with the CalFUSE Pipeline Version
2.2.1. Data from all exposures are processed through the pipeline
and then co-added following the FUSE Data Analysis Cookbook and
The FUSE Observer's Guide. The data were selected to contain only
the night observations. This greatly reduces the geocoronal
background. The night-only exposure times were 32,287 s for the
Northwest and 39,386 s for the Northeast.

\goodbreak
\section{Analysis and Results}
\smallskip

    As mentioned above, the background consists of detector noise,
geocoronal lines, diffuse galactic O VI and an astrophysical UV
continuum. The first two sources were explained in the previous
section, but the additional diffuse UV continuum must be treated
separately. It does not originate from SN1006, as it is seen in
both of the entirely different regions of the remnant; the NE
shock and the NW shock. The diffuse background is attributed to
light from hot stars scattering on dust.  The diffuse UV continuum
is especially bright in this region of the sky according to models
by \cite{mur95}.  A value of $8.4\times 10^{-15}$ erg cm$^{-2}$
s$^{-1}$ \AA$^{-1}$ was quoted by \cite{ray95} while we are seeing
approximately $6.5\times 10^{-15}$ erg cm$^{-2}$ s$^{-1}$
\AA$^{-1}$ through an aperture one quarter the size of the HUT
observation.

\begin{figure}[h]
  \def\baselinestretch{1.0}
\centering
\rotatebox{90}{\resizebox{!}{12cm}{\includegraphics{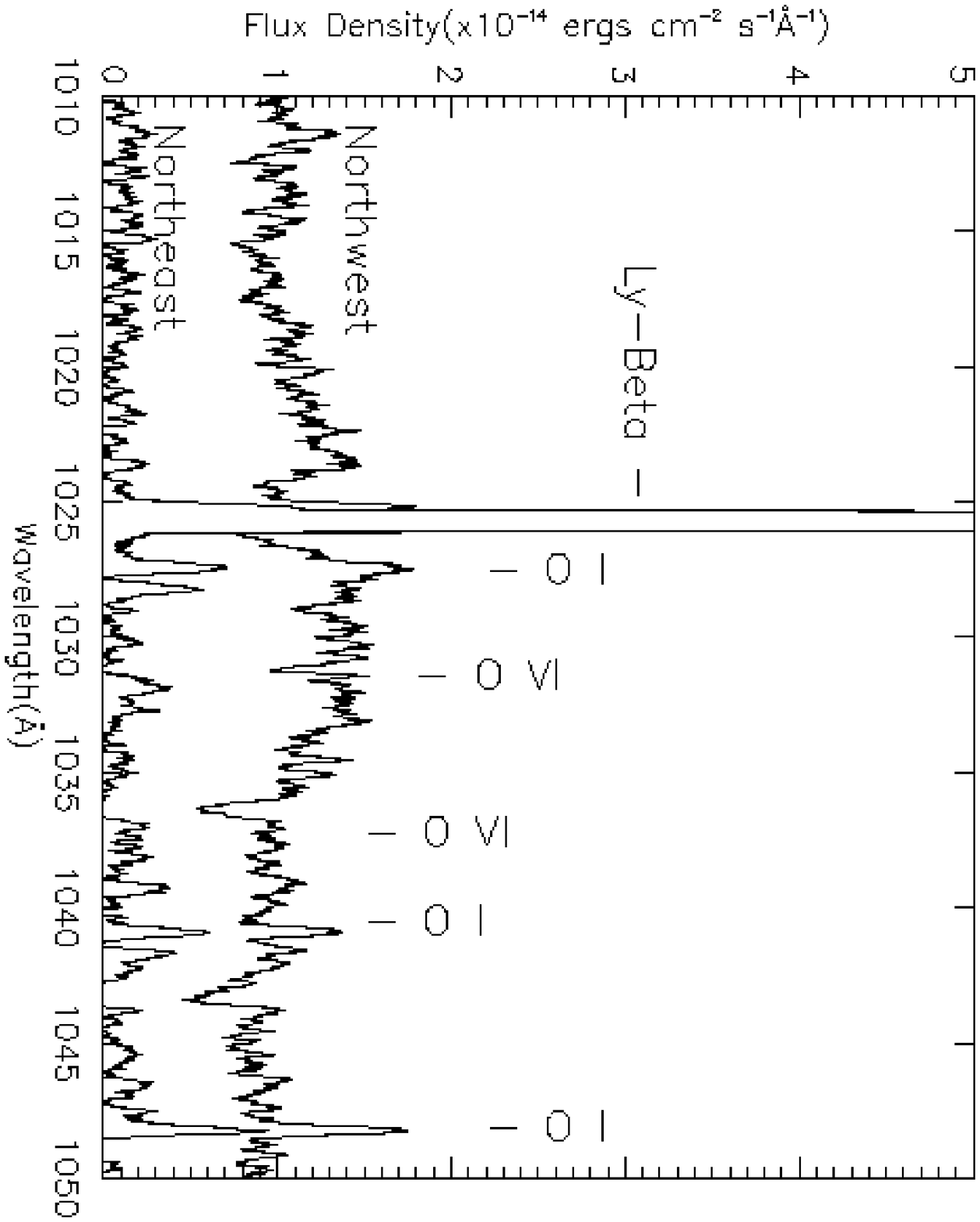}}}
\caption{Raw FUSE spectra from the Northeast and Northwest region
of SN1006.  The NW is offset from the NE by 0.5 for clarity.
Geocoronal lines are marked.  The Ly-$\beta$ peak dominates in
both spectra. The NE is consistently fainter than the NW, but the
geocoronal line intensities are similar.}
\label{fullspectra}
  \def\baselinestretch{2.0}
\end{figure}

    In addition to the diffuse UV continuum, Shelton et al. (2001,2002) and
\citet{ott03} have found a diffuse O VI background.  The
brightness of the O VI background is 4700 $\pm$ 2400 photons
cm$^{-2}$ s$^{-1}$ sr$^{-1}$ \citep{ott04}. The widths of the
diffuse O VI lines fall between 10 and 160 km s$^{-1}$.  In the
current NE spectrum diffuse O VI emission has a width of $\le$ 200
km s$^{-1}$ and a brightness of 3500 photon cm$^{-2}$ s$^{-1}$
sr$^{-1}$.  We attribute the NE emission to the diffuse galactic O
VI background. This enabled us to subtract the NE as a background
from the NW data to further eliminate airglow lines, the diffuse
UV emission and the galactic O VI background. The intensities of
the airglow lines at 1042\AA\ and 1048\AA\ are quite similar in
both the NE and NW, further allowing this subtraction.  The raw
spectra of the NW and the NE regions are shown in Figure
\ref{fullspectra}, with airglow lines marked.

    For the NW region, a nonlinear chi-squared minimization routine
was used to fit Gaussian line profiles to the spectra. The
wavelengths considered for analysis were restricted to 1010-1050
\AA\ to minimize spurious background effects near the ends of the
detector's spectral range. The data were binned by 0.1 \AA\ to
increase the number of counts per bin without losing resolution,
as the line widths were several Angstroms wide. The width of the
broad H$\alpha$ line, from \citet{gha02}, is v$_{H}$=2290 km
s$^{-1}$. Since the Ly-$\beta$ line is formed by the same process
\citep{kcr}, its line width was set equal to the H$\alpha$ broad
component width. The shift of the centroid of the broad and narrow
component of H$\alpha$, v=29 km s$^{-1}$, is effectively
negligible (implying that the shock is viewed completely edge-on)
so the broad Ly-$\beta$ line centroid was fixed at its rest
wavelength. Only the intensity of the line was a free parameter.
The blue wing of the line was fit from 1010 \AA\ to 1024.5 \AA.
Due to the extinction from interstellar dust, a correction factor
must be applied to deredden the observed flux.  Using the
extinction curves of Cardelli, Clayton, \& Mathis (1989), the
resulting dereddened Ly-$\beta$ flux is 2.3 $\pm$ 0.3 x
10$^{{-13}}$ erg cm$^{-2}$ s$^{-1}$.

    After subtracting the fitted broad Ly-$\beta$ line profile, the
wavelength range from 1022-1028 \AA\ was excluded from the fitting
routine in order to avoid negative fluxes and residual airglow
that would skew the gaussian fits of the O VI lines. At
$\sim$1037.0 \AA\ there were absorption features present that
coincided with a C II line and molecular hydrogen lines, along
with an O I airglow line. The absorption feature with the spectral
range from 1035 - 1038 \AA\, was therefore excluded from the fit.

\begin{figure}[h]
  \def\baselinestretch{1.0}
\centering
 \rotatebox{90}{\resizebox{!}{12
cm}{\includegraphics{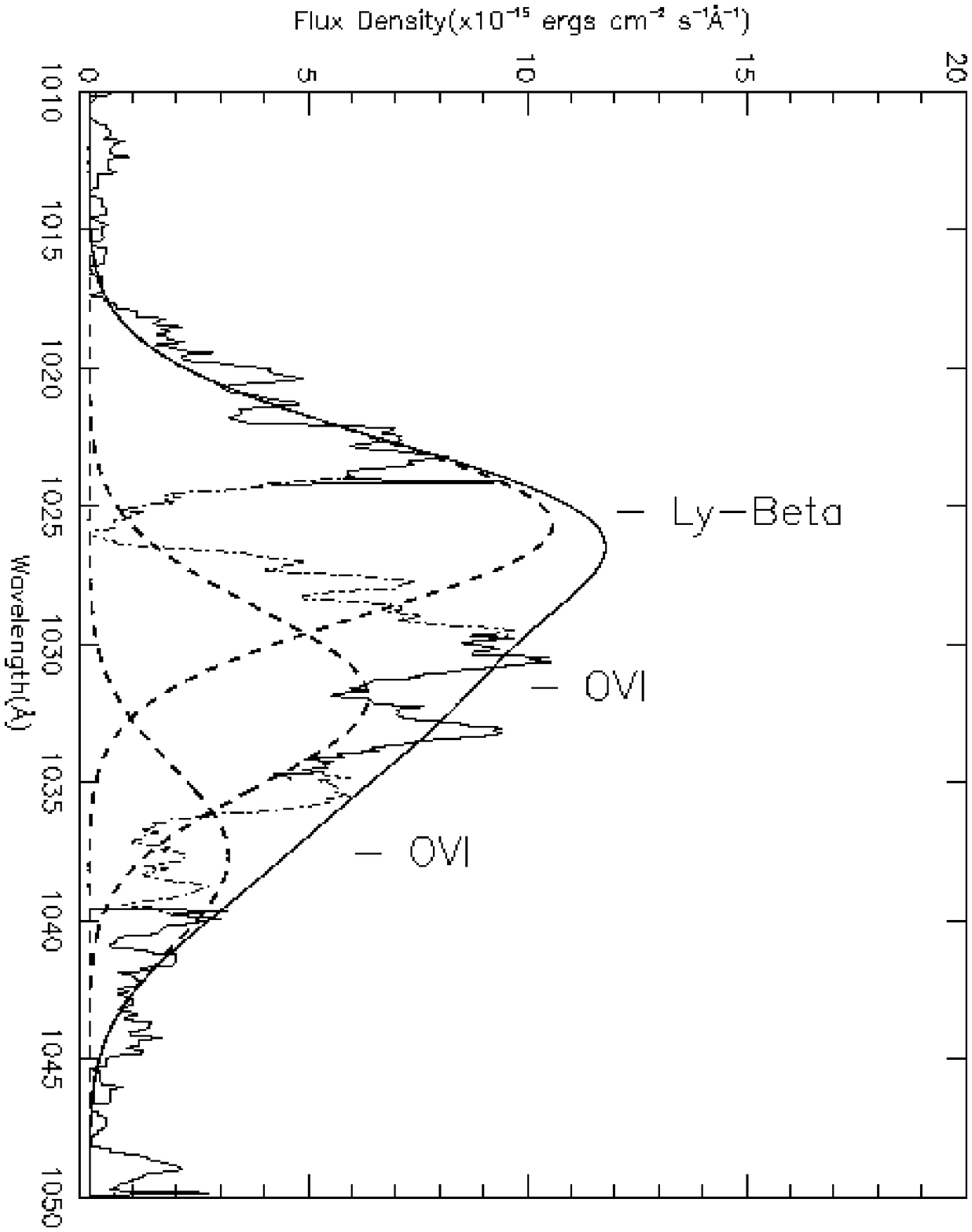}}}
 \caption{FUSE spectra from the NW, binned at 0.1 \AA\, with
the NE subtracted as background.  The dotted dashed lines are the
regions of the spectra that were excluded from the fits.  The
dashed lines are the fits for the Ly-$\beta$, O VI 1032 and 1037
\AA\ lines, with FWHM of 2290 and 2100 km s$^{-1}$ respectively.
The solid line represents the addition of the fits of the three
spectral lines.}\label{fit1}
  \def\baselinestretch{2.0}
\end{figure}

    The O VI doublet was fit with two gaussians with fixed centers at
1031.91 and 1037.61 \AA\ respectively corresponding to the
centroid of H$\alpha$. The doublet line intensities were forced to
have a 2:1 ratio but the magnitude of the intensities were allowed
to vary. The observed flux is 6.7 $\pm$ 0.1 x 10$^{-17}$ erg
cm$^{-2}$ s$^{-1}$ arcsec$^{-2}$. Total dereddened flux for the O
VI doublet lines was 1.8 $\pm$ 0.2 x 10$^{-13}$ erg cm$^{-2}$
s$^{-1}$. The O VI line widths were measured to be 7.2 $\pm$ 0.4
\AA\ FWHM, or equivalently 2100 $\pm$ 100 km s$^{-1}$. The formal
error on the fit is 100 km s$^{-1}$. However, due to systematic
error a more conservative error of $\pm$ 200 km s$^{-1}$ is used.
The fits are shown in Figure \ref{fit1}.

    The width is within the limiting estimate of \cite{ray95} of $\le$
3100 km s$^{-1}$ and is within 1$\sigma$ error of the H$\alpha$
width of 2290 km s$^{-1}$. Although the faint signal in the NE did
not allow for a statistically significant fit, an upper limit of O
VI intensity was found assuming a width of 2000 km s$^{-1}$. The
observed upper limit on the O VI line is 1.6 x 10$^{-17}$ erg
cm$^{-2}$ s$^{-1}$ arcsecond$^{-2}$.  An upper limit on the
dereddened intensity of O VI in the NE is 4.2 $\times$ 10$^{-14}$
erg cm$^{-2}$ s$^{-1}$. The upper limit of flux for Ly-$\beta$ in
the NE is 1.6 x 10$^{-17}$ erg cm$^{-2}$ s$^{-1}$
arcsecond$^{-2}$. An upper limit to the dereddened Ly-$\beta$
intensity in the NE region is 4.4 $\times$ 10$^{-14}$ erg
cm$^{-2}$ s$^{-1}$.

    Past observations of SN1006 line widths and intensities are
summarized in Table 1.
    In order to compare past measurements made with varying aperture
sizes, we use intensity per arcsecond measured along the length of
the filament.  The Ly-$\beta$ from the HUT and the current FUSE
observation are consistent.  We can use the various measurements
to study the ion heating.  The proton temperature was found using
the shock speed of 2890 km s$^{-1}$ from \citet{gha02}.  This
proton temperature was then multiplied by m$_{ion}$/m$_{p}$ to
calculate the mass proportional temperatures. These calculated
temperatures were then compared to the temperatures given by using
the FWHM of each ion line.  The temperature of O VI as indicated
by its FWHM is less than mass proportional by 48\%.  For the other
ions, He II, C IV the heating was also less than mass
proportional, by 21\% and 18\% respectively.

\clearpage
\begin{table}[h]
  \def\baselinestretch{1.0}
\centering
\caption{Summary of UV Emission Lines in NW Filament of
SN1006}
\medskip
\begin{tabular}[h]{|c|c|c|c|c|c|c|}
\hline
Ion&Intensity\footnotemark[1] & Filament& FWHM &Temperature & m$_{ion}$/m$_{p}$T & \% \\
& &Length & (km s$^{-1}$) &(Kelvin)&(Kelvin) &Mass\\
&(x10$^{-4})$ &(arcsec) &Observed&from FWHM & & Prop\\
\hline
 H-$\alpha$\footnotemark[2] &2.1 &51 & 2290 $\pm$ 80&(1.8 $\times$
10$^{8}$ )\footnotemark[3]& - & - \\ \hline

Ly-$\beta$ &4.0 &30 & 2290(fixed)&   &  &  \\ \hline

He II\footnotemark[4] &0.99 &197 & 2558$\pm$618& 5.7 $\times$
10$^{8}$ & 7.2 $\times$ 10$^{8}$  & 79\% \\\hline

C IV\footnotemark[4] &1.7& 197& 2641$\pm$ 355& 1.8$\times$
10$^{9}$ &2.2$\times$ 10$^{9}$ &82\% \\\hline

O VI &3.1&30 &2100 $\pm$ 200&1.5$\times$ 10$^{9}$ &2.9$\times$
10$^{9}$ &52\% \\ \hline

O VII\footnotemark[5]& & 60 &1775$\pm$261& 1.1 $\times$ 10$^{9}$
&2.9$\times$ 10$^{9}$ & 38\% \\\hline
\end{tabular}
  \def\baselinestretch{2.0}
\end{table}
\def\baselinestretch{1.0}
 \footnotetext[1]{photons cm$^{-2}$s$^{-1}$arcsec$^{-1}$}
\footnotetext[2]{Ghavamian et al. 2002}
\footnotetext[3]{Temperature derived from shock speed of 2890 km
s$^{-1}$.} \footnotetext[4]{Raymond et al. 1995}
\footnotetext[5]{Vink et al. 2003}
\def\baselinestretch{2.0}

    The brightness of the O VI lines is proportional to density,
n$_{0}$, and the depth of the filament along the line of sight.
Therefore, an upper limit to the density in the NE can be found by
the ratio of intensities provided that the depths along the line
of sight are known. \cite{lon03} calculated a density ratio of
n(NW)/n(NE) = 2.5.  From the thermal component of the Chandra
X-ray spectra \cite{lon03} estimated a pre-shock ISM density of
0.25 cm$^{-3}$ in the NW. Using the limit to the O VI intensity
ratio of the NW and NE a ratio of the densities is found to be
n(NW)/n(NE) $\ge$ 4, which is within the uncertainties of the Long
et al. calculations. Therefore, assuming a pre-shock density in
the ISM of 0.25 cm$^{-3}$ in the NW, the pre-shock NE density
$\le$ 0.06 cm$^{-3}$.  This density calculation depends on the
assumptions of similar depths along the line of sight in the NE
and the NW and of similar numbers of O VI photons per atom passing
through the shock.  The amount of electron-ion equilibration in
the NE would affect these assumptions.  Greater electron-ion
equilibration in the NE would increase the number of O VI photons
per atom \citep{lam96}, so the limit on the density in the NE
would be even smaller.  We attribute the low upper limit on the O
VI intensity in the NE to the low density medium into which the
remnant is expanding.

\goodbreak
\section{Discussion}
\smallskip

    O VI lines were not conclusively observed in the faint
non-radiative non-thermal NE shock indicating that the two
distinct shock regions heat ions differently.  Ion heating is
important to cosmic ray acceleration and the overall energy
distribution of the system.  The ions have most of their kinetic
energy in a broad distribution which is generally non-Maxwellian
as the time to equilibrium via Coulomb collisions for ions and
protons is 1.2 $\times$ 10$^{5}$ years \citep{spi56}.  To
understand the heating at the shock front, turbulence, line
widths, methods of calculating heating, and the role of neutrals
at the shock front will be discussed.

\subsection{Small Scale Turbulence}
\smallskip

    Turbulence plays a role in the evolution of fast shocks in
supernova remnants \citep{rey04,ell92}.  Small scale turbulence
spreads the line profile of an ion much like thermal broadening of
a line profile.  Since some of the shock energy must be used for
bulk flow, we will examine turbulence with a velocity of 1500 km
s$^{-1}$ which is large enough to affect the spectra but not
contain all the energy of the flow. Turbulence decays on a time
scale proportional to the characteristic length of the turbulence
divided by the velocity of the turbulence $\sim$ $\ell$/v
\citep{ten57}. The width of the H$\alpha$ filament is at most
10$^{16}$ cm based on its 1" apparent width on the sky
\citep{gha02}, making the time scale of the turbulence 10$^{8}$ s
$\sim$ 3 years. Using this decay time and the post-shock speed of
750 km s$^{-1}$, one quarter of the shock speed, the post-shock
region affected by turbulence would be 7.5 $\times$ 10$^{15}$ cm.
The O VI filament with an observed width of 3 $\times$ 10$^{17}$
cm, assuming the 30"  FUSE aperture is filled, is also too wide to
be dominated by turbulence.  Thus the turbulence that is present
in the shock of SN1006 is short-lived and not a major source of
line broadening.
\medskip
\subsection{Line Widths of O VI, UV lines and H$\alpha$}
\smallskip
    The UV line profiles of the current observations can be compared
with past observations of various ion species. The currently
observed O VI line width in the NW shock is within 1$\sigma$ of
the H$\alpha$ line width previously measured by \citet{gha02}.
\cite{vink03} measured an O VII line width of 3.4 $\pm$ 0.5 eV, or
approximately 1775 $\pm$ 261 km s$^{-1}$ from a different
northwest region. This line width is substantially narrower than
those of other ion species measured thus far, although the region
of observation for this measurement is different from the position
of our observations. Along a 124" slit, \cite{smi91} found little
variation in the H$\alpha$ profiles, indicating that the oxygen
temperature does not vary significantly along the length of the NW
filament.
    This implies one of two processes. First, the line
width could decline with ionization state and distance behind the
shock due to Coulomb collisions, as Coulomb collisions would
transfer heat to other species.  In Section 2.4.1, we found the
Coulomb collision time to be far too long for this process to be
important.  The second more probable scenario is that some of the
lower temperature O VII is from the reverse shock in the supernova
ejecta. The detection of Si XIII and Mg XI X-ray lines
\citep{lon03} in the NW region of the remnant agrees with the
hypothesis that the emission is coming from ejecta near the shock
front.

    The proton temperature quoted thus far used the width of
the H$\alpha$ line.  However, the proton thermal speed is not
simply equal to the velocity derived from the width of the
H$\alpha$ line at high temperatures.  The cross section for
neutral-proton charge transfer, the process that produces the
broad H$\alpha$, falls off at high energies allowing for the
neutral hydrogen distribution function to be narrower than that of
the protons \citep{kcr}. This results in an H$\alpha$ profile that
would incorrectly indicate a lower temperature than the actual
proton temperature.

\medskip
\subsection{Heating at the Shock Front}
Using the current observations the temperatures of the ions are
calculated in two ways.  The first method to calculate the
temperature is based on the thermalization of the bulk velocity of
the shock.  The second method uses the FWHM of the gaussian line
fits as the thermal velocity that can be used to find the
temperature.  The shock species is heated by bulk thermalization
to a kinetic temperature described by the following equation,
\begin{equation}
kT_{i}=\frac{3}{16}m_{i}v_{shock}^{2}
\end{equation}

\noindent where the subscript $i$ indicates the species, $k$ is
the Boltzman constant, $T$ is temperature, $m_{i}$ is the mass of
the species and $v_{{shock}}$ is the shock speed = 2890 km
s$^{-1}$ \citep{gha02}. This gives a temperature for O VI of 2.9
$\times$ 10$^{9}$ K and for the protons of 1.8 $\times$ 10$^{8}$
K.  The ratio of the temperatures is mass proportional,
T$_{{oxygen}}$=16T$_{{proton}}$, which is expected using this
method. This heating occurs when some fraction of the energy of
the shock speed is transferred to the thermal velocity of the
protons or ions.

The width of the O VI lines determines the temperature to be 1.5
$\times$ 10$^{9}$ K.  The O VI temperature derived from the
observed line width is less than that predicted by the kinetic
temperature equation for no equilibration among particle species.
The ratio of the temperatures indicates that O VI is heated to a
temperature 48\% less than the value predicted for mass
proportional heating. Ions are being heated by a process other
than the bulk fluid velocity thermalization or there is a heat
loss mechanism for the ions.

Heating of ions in collisionless shocks has been studied by
Berdichevsky et al.(1997) using heliospheric shock data.  In
examining O VII, it was found that the oxygen was preferentially
heated 19-39 times more than the protons.  In studying the solar
wind, Lee \& Wu (2000) assume greater than mass-proportional
heating as part of the coronal heating process. As a consequence,
ions non-adiabatically expand upstream (not being reflected by the
shock front) and move with a velocity equal to their gyration
velocity as they go upstream. These hot highly energized ions
could act as a precursor that takes away a significant amount of
energy.

    The current supernova observation of less than mass proportional
heating lies in stark contrast to the heliospheric collisionless
shocks.  Several factors and processes determine the extent of ion
heating.  The first comparison to be made is the speed of the
shock relative to the local Alfv\'{e}nic speed.  The solar shocks
propagate at 400-1000 km s$^{-1}$.  SN1006's shock is propagating
at almost 3000 km s$^{-1}$.  The Alfv\'{e}nic Mach number, the
ratio of the shock speed to the square root average of the thermal
and local Alfv\'{e}nic speed, is $\le$ 10 for solar shocks but
upwards of 200 for the supernova shock. The orientation of the
magnetic field with respect to the normal of the shock front is
also of importance as quasi-perpendicular shocks and
quasi-parallel shocks are quite different.  If the current
magnetic field orientation for SN1006 is correct, the NW is
propagating parallel to the ambient magnetic field, while both
parallel and perpendicular shocks are observed in the solar wind.

A measure of the importance of the magnetic field is the parameter
$\beta$.  The plasma $\beta$, the ratio of thermal to magnetic
pressure upstream, is small ($\le 1$) for heliospheric shocks.
Using the parameters for SN1006 and the general value for the
Galactic plane ISM magnetic field ($\sim$3 $\mu$G), the NE has
$\beta=0.02$ and in the NW $\beta = 0.1$.  The magnetic field
pressure dominates the thermal pressure at the ISM/remnant
boundary as in the solar wind, in contrast to the ISM which is
assumed to have a $\beta$ of unity.  It is possible that the
change in density from pre-shock to post-shock conditions is an
important characteristic in the propagation and heating of ions in
the collisionless shock fronts.

In order to determine the cause of the different heating found in
the heliosphere and supernovae further investigation of the
influence of pressure, density, Mach number and velocity on ion
heating by the shock is necessary.

\medskip
\subsection{Neutrals at the Shock Front}
In the analysis scheme used here from Chevalier \& Raymond(1978) and\\
Chevalier, Kirshner, \& Raymond(1980), neutrals play a vital role.
Neutrals undergo charge exchange or emit line radiation to produce
the H$\alpha$ and Ly-$\beta$ emission.  Shocks produce fast
neutrals, as evident by the broad components of the H$\alpha$ and
Ly-$\beta$ lines. This could create a neutral precursor for the
shock \citep{smi94, lim96}. The hydrogen and oxygen neutral
fractions are tightly coupled by charge transfer, thus information
about the neutral fraction of hydrogen can be used to diagnose the
neutral fraction of oxygen. These neutrals should become pickup
ions like those seen in the solar wind \citep{val76} when they
pass through the shock and become ionized. Pickup ions like those
in the heliosphere can then act as a high energy seed population
for Fermi acceleration just as heliospheric pickup ions are the
seed population for anomalous cosmic rays \citep{fis74}.

    The He II 4686\AA\ line can be used as an indicator of neutral
fraction due to its insensitivity to pre-shock neutral fraction
and electron-ion pre-shock temperature equilibrium \citep{gha02}.
In the NW, observations have shown He II emission lines \\
\citep{ray95,gha02}.  The ratio of HeI/HeII can then be used to
find an H neutral fraction which is a parameter in the relation of
the H$\alpha$ two component intensity ratio,
I$_{{broad}}$/I$_{{narrow}}$, and the electron-ion temperature
ratio.  In addition, \citet{gha02} calculated the pre-shock H
population to be 90\% ionized but the pre-shock He population is
70\% neutral. Using the H$\alpha$ broad-to-narrow intensity ratio
calculated for 90\% pre-ionized medium, the temperature ratio,
T$_{{electron}}$/T$_{{proton}}$, was found to be $\le$ 0.07,
showing little to no equilibration between protons and electrons.
Using the ratio of T$_{{electron}}$/T$_{{proton}}$, we find an
electron temperature of $\le$ 1.2 $\times$ 10$^{7}$ K,
approximately 1 keV, which is an upper limit that agrees with the
value found by \cite{lon03} of T$_{{electron}}$ $\le$ 0.6 keV, but
significantly less than the oxygen and proton temperatures found
for this observation(1.5 $\times$ 10$^{9}$ K and 1.8 $\times$
10$^{8}$ K, respectively).

\goodbreak
\section{Summary}
    In summary, the two shock regions of SN1006 studied here provide a
unique cosmic laboratory for shocks and their acceleration
processes.  Clearly, the properties of the interstellar medium
play a crucial role in shaping these shocks.  We conclude with the
following summary of our observations and interpretations.
\smallskip
\begin{enumerate}
\item The material that the NE shock front is encountering is less
dense than in the NW region, with a ratio n(NW)/n(NE)$\ge$4, and is best seen in the
X-ray or radio wavelengths.  The NW shock front could be moving
into a diffuse H I cloud or similarly dense region.

\item The O VI line width of the NW shock indicates that oxygen ions are
heated to a temperature less than 48\% of the value predicted by
mass proportional heating.  This differs from the observations of
other non-radiative collisionless shock fronts such as those in
the heliosphere which found ion temperatures 20-40 times in excess
of the values predicted by mass proportional heating.  The roles
of density, pressure, magnetic field orientation with respect to
the shock normal, velocity and Mach number should be examined to
better determine the ion heating mechanisms.

\item The plasma at the shock front has not had time to come to
equilibrium via Coulomb collisions.  The plasma is in a
non-equilibrium state with energy distributed differently between
species of the plasma. This is in agreement with the work on
temperature equilibrium done by \cite{gha02} who found the ratio
of proton to electron temperature to be $\le$ 0.07 indicating a
plasma far from equilibrium.
\end{enumerate}

    The role of the neutral fraction of the ISM population in the
charge exchange interaction should be examined in detail as it may
greatly affect the outcome of the shock-ISM interaction.  The rate
at which the plasma becomes isotropic, the particle distribution,
and the time scale to reach isotropic and Maxwellian conditions
are in need of examination to understand the heating process
present in the collisionless shock. Further work will be done to
model the plasma conditions in collisionless shock fronts to
include neutral fraction as well as examine the role of electron
population on O VI formation.  This work should help advance the
understanding of the shock acceleration of particles and the
physics of a collisionless shocks in varying environments.



\def\baselinestretch{1.0}

\chapter{Ion Heating by Collisionless Shocks in Front of Coronal Mass Ejections}

\def\baselinestretch{2.0}

\goodbreak
\section{Introduction}
The physical properties of the heating mechanism in collisionless
shocks and its dependence on the properties of ions are important
to understanding the dissipation of energy as a shock evolves as
well as the injection of particles into shock acceleration.
Coronal Mass Ejections (CMEs) propagating through interplanetary
space (IPS) form such collisionless shocks ahead of their ejecta
and are ideally suited for this study.  CME shocks interact with
the solar wind heating and accelerating all solar wind ions. These
shocks have been studied both theoretically \citep{zha91} and
through analysis of data from plasma instruments aboard spacecraft
in the solar wind \citep{ogi80,zer76,ber97}. The heating
mechanisms and their dependencies on mass and charge are not well
understood. Heating of the plasma can occur in several ways to be
discussed later. The most prevalent heating mechanism is bulk
thermalization of kinetic energy.  However, because there is a
magnetic field ever present in the solar wind, interactions of
particles via waves and other electromagnetic interactions in the
vicinity of the shock must be taken into consideration. The types
of waves generated and their effectiveness in heating the plasma
vary with particle species and the orientation of the magnetic
field to the shock normal \citep{pap85}.

The data on CMEs from the Advanced Composition Explorer (ACE)
satellite afford us the opportunity to study these shocks and
their heating processes in greater detail. This study focuses on
thermal velocities from the ACE satellite for 21 shocks which are
well characterized and for which good data for heavy ions exist.
In addition to the proton thermal data, helium, He$^{2+}$, oxygen
(O$^{6+}$, O$^{7+}$), carbon (C$^{5+}$, C$^{6+}$), and Fe$^{10+}$
thermal data were available to study. Temperatures of these ions
are analyzed in order to study in detail the heating that has
occurred in connection with the shock.

\goodbreak

\section{Observations}

The ACE SWICS team has been providing data to the ACE Science
Center (ASC) since 1998 \citep{gar98}, shortly after ACE launched
into its orbit about the L1 point at the end of 1997.  The ASC
provides the science community with select elemental abundance and
ionic charge state measurements .  The SWICS instrument
\citep{glo98} is composed of an electrostatic analyzer, which
measures an ion's energy per charge, and a time of flight mass
spectrometer, which measures an ion's velocity and total energy.
Data obtained from SWICS are analyzed using a numerical code that
identifies and characterizes the properties of ions from He to Ni.
Triple coincidences (combined start, stop, and solid state
detector detection signals) provide identification of the mass
(M), charge (Q), and energy (E) of the ions that enter the
instrument.  However, neighboring peaks overlap due to the
resolution limitations of SWICS.  The ions are identified and
processed in energy-time of flight measurement space using a
forward model \citep{hef98}. The forward model is based on the
pulse height analysis (PHA) data accumulated in energy-time of
flight matrices. This parameterized model identifies and assigns
ions to the appropriate peaks in this measurement space. These
peaks or centers are predicted from the residual energy measured
in the solid state detector while accounting for the appropriate
losses. Due to the limited resolution of the SWICS instrument,
some overlap of peaks in measurement space between species may
exist. To remove this overlap, gaussian fits are computed for the
ion peaks and centered according to the parameterized forward
model. Using the gaussian fits, a spillover is calculated and then
removed using probabilistic methods, thereby eliminating any
statistical biases. The corrected counts are then tallied and
assigned to the individual species. The resulting observed
distribution functions are then calculated from the ions energy
spectra.  The distribution functions are then corrected for
instrument efficiencies, and sensor duty cycle. From these
distribution functions, physical quantities including density,
velocity, and thermal speeds can be calculated by taking the
0$^{th}$, 1$^{st}$, and 2$^{nd}$ moments respectively.

\subsection{Error Analysis}

The sources for error in the data were both systematic and
statistical.  The systematic error stems from the measurement
technique itself.  We assume that all values measured in the
instrument are accurate and that statistical counting errors are
dominate.  Velocity determination was done by taking moments of
the phase space density.  The accuracy of these moments depends on
the counts for the point spread distribution. We use the moments
from the gaussian fits to the distribution to get v$_{th}$.  These
real distributions are not always perfectly fit by a gaussian
introducing uncertainties. In order to ensure the accuracy of the
thermal velocities, the SWICS data were compared to the gaussian
fit of the ion distributions for the time period one hour prior
and one hour after the shock passage. The detector that is used
has velocity resolution or channels that are summed to get the
distribution.  If the difference between the measured thermal
velocity and the fit thermal velocity is less than 3 times the
velocity resolution or channel width of the detector, the ion
thermal velocity for that shock was considered accurate. Equation
\ref{thermalacc} served as the data filter to indicate the
accuracy of the thermal velocities.

\begin{equation}\label{thermalacc}
\frac{2 |v_{th}-v_{th,fit}|}{(0.064v_{th})} \le 3
\end{equation}

where\\
v$_{th}$ is the thermal velocity, v$_{th,fit}$ is the gaussian fit
to the thermal velocity data and the factor of 0.064 is the energy
resolution of the SWICS detector.  Three channel widths were
considered the maximum deviation allowed in order to maintain an
accurate measure of the distribution function.  The systematic
error was then considered as the difference in the fit velocity
and the measured peak velocity divided by the measured peak
velocity by Equation \ref{error}.

\begin{equation}\label{error}
\sigma^{2}=\frac{|v_{th}-v_{th,fit}|}{(v_{th})}
\end{equation}

Next statistical errors were considered based on the number of
counts per ion.  The data files available had a density count
error.  The density being the 0$^{th}$ moment of the distribution
allows for the density counts to be related to the error in the
thermal velocity. The average error for He was about 1\% whereas
the other ions were higher ranging form 1-30\%.  The statistical
error was assumed to be 20\% of the value in order to cover the
uncertainty in measurements. This statistical error was added to
the systematic uncertainty. These errors added together in
quadrature were then propagated to find the error on calculated
ratios or differences presented in the results and discussion
sections.

The other measurements obtained from SWEPAM and MAG such as the
proton thermal velocity and the magnitude of the magnetic field
were averaged for one-half hour upstream and one-half hour
downstream from the time of the shock. This time was chosen to
guarantee that the instrument was collecting data on shocked
interplanetary medium not the ejecta associated with the CME.
There were up to 30 data points available per half hour.  Since
the website where the date was obtained did not provide errors, we
relate the error to the number of data points averaged for our
values.  The error in this value was then taken to be the inverse
of the square root of the number of data points used to compute
the mean value.

\subsection{Shock and Data Selection}
 A recent article by \citet{cra03} detailed a list of shocks
associated with CMEs from 1996-2002. \citet{cra03} used low proton
temperature and magnetic field rotation data to identify the
shock.  This shock list was correlated with the shock list kept on
the ACE website that details time, magnetic angle, and the
Alfvenic Mach number.  Observational data from upstream and
downstream of the shock were used to calculate plasma parameters
using a least-squares fitting of the Rankine-Hugoniot relations.
By fitting the observed data the following parameters were
available:  the shock speed in the spacecraft frame and in the
upstream plasma frame in km/s, the angle between upstream magnetic
field vector and the shock normal, $\theta_{Bn}$, in degrees, and
the upstream Mach number, M$_{A}$.

The first criterion for selecting a shock for the study was to
have all key data for two hours before and two hours after the
shock passage.  The next step in data analysis was to determine
the structure of the shock based on magnetic angle orientation and
the laminar flow of the shock.  The temperature of the solar wind
provides a measure of the shock and the plasma characteristics
upstream and downstream of the shock.  An increase in temperature
before the time of the shock indicates pre-heating and a possible
ramp or reverse shock which needs to be excluded as they heat ions
differently. If the temperature increased between the hour and
half hour before the shock by more than 30\% of the value of the
mean of a half hour before the shock, the shock was considered to
have a ramp structure and not used in the analysis.

The SWICS, MAG \citep{smi98}, and SWEPAM \citep{mcc98} instruments
that collected the data on ion speed, density, and magnetic field
were described in Chapter 1. From the ACE data files, the
following parameters were selected:
\begin{itemize}
\item Temperature
  \item Proton thermal velocity
  \item Proton number density
  \item He$^{2+}$ ion thermal velocity
  \item O$^{6+}$ ion thermal velocity
  \item O$^{7+}$ion thermal velocity
  \item C$^{5+}$ ion thermal velocity
  \item C$^{6+}$ ion thermal velocity
  \item Fe$^{10+}$ ion thermal velocity
  \item Magnetic Field Magnitude and Direction
\end{itemize}

The SWICS data have a 12 minute accumulation time.  The
accumulation cycle that coincides with the time of the shock
passage at ACE is not used because the observed data would mix
upstream and downstream distributions, exhibiting a two population
distribution that would not be able to be fit to a single
gaussian.

The proton thermal velocity, the proton number density, and the
magnetic field data were taken from the SWEPAM/MAG data set at the
ACE Science Center. For the parameters that were taken from the
SWEPAM/MAG data set, 64 second averages of data were used.  The
statistics for these parameters are higher due to the abundance of
data for one hour prior and one hour after the shock passage at
the ACE satellite.

\begin{figure}
\def\baselinestretch{1.0}
  \centering{ \epsfig{file=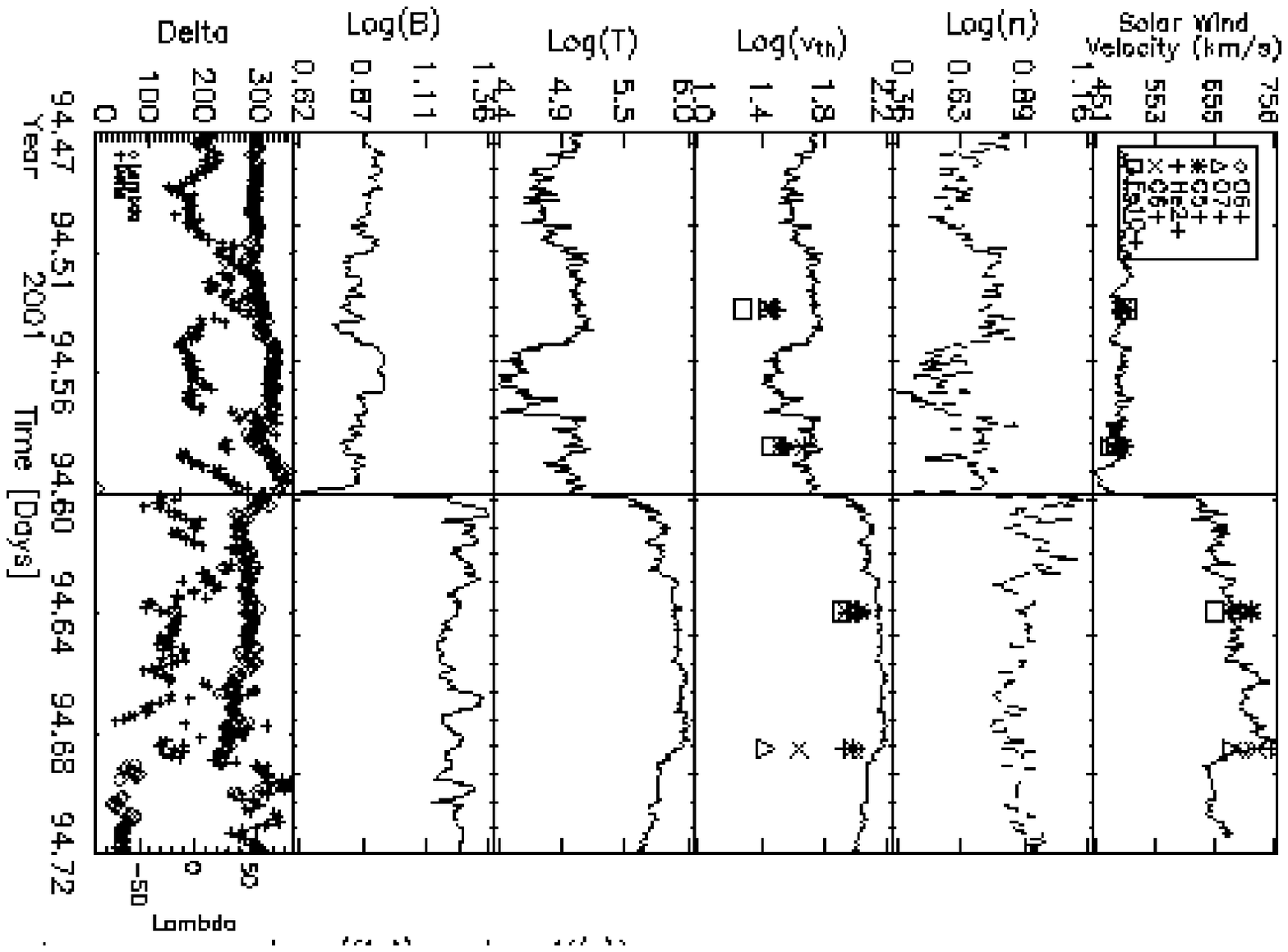,angle=90,clip=,width=11cm} \caption{Plot of
  ACE magnetic and temperature data versus time, in fraction of a
  day, for a parallel shock.  The top panel plots the solar wind
  velocity as the solid line and the speed of each ion is included
  as a symbol.  The second panel is plot of the number density in
  the solar wind.  The third panel is a plot of the thermal speed
  of protons with the symbols representing the thermal speed of
  individual ions. The fourth panel plots the solar wind
  temperature versus time.  The fifth panel contains the magnitude
  of the magnetic field versus time.  The bottom panel is a plot
  of the magnetic latitude,delta, and longitude, lambda, versus
  time.\label{goodpa}} }
  \def\baselinestretch{2.0}
\end{figure}

\begin{figure}
\def\baselinestretch{1.0}
  \centering{ \epsfig{file=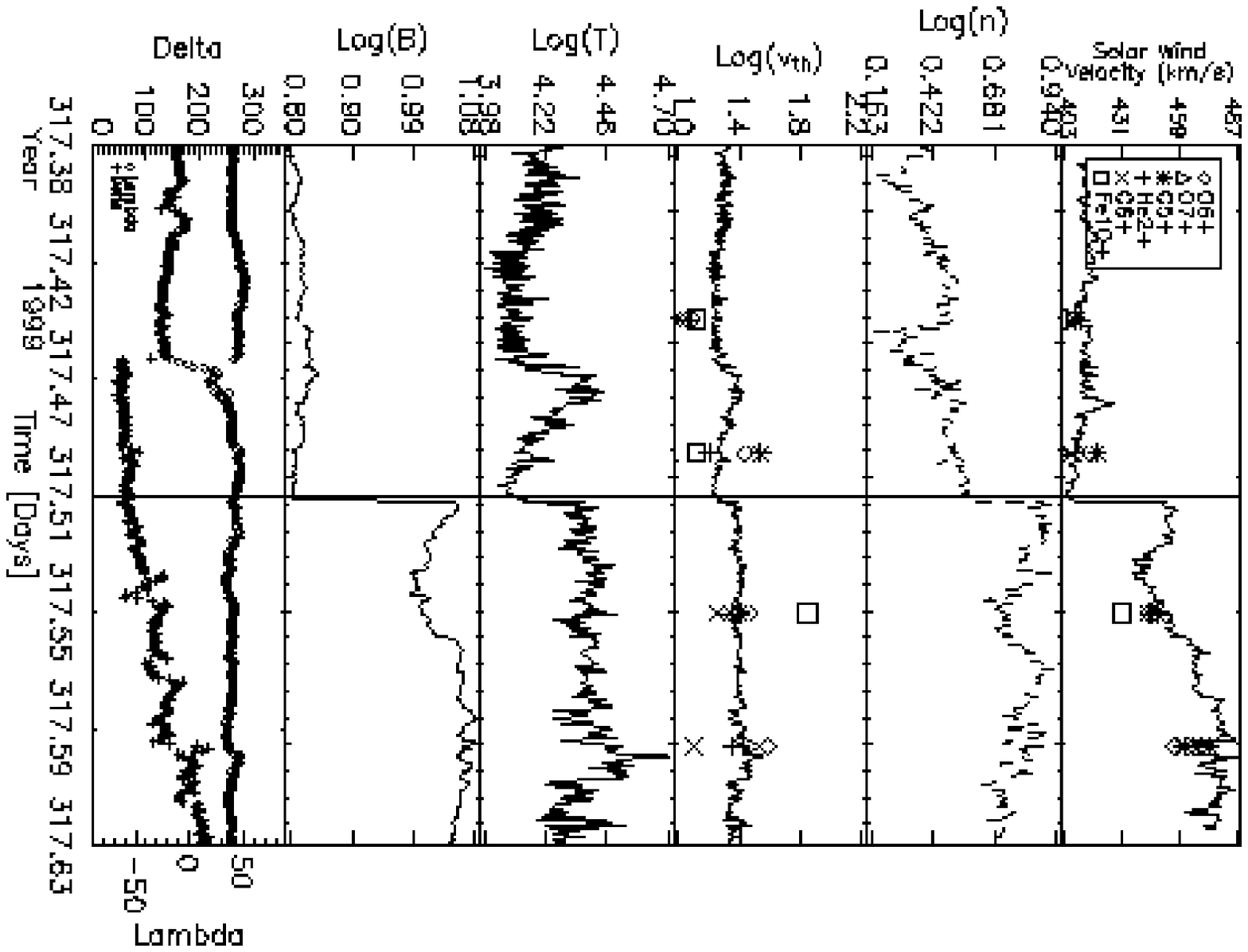,angle=90,clip=,width=11cm} \caption{Plot of
  ACE magnetic and temperature data versus time, in fraction of a
  day, for a perpendicular shock.  The top panel plots the solar
  wind velocity as the solid line and the speed of each ion is
  included as a symbol.  The second panel is plot of the number
  density in the solar wind.  The third panel is a plot of the
  thermal speed of protons with the symbols representing the
  thermal speed of individual ions. The fourth panel plots the
  solar wind temperature versus time.  The fifth panel contains
  the magnitude of the magnetic field versus time.  The bottom
  panel is a plot of the magnetic latitude,delta, and longitude,
  lambda, versus time.\label{goodpe}} }
  \def\baselinestretch{2.0}
\end{figure}

\begin{figure}
\def\baselinestretch{1.0}
  \centering{ \epsfig{file=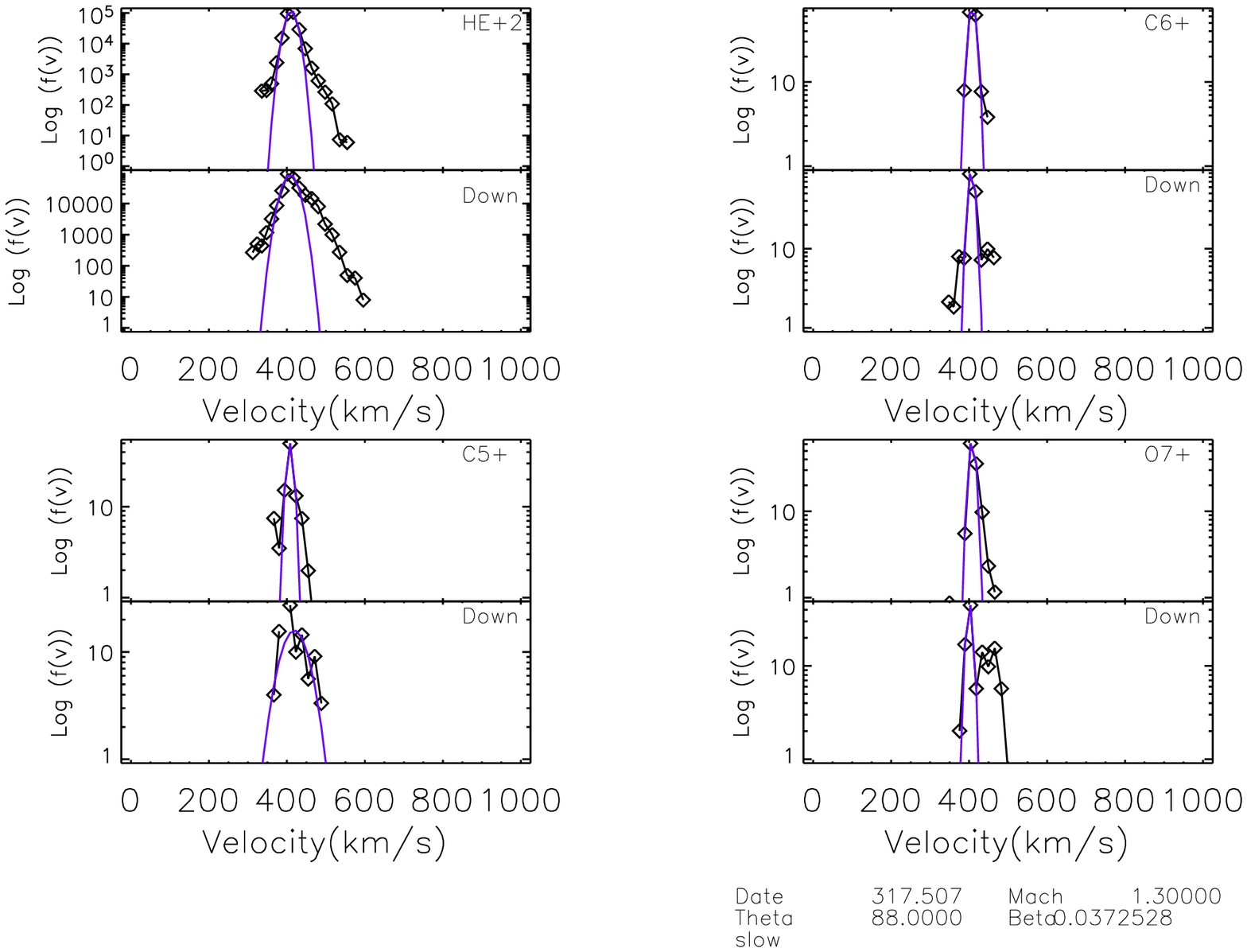,angle=90,clip=,width=11cm} \caption{Plot of
  ion distributions for a perpendicular shock.  Each ion is
  plotted with downstream data in the bottom panel and upstream
  data in the top panel. The diamonds indicate the observed data
  and the solid line is the gaussian fit of that data. Note the
  drop in magnitude of counts from He$^{2+}$, a major ion in the
  solar wind, to minor ions such as C$^{5+}$.\label{ions}} }
  \def\baselinestretch{2.0}
\end{figure}

\begin{figure}
\def\baselinestretch{1.0}
  \centering{ \epsfig{file=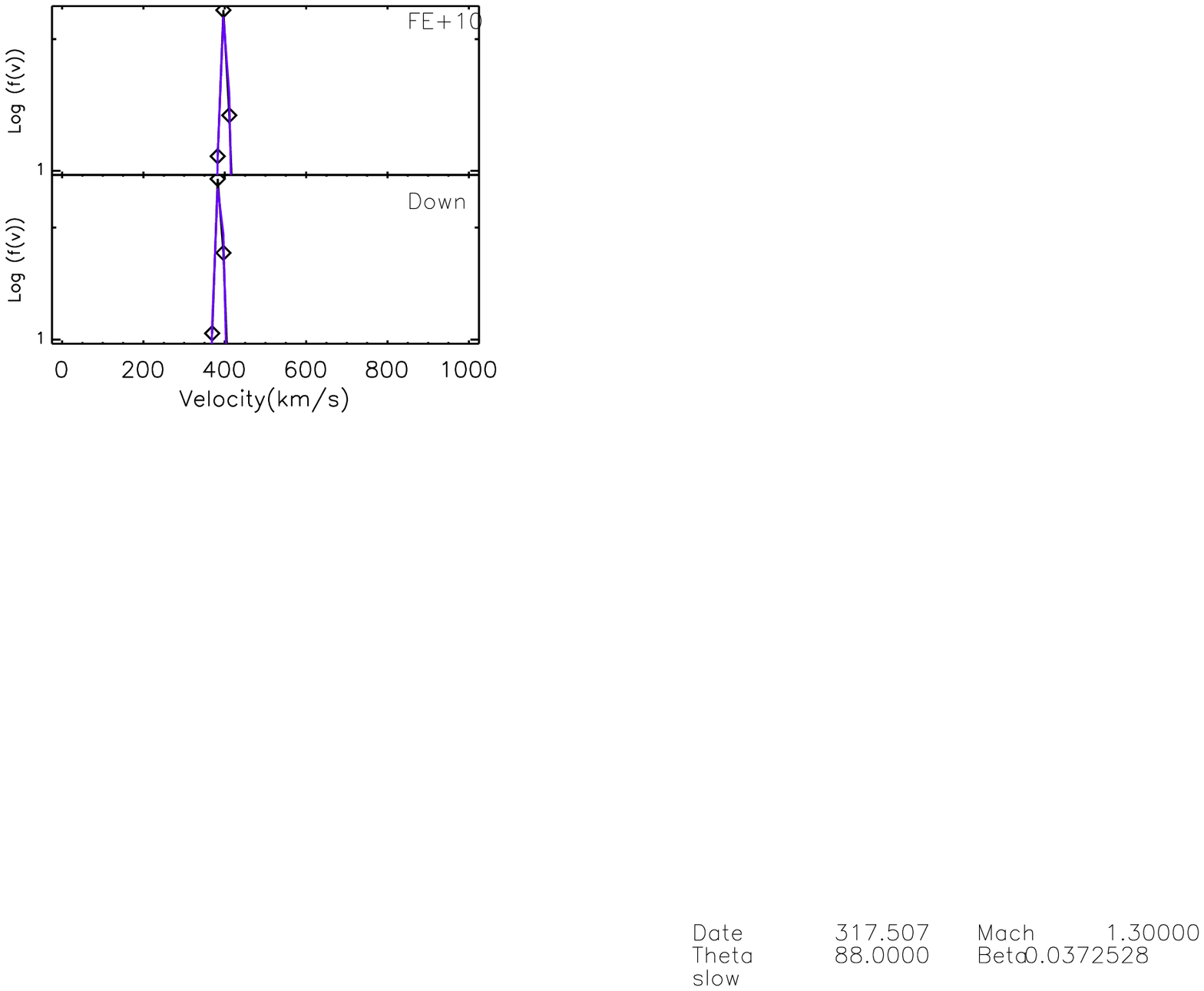,angle=90,clip=,width=7cm} \caption{Plot of
  Iron Distribution for a perpendicular shock.  The upstream
  distribution is plotted in the top panel and the downstream
  distribution is plotted in the bottom panel.  The diamonds
  represent the observed data and the solid line is the gaussian
  fit to that data. Note the low distribution versus a major ion
  such as He$^{2+}$.  \label{iron}} }
  \def\baselinestretch{2.0}
\end{figure}

\begin{figure}
\def\baselinestretch{1.0}
  \centering{ \epsfig{file=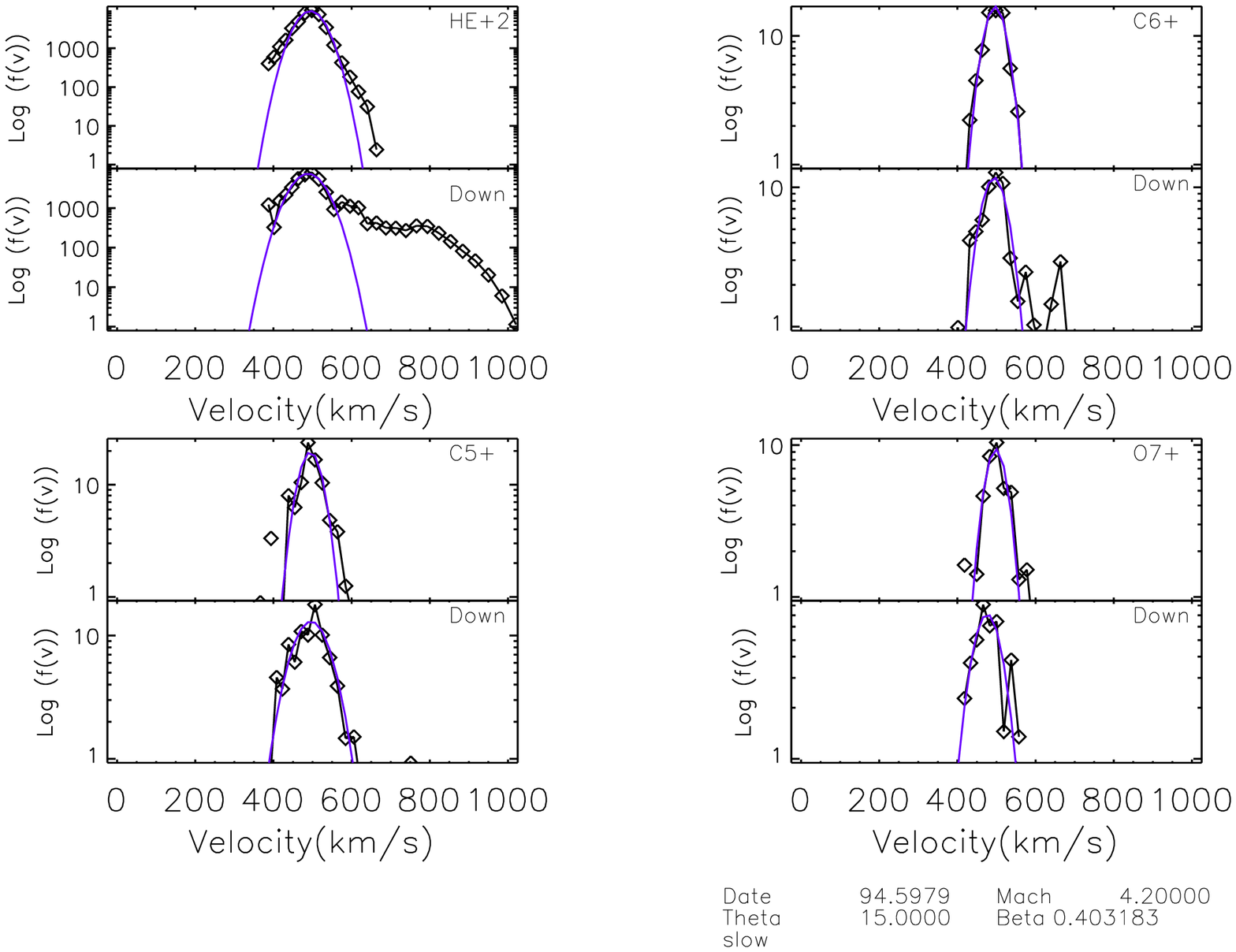,angle=90,clip=,width=11cm} \caption{Plot of
  ion distributions for a parallel shock.  Each ion is
  plotted with downstream data in the bottom panel and upstream
  data in the top panel. The diamonds indicate the observed data
  and the solid line is the gaussian fit of that data. Note the
  drop in magnitude of counts from He$^{2+}$, a major ion in the
  solar wind, to minor ions such as C$^{5+}$.\label{ionspar}} }
  \def\baselinestretch{2.0}
\end{figure}

\begin{figure}
\def\baselinestretch{1.0}
  \centering{ \epsfig{file=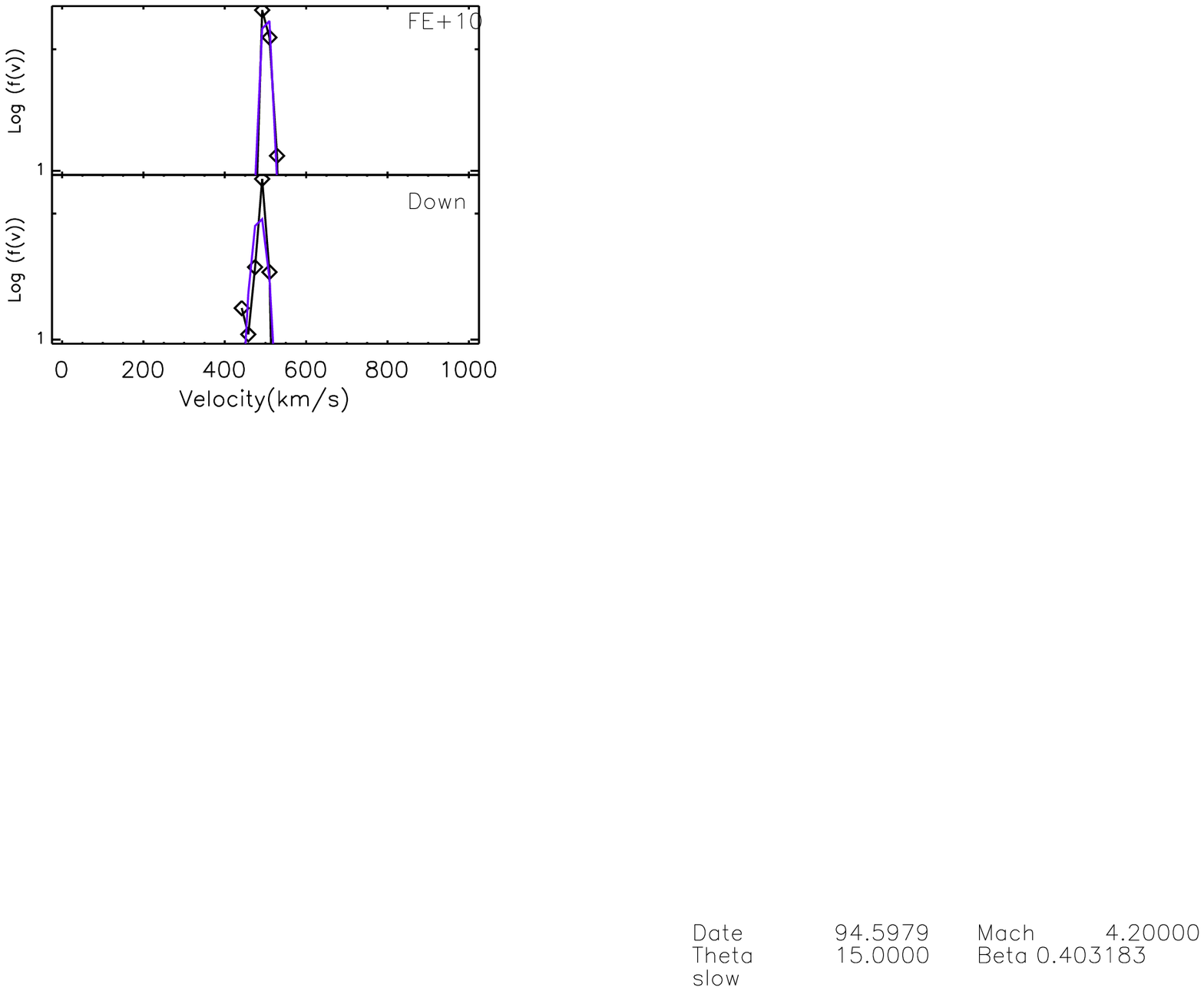,angle=90,clip=,width=7cm} \caption{Plot of
  Iron Distribution for a parallel shock.  The upstream
  distribution is plotted in the top panel and the downstream
  distribution is plotted in the bottom panel.  The diamonds
  represent the observed data and the solid line is the gaussian
  fit to that data. Note the low distribution versus a major ion
  such as He$^{2+}$.  \label{ironpar} } }
  \def\baselinestretch{2.0}
\end{figure}

A plot of representative data for a quasi-parallel and a
quasi-perpendicular shock is shown in Figure \ref{goodpa} and
Figure \ref{goodpe}.  The top panel plots the solar
  wind velocity as the solid line and the speed of each ion is
  included as a symbol.  The second panel is plot of the number
  density in the solar wind.  The third panel is a plot of the
  thermal speed of protons with the symbols representing the
  thermal speed of individual ions. The fourth panel plots the
  solar wind temperature versus time.  The fifth panel contains
  the magnitude of the magnetic field versus time.  The bottom
  panel is a plot of the magnetic latitude, delta, and longitude,
  lambda, versus time.  In addition, plots of the ion distribution
function that were used to determine the thermal velocity,
v$_{th}$, are included in Figures \ref{ions}-\ref{ironpar}.  Each
set of plots is labelled with the ion and the charge state.  The
lower plot of each set is the downstream distribution.  The upper
panel is the upstream distribution function.  Most of the
downstream distributions have "tails" that indicate the
distribution is non-gaussian.



\begin{figure}
\def\baselinestretch{1.0}
  \centering{ \epsfig{file=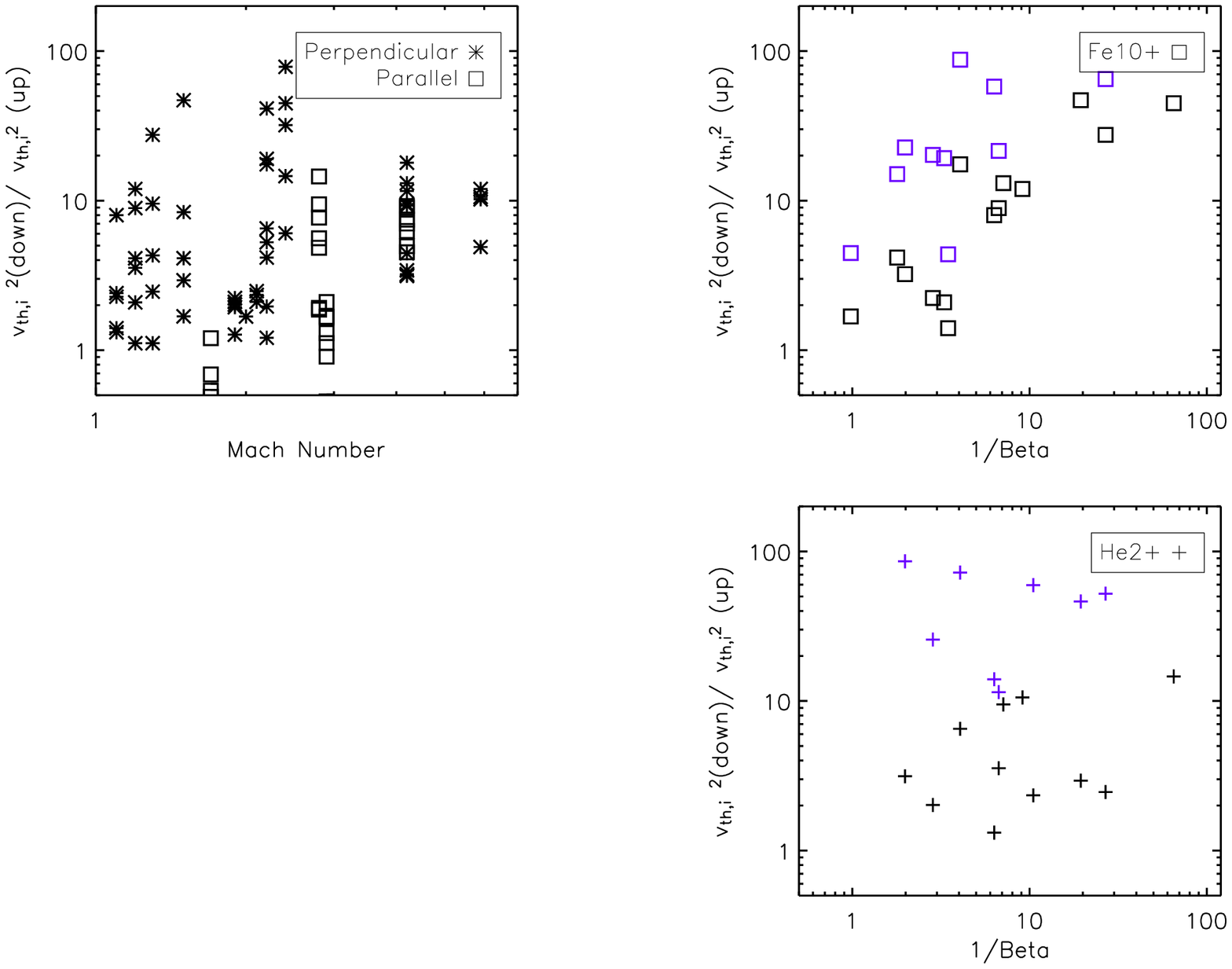,angle=90,clip=,width=7cm}
  \caption{Plot of heating versus Mach number for all shocks.  Note that the
  perpendicular values and the parallel values follow separate
  trends.  The data set is split into perpendicular and parallel
  data in order to understand the trends in the data.
  \label{poortheta} } }
  \def\baselinestretch{2.0}
\end{figure}

 In order to examine the dependence of heating on shock
properties most efficiently, the shock list was broken up into
quasi-parallel and quasi-perpendicular shocks. It is well known
that the orientation of the magnetic field greatly affects the
shock characteristics.  For example, in Figure \ref{poortheta},
the magnetic angle is plotted versus Mach number.  There seems to
be a trend within each of the divisions of parallel and
perpendicular shocks but not an overall trend for all the data.
Hence, the data set was separated into perpendicular and parallel
shocks. There were 16 quasi-perpendicular shocks available for
study and 3 quasi-parallel shocks.

From this shock data, calculations of ion heating can be obtained.
The heating will be defined as follows for this analysis: the
ratio of the change in thermal energy of the species from the
upstream to downstream shock region.
\begin{equation}\label{heatingeq}
H=\frac{v_{th_{d}}^{2}}{v_{th_{u}}^{2}}=\frac{3kT_{s,d}/m_{s}}{3kT_{s,u}/m_{s}}
\end{equation}
where\\
T$_{s}$=Temperature of the species\\
m$_{s}$=mass of the species in units of proton mass\\
v$_{th_{d}}$=one hour downstream average thermal velocity of the species\\
v$_{th_{u}}$=one hour upstream average thermal velocity of the species\\
k=Boltzman coefficient\\

Heating rates are calculated for each ion including protons at
each shock event.  Heating is summarized in Table
\ref{shocksummary} for perpendicular shocks and in Table
\ref{shockspar} for parallel shocks. The tables include the Mach
number, the magnetic angle, $\theta_{Bn}$, plasma $\beta$ -
calculated for upstream protons, and the heating as described by
Equation \ref{heatingeq}.

\begin{table}
   \centering
\caption{Summary of Ion Heating in Perpendicular Collisionless
Shocks \label{shocksummary} }
\medskip
\begin{tabular}[h]{|c|c|c|c|c|c|c|c|c|c|c|c|c|}
\hline
 Year&DOY&Time&M$_{A}$ & $\Theta_{B_{n}}$
&\multicolumn{7}{c|}{Amount of Heating}\\ \hline
&&UT&&&p$^{+}$& He$^{2+}$& C$^{5+}$ & C$^{6+}$& O$^{6+}$ & O$^{7+}$ &  Fe$^{10+}$\\
\hline

1999&48&06:20&2.0 &  100&1.7 &0.8 &0.8& 0.5 &0.4& 0.2& 0.9\\\hline

1999&63&11:00&2.2& 88& 1.4 &1.1 &1.2 &1.2 &1.0 &2.0 &4.2\\\hline

1999&218&06:44&1.5&82&2.4& 2.9& 9.7& 1.7& 8.4& 4.1& 46.9 \\
\hline

1999&301&11:30&1.2 &83 & 1.4 &0.9 &1.2 &1.0& 1.1 &0.7& 2.1\\\hline

1999&317&12:10&1.3 &88 & 1.6& 2.5& 0.4& 4.3& 1.1& 9.5&
27.5\\\hline

1999&360&09:30&1.1& 82&  1.3& 0.9& 0.7& 0.7& 0.7 &0.8& 1.4\\\hline

2001&62&10:41&1.9& 82& 2.7& 2.0& 1.0& 1.3& 1.9& 2.1& 2.2\\\hline

2001&98&10:32&4.2& 90&  4.6& 3.1& 1.4& 4.5& 3.4& 3.2& 3.2\\\hline

2001&101&15:28&2 &98 &1.3 &0.9 &1.4& 0.6 &1.0 &0.8 &1.7\\\hline

2001&118&04:31&5.9& 92& 10.9& 10.6& 46.0& 0.0& 10.2& 4.9
&12.0\\\hline

2001&132&09:20&1.2 &84& 1.0& 3.6& 2.1& 12.0 &4.1& 0.0& 8.9\\\hline

2001&217&11:55&1.6 &88& 1.2& 0.9& 0.4& 0.4& 0.9& 0.7& 0.3
\\\hline

2002&107&10:07&2.2& 90 &1.7 &6.5& 0.3& 5.3 &19.0 &41.3
&17.5\\\hline

2002&143&10:15&4.2 &96  &3.6 &9.5& 8.9 &17.9& 11.6 &9.2 &13.1
\\\hline

2002&250&16:10&2.4& 89  &22.2 &14.6& 28.3& 6.1 &78.4 &32.0&
44.8\\\hline

2002&313&17:54&2.1 &96&1.8& 2.3 &86.2 &0.2& 2.5& 2.1& 1.0\\\hline
\hline
\end{tabular}
\end{table}

\begin{table}
\begin{center}
\caption{Summary of Ion Heating in Parallel Collisionless Shocks
\label{shockspar}}
\begin{tabular}[h]{|c|c|c|c|c|c|c|c|c|c|c|c|c|}
\hline

Year & DOY & Time&M$_{A}$ & $\Theta_{B_{n}}$ &
\multicolumn{7}{c|}{Amount of Heating}\\ \hline
&&UT&&&p$^{+}$& He$^{2+}$& C$^{5+}$ & C$^{6+}$& O$^{6+}$ & O$^{7+}$ &  Fe$^{10+}$\\
\hline

2001&23&10:06&2.8& 3&5.6 &1.9& 9.4 &14.5&4.8 &1.9 &7.7\\\hline

2001&94&14:21&4.2&15&  4.5& 6.3& 4.9& 7.0& 8.9& 9.2& 7.5\\
\hline

2001&272&09:07&2.9 &  19&1.7&1.1&2.0 &0.9& 1.3 &0.5 &1.6
\\\hline

 \hline
\end{tabular}
\end{center}
\end{table}
\def\baselinestretch{2.0}

\begin{figure}
\def\baselinestretch{1.0}
  \centering{ \epsfig{file=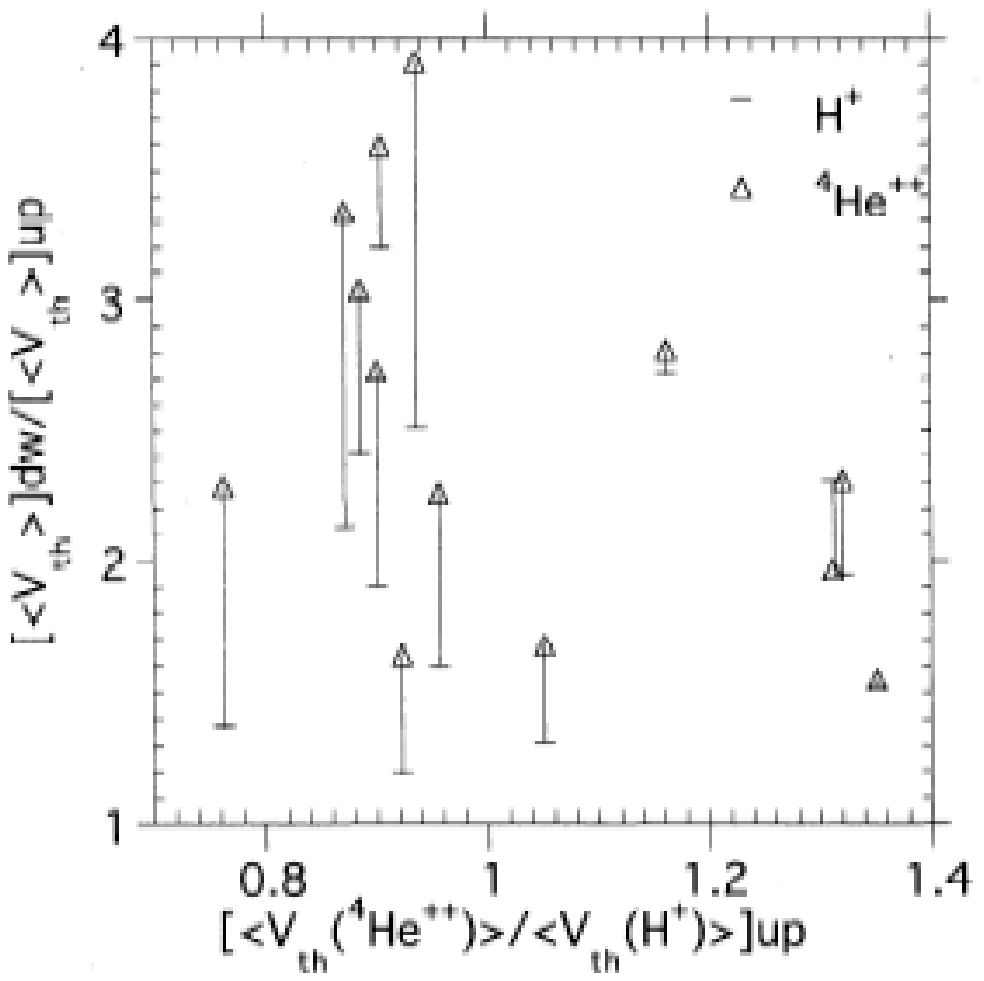,width=10cm} \caption{Figure 2
  from \citet{ber97}.  The x-axis is the ratio of the thermal
  speed of He$^{2+}$ upstream to the proton thermal speed
  upstream.  This is a measure of the initial conditions of the
  plasma.  The y-axis is the ratio of the downstream to upstream
  thermal speed of an ion.  The triangles represent the ratio for
  He$^{2+}$ and the dash indicates the proton.  The vertical line
  connecting two symbols indicates that they are from the same
  shock.\label{mybpic}
}    }
  \def\baselinestretch{2.0}
\end{figure}

Heating of an ion depends on the initial conditions in the
upstream plasma. Figure \ref{mybpic} is taken from Figure 2 of the
\citet{ber97} to compare the results of the current ACE data set
with the results of \citet{ber97}. The x-axis is the ratio of the
thermal
  speed of He$^{2+}$ upstream to the proton thermal speed
  upstream.  This is a measure of the initial conditions of the
  plasma.  The ratio
($v_{th,i}$/$v_{th,p+}$)$_{up}$ indicates how close to equilibrium
the plasma started.  In equilibrium, the ratio of the ion to
proton velocity would be proportional to the inverse square root
of atomic mass of the ion.  Therefore, if v$_{th, i}$ is less than
v$_{th, p}$ and approximately 1/$\sqrt{m_{ion}}$, the plasma is
close to equilibrium.  The y-axis is the ratio of the downstream
to upstream
  thermal speed of an ion.  The triangles represent the ratio for
  He$^{2+}$ and the dash indicates the proton.  The vertical line
  connecting two symbols indicates that they are from the same
  shock.  The thermal velocities are the square root of the
  temperatures of the species.  The current data confirms the \citet{ber97} observations where the
heating was most prevalent when the initial velocity ratio or
x-value was less than 1.0. For ratios of upstream ion to proton
velocities greater than 1.0 moderate to no heating occurred.

\begin{figure}
\def\baselinestretch{1.0}
  \centering{
  \epsfig{file=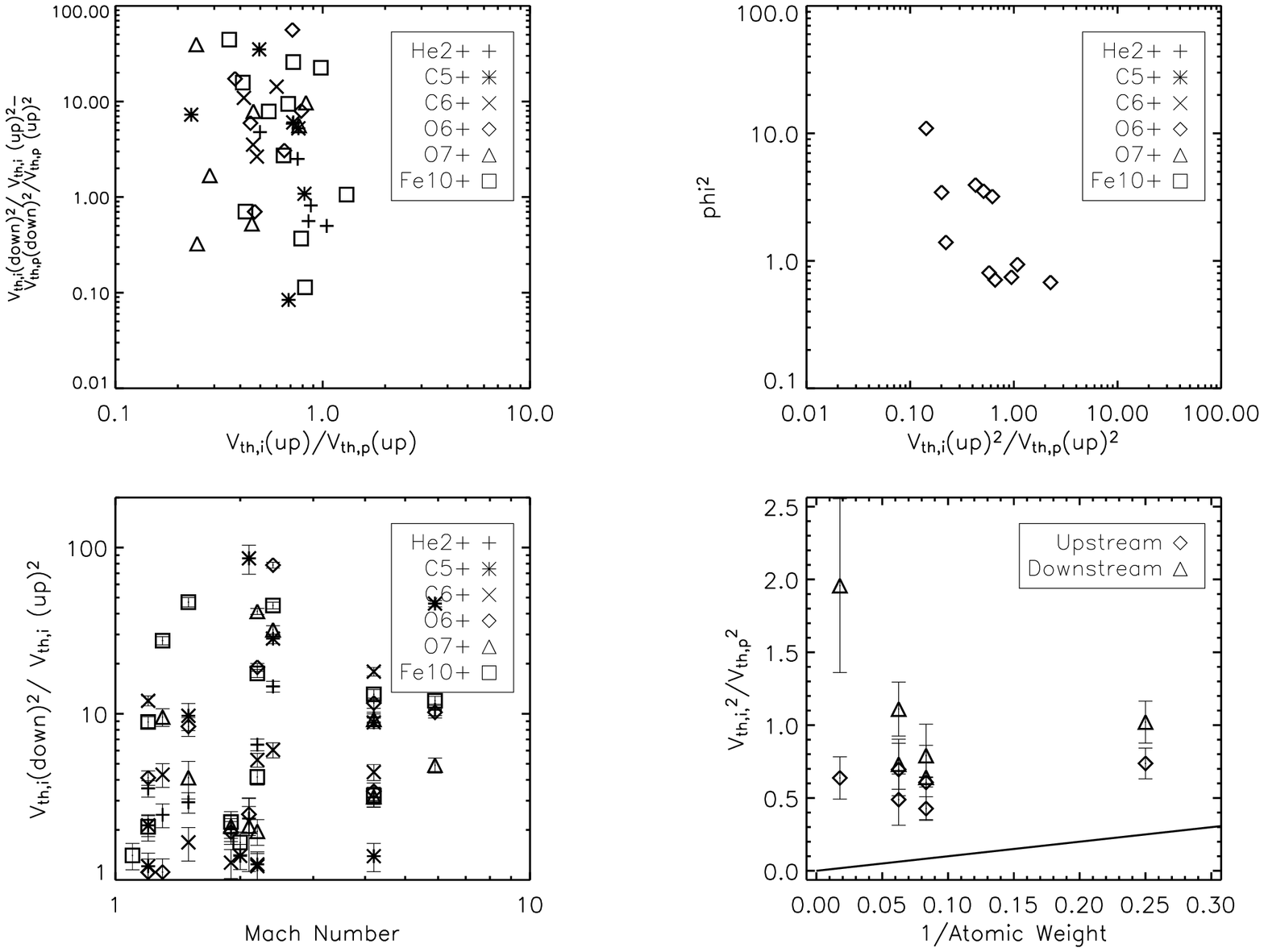,angle=90,clip=,width=10cm}
  \caption{Quasi-Perpendicular Shock Heating Ratios versus upstream
thermal temperature ratio.  The upstream ratio of ion thermal
speed to proton thermal speed is the x-axis.  The difference in
the ratio of temperature increase between the ion and the proton
is the y-axis.\label{berdnewperp}}
  }
  \def\baselinestretch{2.0}
\end{figure}

\begin{figure}
\def\baselinestretch{1.0}
  \centering{
  \epsfig{file=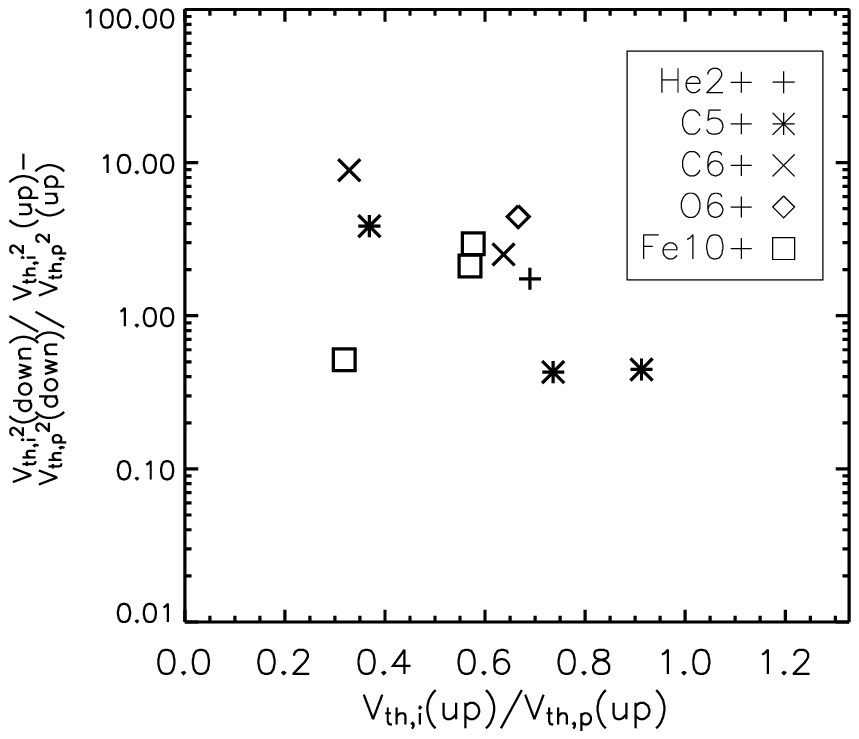,angle=90,clip=,width=10cm}
  \caption{Quasi-Parallel Shock Heating Ratios versus upstream
thermal temperature ratio.  The upstream ratio of ion thermal
speed to proton thermal speed is the x-axis.  The difference in
the ratio of temperature increase between the ion and the proton
is the y-axis.\label{berdnewpar}}
  }
  \def\baselinestretch{2.0}
\end{figure}

While Berdichevsky et al. 1997 included helium, oxygen (O$^{6+}$),
and protons, the current data include O$^{7+}$,C$^{5+}$,C$^{6+}$,
and Fe$^{10+}$ ions extending the mass-to-charge ratio to 5.6.  In
order to test the mass proportional heating along with the initial
conditions of the plasma, Figure \ref{berdnewperp} and
\ref{berdnewpar} were created.  The x-axis is the same upstream
ratio described above for Figure \ref{mybpic}.  The y-axis was
constructed as a function of heating.  The heating of the ion
minus the heating of the proton describes the relative heating of
the ions in the shock. As previously stated, the current data
confirms a greater heating for x-values less than 1.0.  The
perpendicular shocks in Figure \ref{berdnewperp} exhibit
considerably more heating than that of the parallel shocks in
Figure \ref{berdnewpar}.

In order to confirm the greater than mass proportional heating,
another plot was constructed.  The x-axis is the same as that in
Figure \ref{mybpic}, however, the y-axis is adjusted to determine
if the heating of the ion is greater than that of the protons. The
y-axis is the difference of the increase in the thermal speed of
the ion compared to the increase in the proton thermal speed,
shown in Figure \ref{berdnewpar} for parallel and Figure
\ref{berdnewperp} for quasi-perpendicular shocks.  If the
difference is greater than zero, the ion is preferentially heated
to the protons. Most of the ions lie above this value confirming
the greater than mass proportional heating for both types of
shocks.

\begin{figure}
\def\baselinestretch{1.0}
  \centering{ \epsfig{file=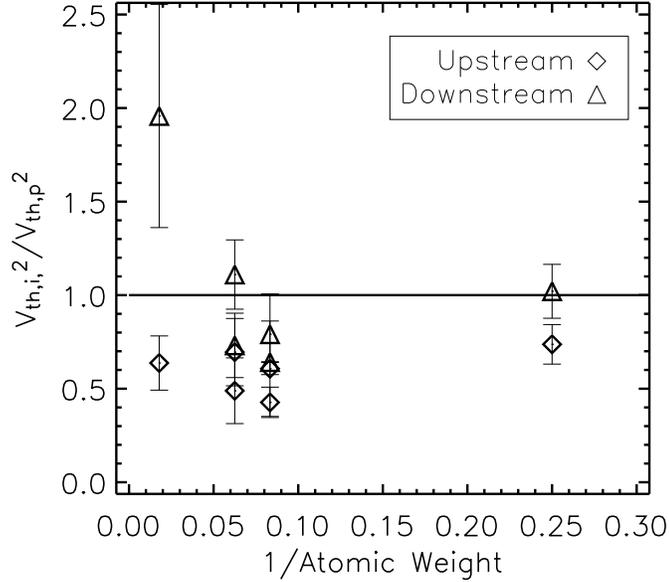,angle=90,clip=,width=10cm}
  \caption{Heating versus atomic weight. The y-axis is the ratio
  of temperature of the ion to the proton. The x-axis is the
  inverse of the atomic weight. The solid line represents the
  heating expected if the mechanism for heating was
  mass-proportional.  Each ion is averaged for an upstream
  temperature ratio and a downstream temperature ratio.  The
  larger the atomic mass the more the ion is heated compared to
  the proton. \label{atomicperp}} }
  \def\baselinestretch{2.0}
\end{figure}

\begin{figure}
\def\baselinestretch{1.0}
  \centering{ \epsfig{file=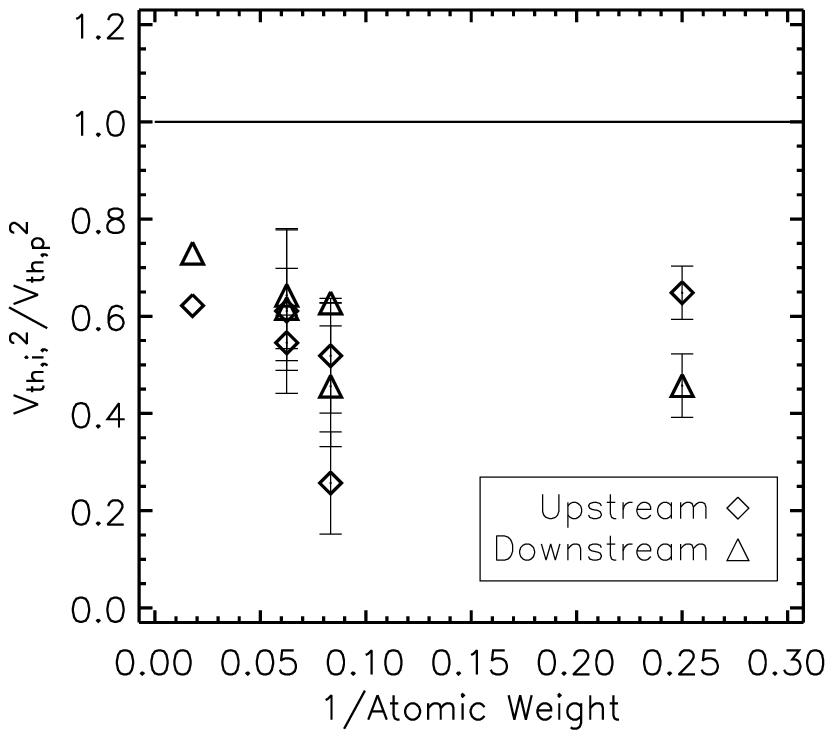,angle=90,clip=,width=10cm} \caption{Atomic
  Weight versus Heating In Quasi-Parallel Shock.  The y-axis is
  the ratio of temperature of the ion to the proton. The x-axis is
  the inverse of the atomic weight. The solid line represents the
  heating expected if the mechanism for heating was
  mass-proportional.  Each ion is averaged for an upstream
  temperature ratio and a downstream temperature ratio.  The
  larger the atomic mass the more the ion is heated compared to
  the proton. Note that in the parallel shocks the ratios are much
  smaller and the differences in values from upstream to
  downstream, a measure of the temperature increase, is smaller.\label{atompar}}
  }
  \def\baselinestretch{2.0}
\end{figure}

To determine the heating according to ion species, Figures
\ref{atomicperp} and \ref{atompar} were created. The x-axis is the
inverse of atomic weight.  The y-axis is the average amount of
heating for upstream and downstream for a particular ion from all
perpendicular (Figure \ref{atomicperp}) and parallel (Figure
\ref{atompar}) shocks studied. The solid line would indicate mass
proportional heating. If the species were in thermal equilibrium,
the ratio of the square of their thermal velocities would be
proportional to 1/$\sqrt{m_{i}}$.  The downstream values for four
of the six ions are mass proportional or greater as they fall on
or above the solid line. However, as atomic weight increases, the
heating experienced by the ions increases. This implies a
mechanism that is based on weight or mass-to-charge ratio rather
than initial temperature. The effect is much greater for the
quasi-perpendicular shocks than the quasi-parallel shock. The
perpendicular shocks have greater heating ratios than that of the
parallel shocks.  This seems to indicate that more heating can
occur at the perpendicular shocks than the parallel.

Although this and past data sets have confirmed the ions are
heated more than the protons in the shock passage, it does not
determine the method of heating of the ion species.  Next, Mach
number and plasma $\beta$ for quasi-parallel and
quasi-perpendicular shocks are examined to better determine
criteria for the heating mechanism.

\section{Perpendicular Shocks}
\subsection{Mach Number versus Heating}

The Mach number of the shock indicates the shock speed with
respect to the Alfven speed in the surrounding material.  As the
Mach number increases, the critical Mach number is achieved. At
this critical Mach number dissipation of energy can no longer be
accomplished by viscosity, scalar resistivity or thermal
conductivity, other mechanisms such as MHD waves must be invoked.

\begin{figure}
\def\baselinestretch{1.0}
  \centering{ \epsfig{file=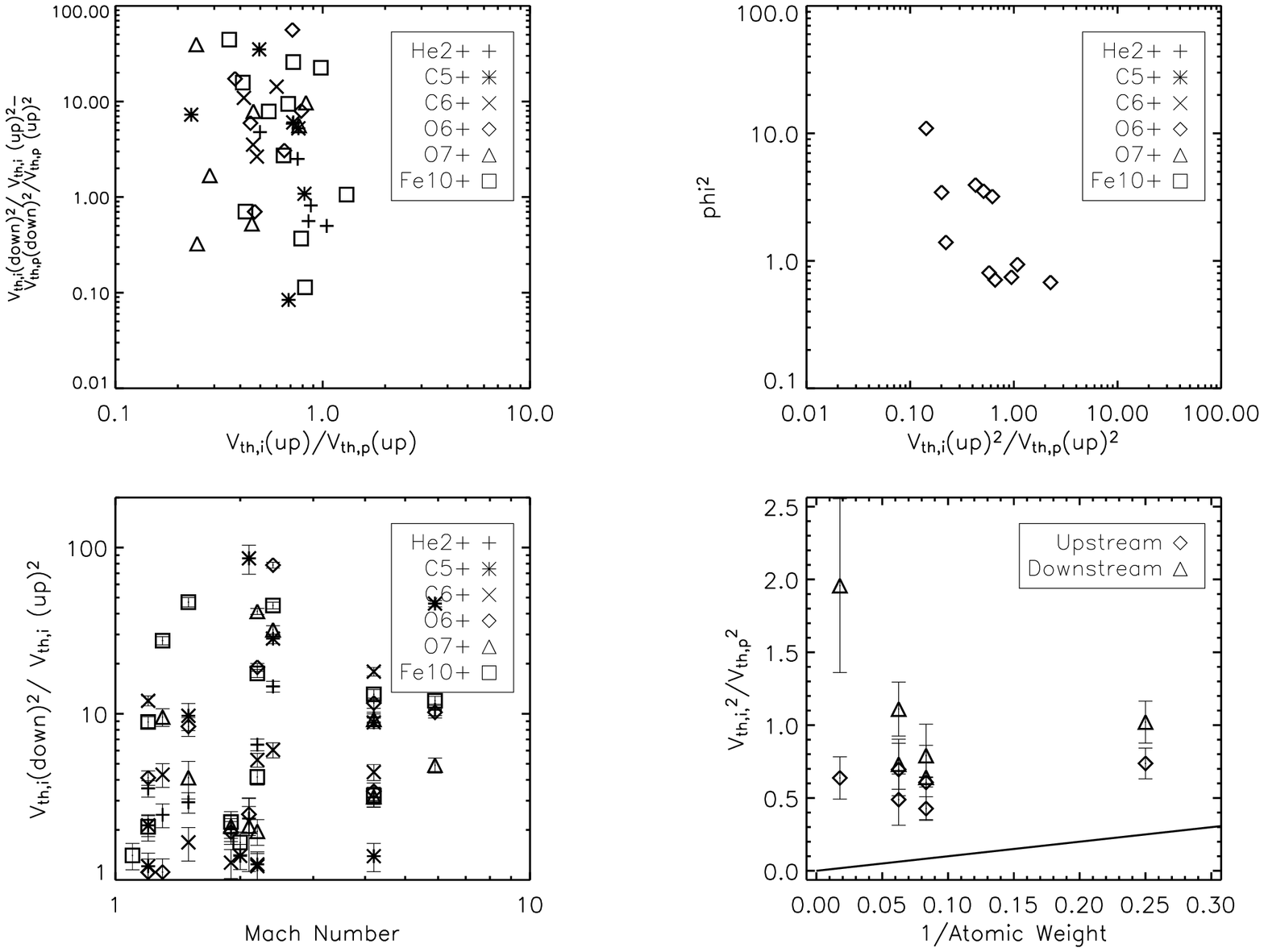,width=10cm,angle=90,clip=} \caption{Plot of
  heating versus Mach number for perpendicular shock.  The Mach
  number is the Alfvenic Mach number.  The temperature ratio is
  the square of the thermal speeds of the ion downstream to upstream.\label{Machperp} } }
  \def\baselinestretch{2.0}
\end{figure}

 Figure \ref{Machperp}, describes the Mach number versus
the heating for the perpendicular shocks studied. There is a large
amount of scatter in any one ion species.  There is no clear trend
in heating with respect to Mach number.  This is an indication
that the bulk thermalization does not play a key role hence the
lack of mass proportional heating.

\subsection{Plasma $\beta$ Effect on Heating}

\begin{figure}
\def\baselinestretch{1.0}
  \centering{ \epsfig{file=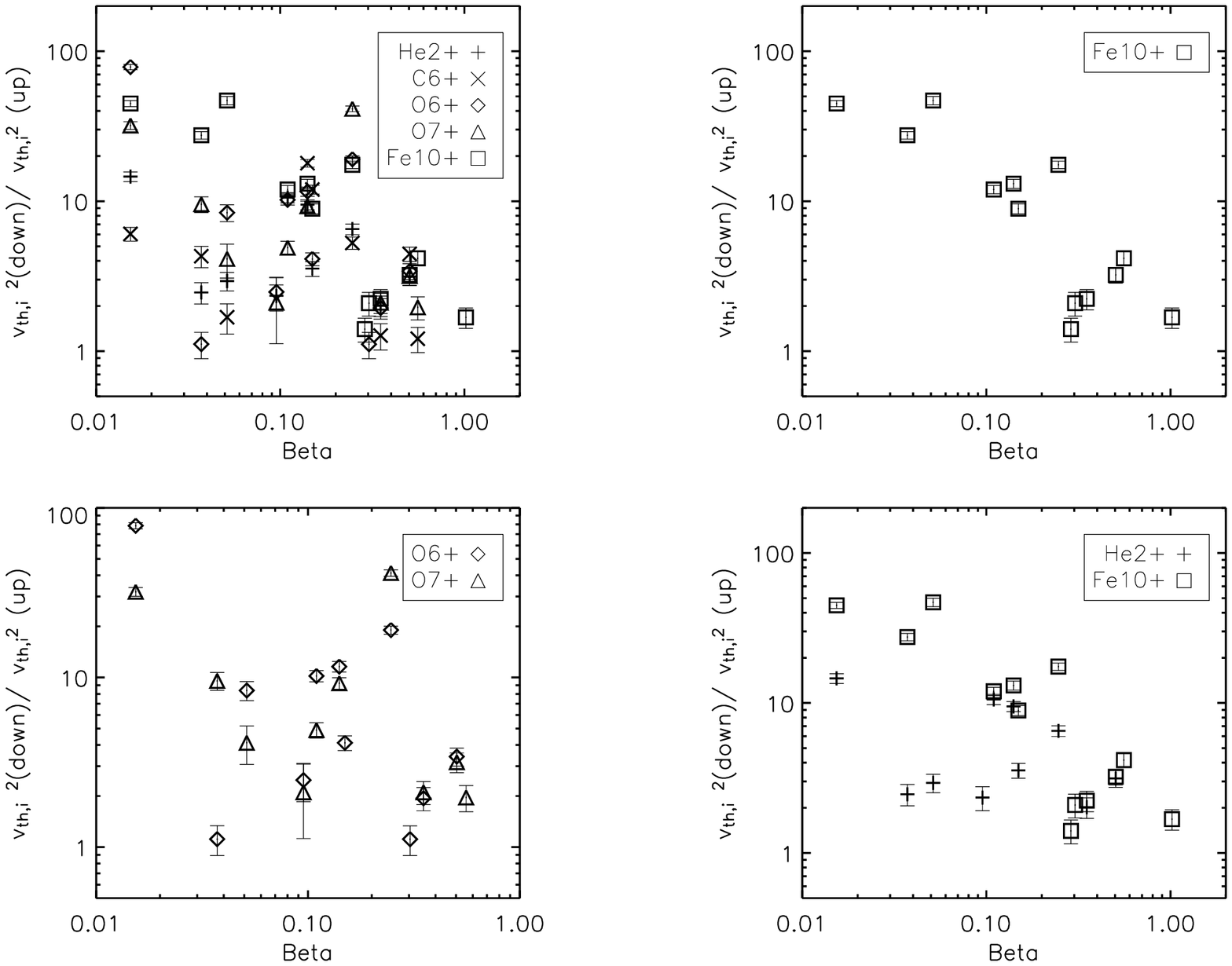,angle=90, width=10cm, clip=} \caption{Plot
  of $\beta$ versus heating for all the ions in perpendicular
  shocks.  The heating is the ratio of the square of the
  downstream ion thermal velocity to the square of the upstream
  ion thermal velocity. $\beta$ is the ratio of thermal to magnetic energies.\label{betaperp}} }
  \def\baselinestretch{2.0}
\end{figure}

Plasma $\beta$, the measure of thermal to magnetic energy, is
plotted versus the heating for each ion species for perpendicular
shocks in Figure \ref{betaperp}.  The heating is the ratio of the
square of the downstream ion thermal velocity to the square of the
upstream ion thermal velocity. $\beta$ is the ratio of thermal to
magnetic energies.  The plot shows that with increasing $\beta$
the heating of the ions decreases. The mechanism for heating of
ions seems to be based on a strong magnetic energy as indicated by
a small $\beta$.

\begin{figure}
\def\baselinestretch{1.0}
  \centering{ \epsfig{file=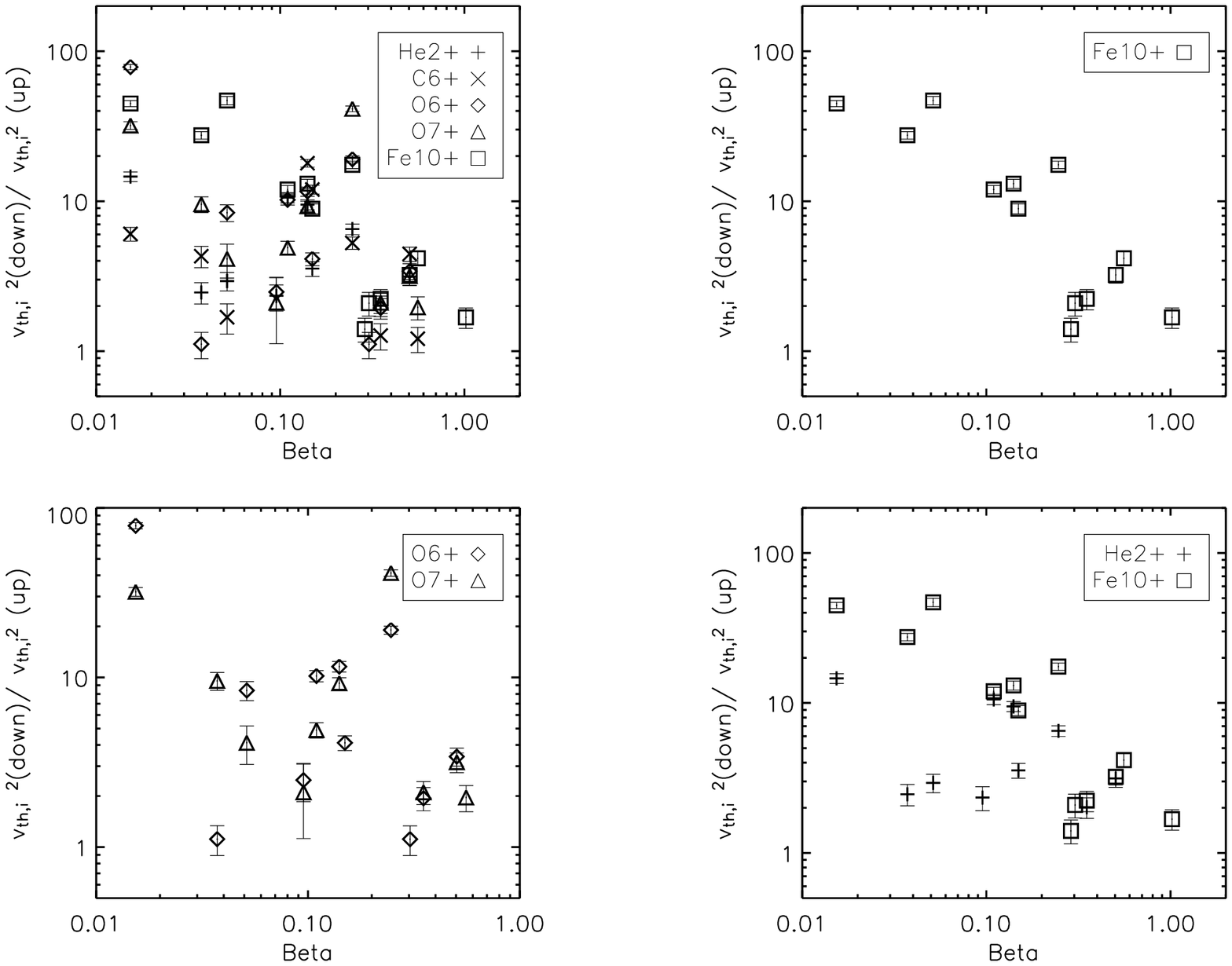,angle=90,clip=,width=10cm} \caption{Plot
  of $\beta$ versus heating for Fe$^{10+}$ ions in perpendicular
  shocks. The heating is the ratio of the square of the downstream
  ion thermal velocity to the square of the upstream ion thermal
  velocity. $\beta$ is the ratio of thermal to magnetic energies.
  Fe$^{10+}$ shows a clear downward trend in heating with
  increasing $\beta$.\label{betairon} } }
  \def\baselinestretch{2.0}
\end{figure}

\begin{figure}
\def\baselinestretch{1.0}
  \centering{ \epsfig{file=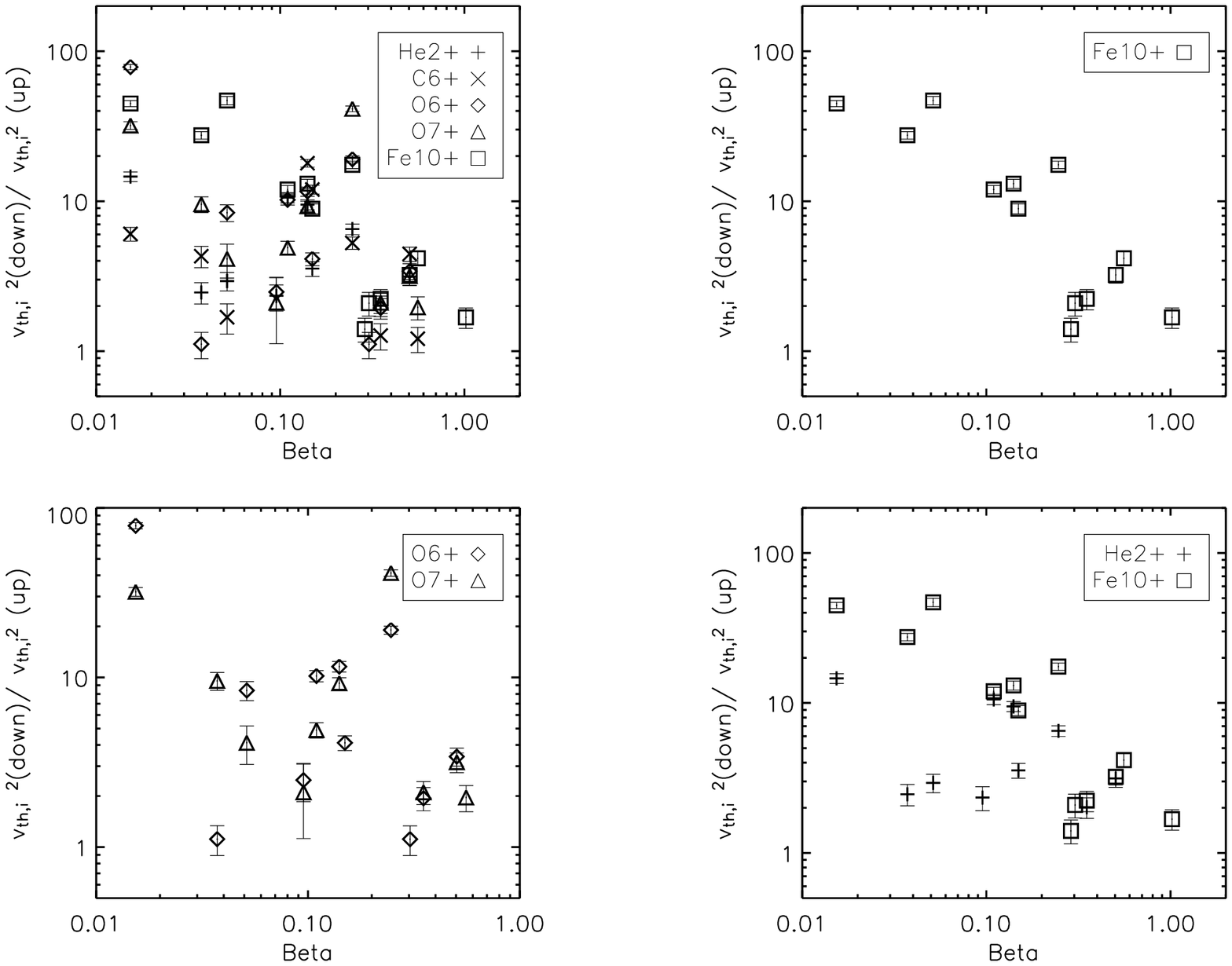,angle=90,clip=,width=10cm}
  \caption{Plot of $\beta$ versus heating for Oxygen ions in
  perpendicular shocks. The heating is the ratio of the square of
  the downstream ion thermal velocity to the square of the
  upstream ion thermal velocity. $\beta$ is the ratio of thermal
  to magnetic energies.  If the heating was based on mass these
  two ions should have identical heating.  However, the heating is
  widely variable for the two ions.\label{betaoxygen}} }
  \def\baselinestretch{2.0}
\end{figure}

\begin{figure}
\def\baselinestretch{1.0}
  \centering{ \epsfig{file=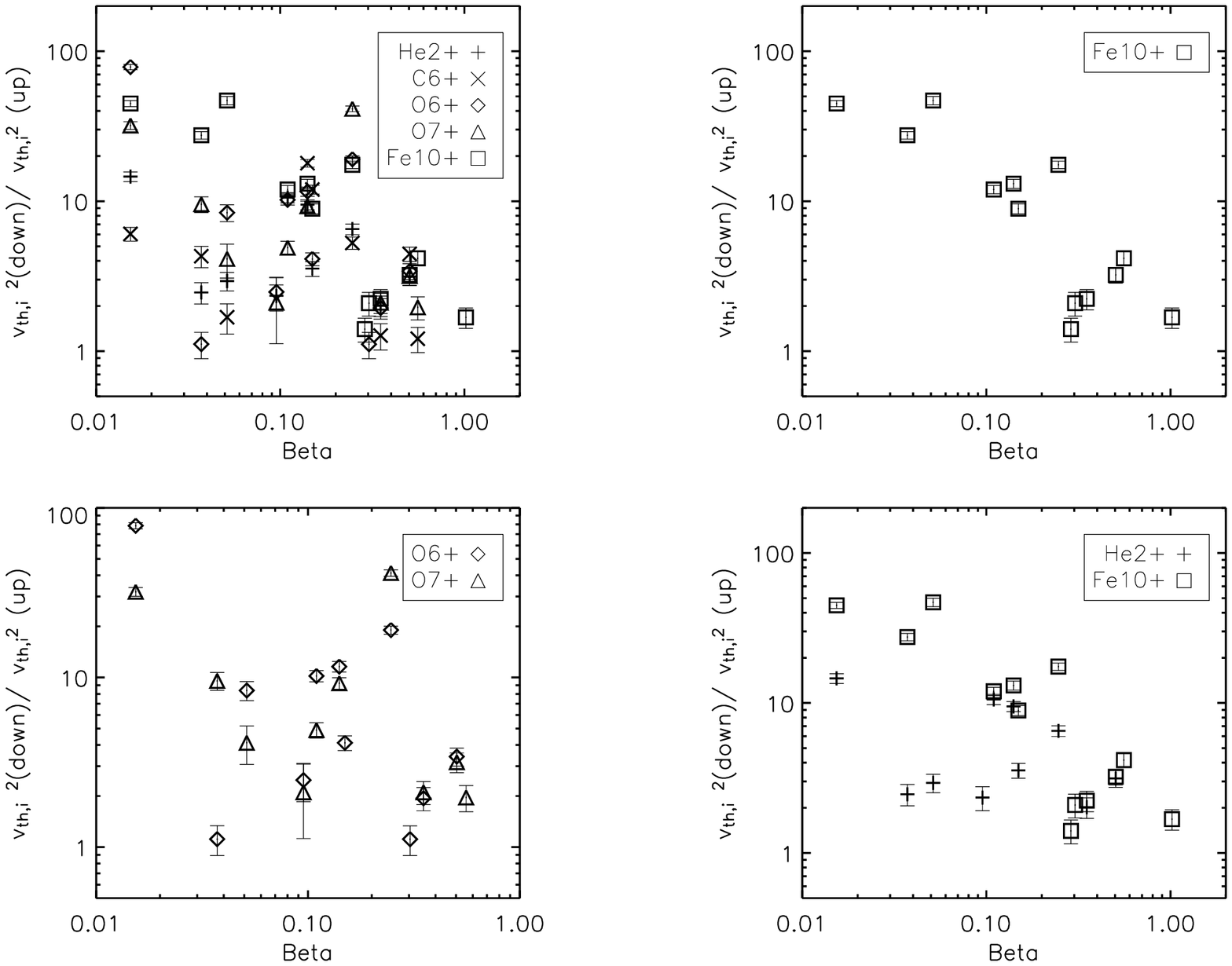,angle=90,clip=,width=10cm} \caption{Plot
  of $\beta$ versus heating for Fe$^{10+}$ and He$^{2+}$ ions in
  perpendicular shocks.  The heating is the ratio of the square of
  the downstream ion thermal velocity to the square of the
  upstream ion thermal velocity. $\beta$ is the ratio of thermal
  to magnetic energies.  Although both show a decrease in heating
  with increasing $\beta$, the Fe$^{10+}$ ions exhibit more
  heating than the He$^{2+}$ ions. \label{betafehe}} }
  \def\baselinestretch{2.0}
\end{figure}

To further illustrate this heating trend, Fe$^{10+}$, the heaviest
ion investigated as well as the one with the highest mass to
charge ratio, was plotted versus $\beta$ in Figure \ref{betairon}.
The iron heating decreases with increasing $\beta$ as it does for
the other ions plotted in Figure \ref{betaperp}. However,
Fe$^{10+}$ shows the most heating of any of the ions studied.  In
order to determine if this heating is linked to the mass or the
mass-to-charge ratio, two other plots were made.  The first,
Figure \ref{betaoxygen}, plots the O$^{6+}$ and O$^{7+}$ ion
heating.  If the dependence on $\beta$ was related to mass only,
then these points should be relatively the same. The second plot,
Figure \ref{betafehe}, graphs the heating of the heaviest ion
studied, iron and the lightest, helium.

The oxygen plot, Figure \ref{betaoxygen}, shows variation between
the two species. This seems to imply that the mass-to-charge ratio
is of importance more than the atomic mass.  There are shock
events where O$^{6+}$ is heated more than O$^{7+}$ and vice versa.
There is always a separation in the amount of heating.  The second
figure, Figure \ref{betafehe}, plots the $\beta$ of helium and
iron that shows the mass is indeed not the only factor as the
helium does seem to have similar heating. One point to consider is
that helium is a major species in the solar wind whereas iron is a
minor species. The major species would correlate well with the
bulk characteristics of the plasma and a minor ion could not be
affected in the same way as it could act as a separate fluid.

\goodbreak
\section{Parallel Shocks}
\subsection{Mach Number versus Heating}

\begin{figure}
\def\baselinestretch{1.0}
  \centering{
  \epsfig{file=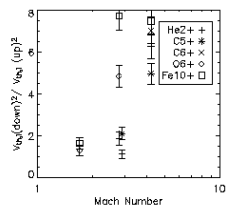,angle=90,clip=,width=10cm}
  \caption{Heating versus Mach Number for Quasi-Parallel Shocks.  The Mach
  number is the Alfvenic Mach number.  The temperature ratio is
  the square of the thermal speeds of the ion downstream to upstream.\label{mpar} }
  }
  \def\baselinestretch{2.0}
\end{figure}

Figure \ref{mpar}, describes the Mach number versus the heating
for quasi-parallel shocks.  The heating is that which is described
by Equation \ref{heatingeq}.   The Mach
  number is the Alfvenic Mach number.  The temperature ratio is
  the square of the thermal speeds of the ion downstream to upstream.  In quasi-parallel
shocks, as the Mach number increases, the temperature increase for
the ions. However, with the small number of data points it is
difficult to define a trend.  More parallel shocks would be
necessary to fit a trend to the data.

\subsection{Plasma $\beta$ Effect on Heating}

\begin{figure}
\def\baselinestretch{1.0}
  \centering{ \epsfig{file=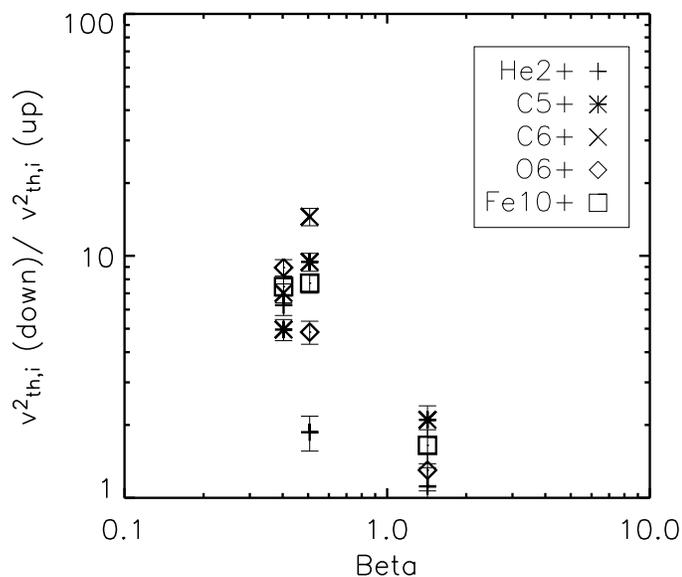,angle=90,clip=,width=10cm} \caption{Heating
  versus Plasma $\beta$ for Quasi-Parallel Shocks.   The heating
  is the ratio of the square of the downstream ion thermal
  velocity to the square of the upstream ion thermal velocity.
  $\beta$ is the ratio of thermal to magnetic energies.  Due to
  the lack of statistics there is no clear trend of heating with
  respect to plasma $\beta$.  \label{betapar}} }
  \def\baselinestretch{2.0}
\end{figure}

The plasma $\beta$ was also examined for the parallel shocks.  The
heating versus plasma $\beta$ was plotted for the parallel shocks
in Figure \ref{betapar}.  The heating
  is the ratio of the square of the downstream ion thermal
  velocity to the square of the upstream ion thermal velocity.
  $\beta$ is the ratio of thermal to magnetic energies.  There was a shock with higher $\beta$
than that of the quasi-perpendicular shocks.  There seems to be a
decrease in heating with increasing $\beta$, however with only a
few shocks to study the trend is unclear.  The Helium has a
downward trend where as the Carbon ions exhibited an increase in
heating with increasing plasma $\beta$. The lack of statistics
makes trend fitting unrealistic.

\subsection{Rankine Hugoniot Conditions for Ions}
In a paper by \citet{bur91}, the Rankine Hugoniot conditions were
rederived for parallel shocks with an assumption that the plasma
was a multi-fluid plasma. The assumption that the ions and the
electrons and neutrals were in equilibrium was relaxed allowing
for each charged species to be a separate fluid.  For each
species, heat flux carried by the ions were neglected because it
was small compared to the kinetic energy and enthalpy fluxes.
Using conservation equations for continuity, momentum, and energy,
the following conditions were found for a specific ion species:

\begin{equation}\label{massrh}
[n_{i} u_{i}]=0
\end{equation}

\begin{equation}\label{momentumrh}
[u_{i}^{2}T_{i,||}]=0
\end{equation}

\begin{equation}\label{energyrh}
[\frac{1}{2}m_{i}u_{i}^{2} + \frac{3}{2}k T_{i,||}]=0
\end{equation}

Equation \ref{massrh} is the continuity equation for each
individual species.  Equation \ref{momentumrh} describes the
dependence of the momentum on the parallel temperature.  The
energy equation, Equation \ref{energyrh}, uses the kinetic energy
of the species as well as the thermal energy in the parallel
direction as the two sources of energy of the fluid.  The quantity
in Equation \ref{energyrh} will be referred to as $\rho$ in future
plots. The data from the parallel shocks in this study were then
fit to the rederived conditions. Below in Figure \ref{rhc}, plots
of the data versus the multi-fluid Rankine Hugoniot conditions are
shown. For the multi-fluid approach to be correct, the downstream
and upstream values should be equal, represented by the line x=y
in the three plots.
\begin{figure}
\def\baselinestretch{1.0}
  \centering{ \epsfig{file=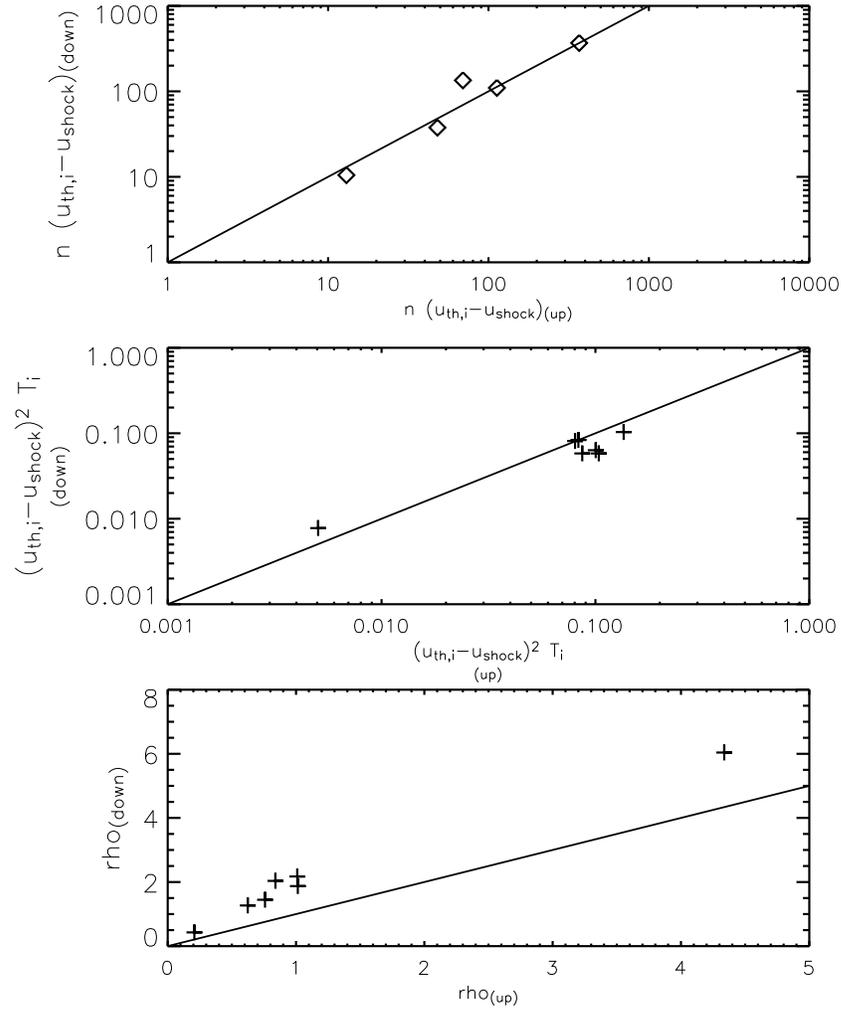,width=11cm} \caption{Fit of CME
  Shock Data to the Rankine Hugoniot conditions derived by Burgi
  1991.  One parallel shock is used.  The speed is the difference
  between the thermal speed and the shock speed. Each symbol
  represents the ratio for a single ion in the shock.  The top
  plot is the continuity equation.  The second is momentum
  conservation.  The third plot is conservation of energy.  Rho is
  defined in Equation 3.5. \label{rhc}} }
  \def\baselinestretch{2.0}
\end{figure}
The mass conservation for the parallel shock is within the errors
of the predicted values as well as that for the momentum
conservation. However, the energy increase observed is greater
than that predicted for these ions.  The increase over equality
measures the potential across the shock.  The acceleration and
heating effect of this potential will be discussed in the
following section.  The data matches the hypothesis of a
multi-fluid plasma treating each ion species separately and
relaxing the condition of equilibration between the protons,
electrons, and ions.

\goodbreak
\section{Discussion of Heating Mechanisms}
Bulk thermalization leading to mass proportional heating has been
ruled out as the lone source of heating at the collisionless
shocks front.  \citet{lee00} discuss three parameters key to
understanding heating in collisionless shocks: Alfvenic Mach
number, magnetic shock angle, and plasma $\beta$. Each dependence
has been examined in the coronal mass ejections shocks.  In
addition to these three parameters there are two other factors to
consider, the mass-to-charge ratio and the turbulence around the
shock.  Also, the thermal state of the plasma ahead of the shock
is important to understand heating. The mass-to-charge ratio is
important to MHD interactions because of lower-hybrid waves that
heat electrons in the shock front \citep{lam04}.  In addition to
the mass-to-charge ratio, turbulence plays a role in the plasma
heating. According to \citet{ler82}, supercritical
quasi-perpendicular shock heating is largely due to reflected ions
at the shock front that cause turbulence.  This turbulence then
energizes the ions that encounter the shock front. Although the
turbulence is likely present at the shock front, the data
filtering process took out shock fronts with detectable ramps that
would be an indication of turbulent processes.

The interest in heating of the thermal population stems from the
cosmic ray acceleration processes.  The acceleration mechanisms
such as First and Second Order Fermi acceleration require a "seed
population" that is energetic enough to enter into the
acceleration process and produce high energy cosmic rays.  Second
Order Fermi Acceleration was first discussed by Fermi in 1949. The
acceleration occurs when particles collide with magnetic eddies or
inhomogeneities.  The inhomogeneities are moving with a velocity,
v.  There is different likely hood of a head-on collision
increasing the energy of the particle compared to that of a tail
collisions decreasing the energy of the particle.  The Fermi
mechanism goes as the power of v$_{sc}$/v$_{part}$, where v$_{sc}$
is the speed of the scattering centers and v$_{part}$ is the
particle speed. Fermi was looking at relativistic particles. The
average increase in energy is second order in v/c where c is the
speed of light. However, this implies that a velocity has to be
fairly large initially (greater than 1000 km s$^{-1}$) to
effectively heat the ion. The speed has to be great enough that
acceleration by scattering happens faster than energy loss by
Coulomb collisions, which become less effective as v$^{-3}$ First
Order Fermi Acceleration is based on a system where the particle
increases energy at each scattering center such as a supernova
shock where there is scattering centers on both side of the shock
front. Each time the particle collides on either side of the shock
there is a head-on collision gaining energy. This process is more
efficient than the second order Fermi Acceleration with an average
energy increase of the order v/c . However the initial velocities
in the data set are small compared to c. This mechanism is based
on initial energy and does not have any clear relationship to the
mass or mass-to-charge ratio heating that is observed in the
current shock study.

Looking for a mechanism that can take a thermal particle to a
suprathermal particle is necessary for these acceleration
processes.  From studies of the earth's bow shock by
\citet{fus97}, a heating method is proposed.  This heating
mechanism is based on the potential that forms around the shock
front.  Protons are the species that creates the shock.  The model
assumes the ion is a test particle and that all the velocity
change is in the direction parallel to the shock normal. This
assumption tells us that the minor ions do not actually form the
shock but just experience its effects.  The effective
electrostatic potential that is set up is created by the change in
the proton flow velocity across the shock.

In the shock crossing, a potential is formed, e$\phi$. This
potential slows down the proton as energy must be conserved. The
potential that was formed by the protons is proportional to the
difference in kinetic energy for the proton in transition from
upstream to downstream.  Therefore, the downstream thermal speed
of the proton is proportional to initial velocity minus any
slowing from the electrostatic potential.

The heavy ions at the shock front also see a potential.  However,
they see a potential based on their mass to charge ratio.  To
conserve energy, the ion speed increases across the shock.  The
ion's downstream speed is as follows in Equation \ref{ionheating}
from \citet{fus97} :

\begin{equation}
\label{ionheating}
v_{down,i}=v_{sw}\sqrt{\frac{(\alpha-1)+c^{2}}{\alpha}}
\end{equation}

where $\alpha$ is the mass to charge ratio, v$_{sw}$ is the solar
wind speed which is assumed to be equivalent to the upstream
proton speed, and c is the ratio of downstream proton speed to
upstream proton speed. This equation for the downstream ion
velocity clearly includes a factor of mass-to charge.  After the
ion passes downstream, it can also be scattered.  If this
scattering is strong enough to push the ion upstream, an
acceleration process can begin. Chapter 4 discusses neutrals and
finds that several percent of them overtake the shock.  This is a
small percent compared to observed cosmic ray flux.

\citet{fus97}  discuss pitch angle scattering and therefore
velocity vectors not thermal speeds.  In order to relate these
velocities to thermal speeds, a simplifying assumption must be
made.  The average speed of the particle is the thermal speed over
the square root of 2 \citep{bau97}.  In addition, the solar wind
speed was assumed to be the same as the ion speed for the upstream
conditions.  This is unrealistic as most of the speeds measured
vary.  Squaring the downstream thermal velocity, comparing it with
the upstream velocity squared and rewriting the equation in terms
of Mach number, Equation \ref{ionheating} becomes:

\begin{equation}\label{Machterms}
\frac{v_{down,i}^{2}}{v_{up,i}^{2}}=\frac{M_{A}^{2}v_{A}^{2}(\alpha-1)+c^{2}}{v_{up,i}^{2}\alpha}
\end{equation}

Substituting plasma beta in terms of the Alfven speed the heating
becomes:
\begin{equation}
\frac{v_{down,i}^{2}}{v_{up,i}^{2}}=\frac{M_{A}^{2}2kT/m_{p}
\beta(\alpha-1)+c^{2}}{v_{up,i}^{2}\alpha}
\end{equation}

\begin{figure}
\def\baselinestretch{1.0}
  \centering{
  \epsfig{file=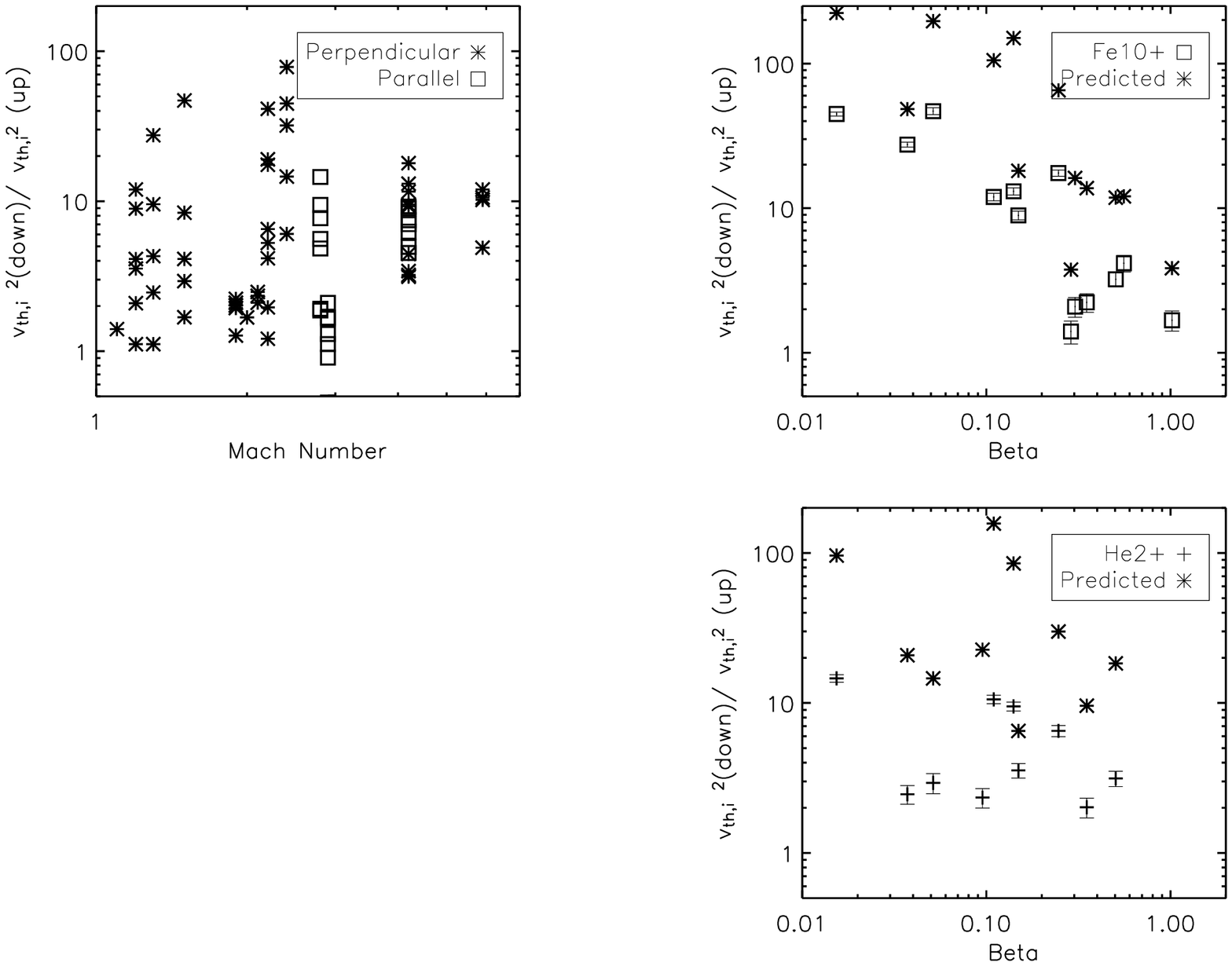,clip=,angle=90,width=8cm}
  \caption{Plotting of the actual and predicted values for heating
  based on plasma $\beta$.  Helium shows similar trends to the
  prediction but the actual values are higher than those observed.
  \label{predicthe}} }
  \def\baselinestretch{2.0}
\end{figure}

\begin{figure}
\def\baselinestretch{1.0}
  \centering{ \epsfig{file=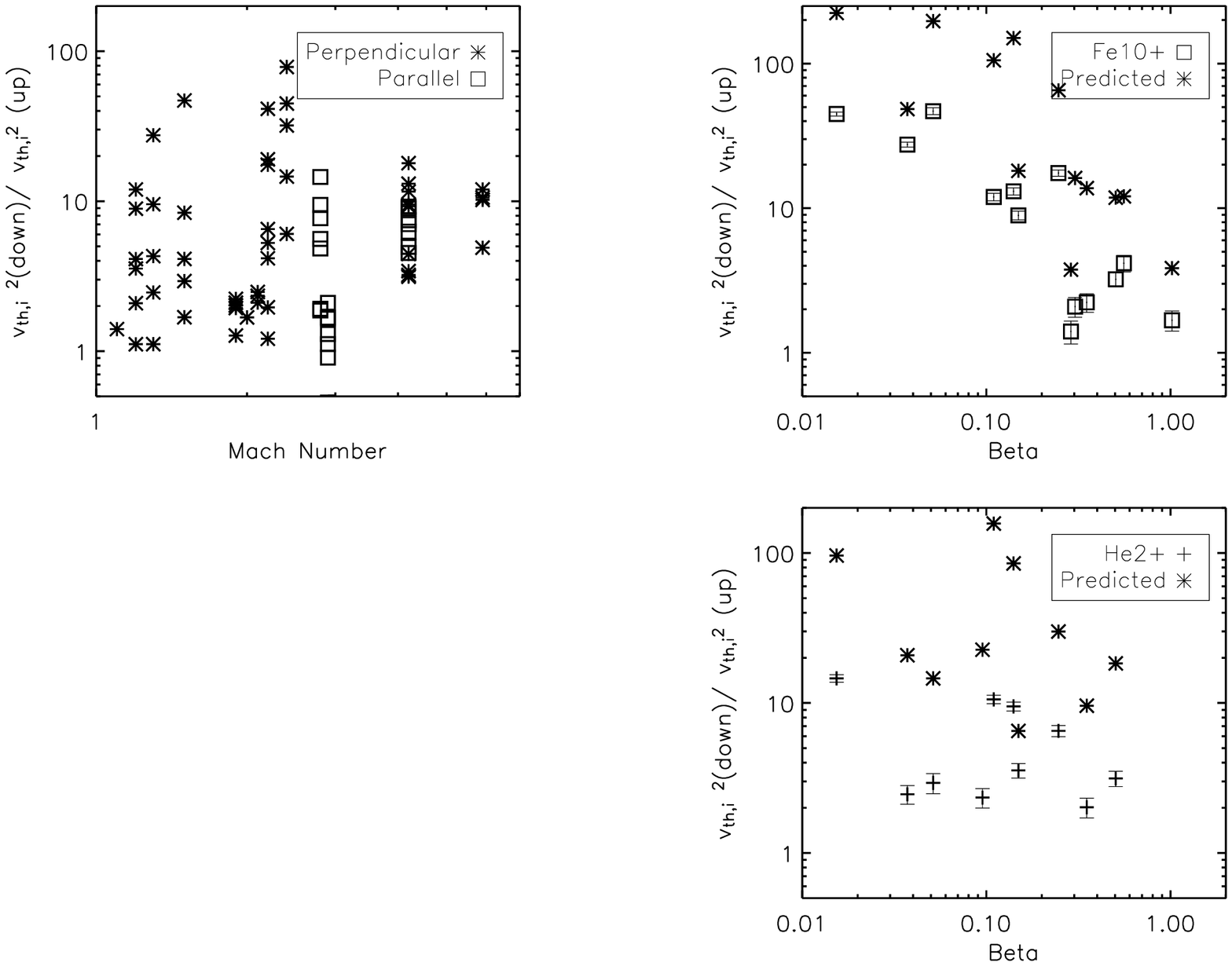,width=7cm,clip=,angle=90} \caption{ Plotting of the actual and predicted values for heating
  based on plasma $\beta$.  Iron shows similar trends to the
  prediction but the actual values are higher than those observed.\label{predictfe}} }
  \def\baselinestretch{2.0}
\end{figure}

There is an inverse dependence on plasma $\beta$ which is similar
to what was found in the analysis of the CME data.  To further
test the relationship, the predicted heating versus plasma $\beta$
was plotted in Figures \ref{predicthe} and \ref{predictfe}.

These figures show similar trends in the heating for both the
observational and the predicted values. However, the absolute
numerical value is higher for the predictions.  The uncertainty in
the measurements and of the calculated Mach number could account
for this difference.  A general dependence on heating of the
inverse of plasma $\beta$ coincides with the trends in the
observational data.

This accounts for the increase in heating due to the mass per
charge ratio of the ion.  However, in perpendicular shocks, there
could be an additional or alternative heating mechanism. The above
model assumed that there are no tangential forces acting on the
ions. The ions however are gyrating around the magnetic field with
a velocity v$_{gyro}$. Again if we assume the relationship between
the speed and the thermal velocity, we can see that this gyration
will add heating to the ions in the perpendicular case. The
perpendicular orientation of the magnetic field to the shock
normal means that the ions are gyrating in the x-z plane.  Most of
the velocity from the gyration is in the x-direction which
increases the thermal velocity.  This gyrating particle sees the
potential and increases the particle's thermal velocity with the
gyration velocity. In a parallel shock the gyration is in the y-z
plane and therefore most of the velocity is tangential to the
potential and does not affect the heating.

This accounts for the differences in heating as seen between the
parallel and perpendicular shocks.  The other dependence
highlighted by the current data set is the affects of plasma
$\beta$ on heating.  It is not obvious how this affects the
heating.

\section{Conclusions}
This study included an unprecedented data set of different ion
species to investigate the heating that occurs at a collisionless
shock front. Based on magnetic angle to the shock normal, each set
of shocks was analyzed for their dependence on Mach number and
plasma $\beta$. In a quasi-perpendicular shock, increasing $\beta$
decreases the heating highlighting the importance of the magnetic
field to the heating process. Mach number has little to no
correlation with ion heating. In a quasi-parallel shock,
increasing $\beta$ again decreases heating. Mach number however
does seem to enter into heating at the parallel shocks either.

An acceleration method based on a potential post-shock described
by \citet{fus97} may account for the heating observed in this
study. The model explains the heating that occurs at a
quasi-perpendicular shock but does not seem to match well with a
parallel shock.  The parallel shock heating is closer to mass
proportional because it is not as affected by the potential as the
ions in the perpendicular shock.

There is also currently similar ion heating data available for
collisionless shocks in supernova remnants showing
non-preferential heating to ions. This begs the question of the
ubiquitous nature of the shock physics: what is the dominant
factor to determine effect heating at a collisionless shock front?
The supernova has a Mach number 10 times that of the CME shocks
however, as seen in this data set, the Mach number does not seem
to play a major role in determining heating.   Density and
magnetic energy seem to be of greater importance.  Although not
examined in this study, a shock precursor can be an addition
source of plasma heating.  Preheating or a precursor would explain
the high heating seen at some of these shock fronts. The next
chapter explores neutrals as a source of a precursor and for their
role in heating.



\def\baselinestretch{1.0}

\chapter{The Effects of Neutrals at Collisionless Shock Fronts}

\def\baselinestretch{2.0}

\goodbreak
\section{Introduction}
Heating at a collisionless shock front was diagnosed by in situ
measurements of CMEs and by UV spectral data in SN1006.  The
thermal speed of pre-shock and post-shock heavy ions in CMEs were
compared to describe heating.  The spectral line width describes
the thermal broadening of the H$\alpha$ and OVI lines in SN1006.
The ISM around a supernova is thought to be highly ionized from
the UV flash that occurs when the supernova detonates. However,
through H$\alpha$ observations and subsequent modeling
\cite{gha02} concluded that the photons from the UV flash from the
initial supernova explosion were not enough to pre-ionize the ISM
to the current shock position. Therefore, the SN1006 shock is
interacting with a partially neutral medium.  The bow shock of the
heliosphere also propagates into a semi-neutral ISM. As a fast
shock propagates, neutrals will be quickly ionized in the hot
plasma downstream of the shock. However, some of the neutrals will
be excited and emit H$\alpha$ radiation before being ionized. This
emission represents the pre-shock conditions before the physical
signatures of the collisionless shock are masked by Coulomb
collisions.

\begin{figure}
\def\baselinestretch{1.0}
  \centering{
  \epsfig{file=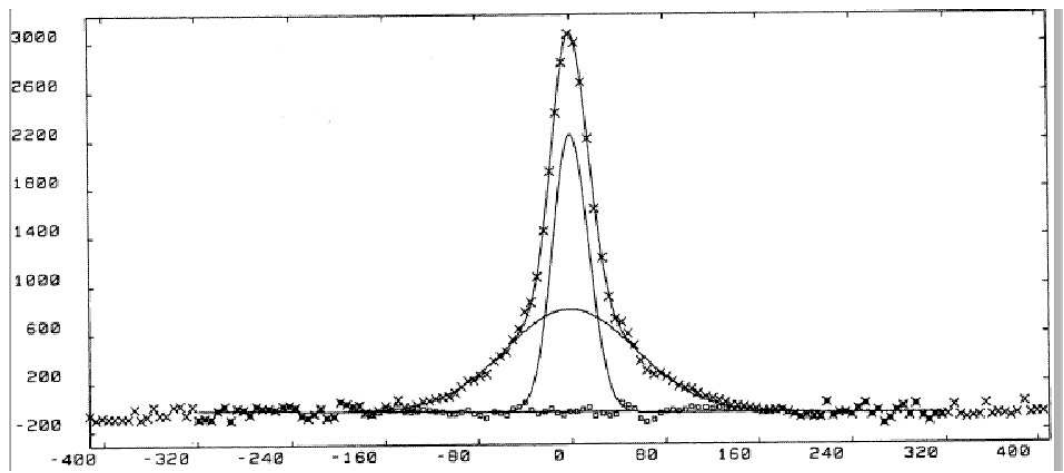,width=14cm}
  \caption{The plot, taken from \citet{hes94}, plots the change is velocity (which indicates change in wavelength) versus intensity. The observed data are ploted as crosses.
  The fit to this spectral line has two
  components shown as the solid line without symbols-the broad component and the narrow component.
  \label{halphaeg}}
  }
  \def\baselinestretch{2.0}
\end{figure}

Neutrals at a collisionless shock front are therefore important to
understand as a diagnostic for the pre-shock conditions as well as
the interpretation of the observed spectral lines.  In addition,
neutrals, because they are not tied to the magnetic fields can run
upstream preheating the upstream plasma.  This can create a
precursor.  The study of neutrals is based on the method for
diagnosing the degree of ion-electron thermal equilibrium of a
shock front using the broad to narrow component intensity ratio of
the H$\alpha$ spectral lines \citep{cr78}. The interactions that
produce the two components, narrow and broad, of H$\alpha$
emission are sensitive to neutral fraction. These two components
are seen in Figure \ref{halphaeg} taken from \citet{hes94}
observations of the Cygnus Loop supernova remnant.

There are two types of neutrals and two types of protons at the
shock front, fast and slow.  Both types of neutrals and protons
are needed to obtain the broad and narrow component of the
H$\alpha$ line.  Fast protons are those that initially start
downstream of the shock. Fast neutrals are fast protons that have
undergone charge exchange to become neutral. Slow protons and slow
neutrals are from the upstream population.  The narrow line is
produced when a slow neutral is excited by an electron or a
proton.  The broad component is formed in a two step process.
First, a fast proton must charge exchange to create a fast
neutral.  The fast neutral must then be excited to emit the
H$\alpha$ broad component.

The theoretical ratio of intensities of the broad to narrow
components can be calculated by considering the three processes
involved in the creation of these components: charge transfer,
ionization, and excitation.  \citet{kcr} derived the following
relation:

\begin{equation}
\frac{I_{b}}{I_{n}}=\frac{(\sigma_{x}v)_{s}}{(\sigma_{i}v)_{f}}[\frac{\epsilon_{A}}{\epsilon_{B}}
+\frac{g_{\alpha}}{\epsilon_{B}}(1+\frac{(\sigma_{x}v)_{f}}{(\sigma_{i}v)_{f}})
]
\end{equation}

where\\
$\epsilon$=efficiency of H$\alpha$ emission\\
case A: optically thin medium for the broad component\\
case B: optically thick medium for the narrow component\\
($\epsilon$ ranges from 0.1-0.8)\\
$\frac{(\sigma_{x}v)_{f}}{(\sigma_{i}v)_{f}}$=the probability a
fast neutral will undergo charge transfer\\ g$_{\alpha}$=the
fraction of emission in H$\alpha$ from charge transfer into an
excited state\\
$\frac{(\sigma_{x}v)_{s}}{(\sigma_{i}v)_{f}}$=ratio of charge
transfer to ionization rates.\\

Fast neutrals created downstream of the shock can migrate upstream
of the shock, preheating the unshocked material.  There are two
steps in computing the pre-shock heating.  First, one must compute
how much energy is carried upstream by these fast neutrals.
Second, the fraction of energy that can be deposited upstream
before the particle is swept back through the shock to the
downstream side must be calculated.

A fast neutral precursor can contaminate the narrow component of
the H$\alpha$ line.  An upstream fast neutral can undergo charge
transfer or ionization and then excite a narrow neutral component.
The number of fast neutrals that can come upstream is still an
open question that this study examines.  In addition to affecting
the intensity profile of the two H$\alpha$ components, neutrals
upstream of the shock can modify the jump conditions of the shock
by travelling from the downstream region to the upstream region
changing the energy, mass, and momentum distribution, physically
broadening the transition region around the discontinuity of the
shock.  This causes a ramp-like structure to form at the shock
front.  The orientation of the magnetic field also affects the
ease of transport of an ion upstream or downstream.  The length
scale involved in heating the upstream plasma is of importance to
understand how much time the shock interacts with the plasma and
at what rate energy is transferred.

The plasma at the shock front in the ISM is essentially a three
species fluid made up of an electron, proton and neutral
components. Each of the three components has a different effect on
the shock and is affected differently by the passage of the shock.
The charged fluids are affected by the magnetic field whereas the
neutral fluid evolves unconstrained by the magnetic field.
Multifluid flows have been studied numerically by \cite{dra80} and
\cite{flo85}. Chevalier \& Raymond (1978) and Chevalier, Kirshner
\& Raymond (1980), studied the neutral fluid of shocks in
supernova remnants through observations of the broad component
line width and the ratio of intensities of the H$\alpha$ broad and
narrow component. Lim \& Raga (1994) simulated the effect of high
pre-shock ionization fraction on strong shocks. However, magnetic
effects, velocity perpendicular to the flow, and the atomic
processes at high energies were not simulated.

In this chapter, a 2-D Monte Carlo particle simulation follows the
neutrals as they pass through the shock front. The pre-shock
ionization fraction is varied to characterize the effect of the
presence of neutrals at the shock front.  This simulation results
in H$\alpha$ broad to narrow intensity ratios with varying shock
speed, magnetic angle, and pre-shock ionization fraction.

\medskip
\section{Model Description}

A 2-D MHD particle Monte Carlo model was created to follow a
neutral particle as it moves through the shock front and interacts
with the surrounding plasma.  The simulation space was chosen to
be four times the ionization mean free path for a hydrogen atom
($\lambda$$_{mfp}$ $\sim$ 1 x 10$^{16}$ cm) assuming a density of
1 particle per cm$^{3}$.  These dimensions ensure that the shock
front related emission is all within the computational grid and
that any relevant physical interactions would occur within the
calculation space.  We followed 1 $\times$ 10$^{5}$ neutral
particles in order to get statistically significant results.

The density was assumed to be one.  The mean free path distance
scales are then proportional to $\frac{1}{n}$.  The intensities
scale as n so that the broad to narrow intensity ratio is
independent of n.  The simulation is done in the frame of
reference where the shock is at rest. Therefore the upstream
plasma is moving towards the shock at the velocity of the shock,
v$_{s}$, and the downstream plasma is moving away from the shock
at a quarter of the velocity of the shock, 1/4 v$_{s}$. The
pre-shock (upstream) plasma is assumed to be in local
thermodynamic equilibration; the protons, electrons, and neutrals
are in thermal contact and therefore a single fluid. The neutral
fluid is assumed to be in a pre-shock Maxwell-Boltzman
distribution.  The post-shock (downstream) characteristics such as
density and temperature are calculated from the Rankine-Hugoniot
conservation conditions across the shock, assuming a strong shock.
These are read in from input files that detail these
characteristics versus distance from the shock front.  The
resulting temperatures for each species separates the plasma into
three species.

\begin{figure}
\def\baselinestretch{1.0}
  \centering{ \epsfig{file=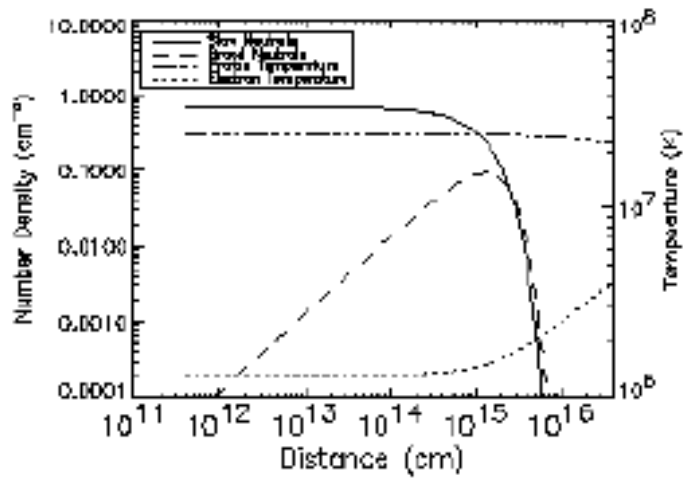,width=7cm,clip=,angle=90}
  \caption{The neutral density and electron and proton temperature
  plotted for a 1000 km s$^{-1}$ shock with 30$\%$ neutral
  fraction for an $\alpha$=0.1.  Note the neutral population decreases significantly within one ionization mean free path. \label{predone} } }
  \def\baselinestretch{2.0}
\end{figure}

The velocity, position, and timing of the atomic interactions were
recorded throughout the code.  Input files specified the
downstream temperature and density as seen in Figure
\ref{predone}. The temperatures for protons and electrons are
plotted along with the density of neutrals and the fast neutrals
which would produce a broad component.
The temperatures are relatively steady throughout the simulation
space of 4 $\times$ 10$^{16}$ cm.  However, the number of slow
neutrals available to produce narrow emission drops off sharply
within one ionization mean free path.  The neutrals that are
created from fast protons, the broad neutrals, are created and can
exist further downstream than the slow narrow emission neutrals.
Neutrals that become protons downstream quickly become thermalized
with the hot downstream population. This implies all of the narrow
emission comes from just behind the shock front within one
ionization mean free path whereas the broad emission is more
dispersed spatially.

\begin{figure}
\def\baselinestretch{1.0}
  \centering{
  \epsfig{file=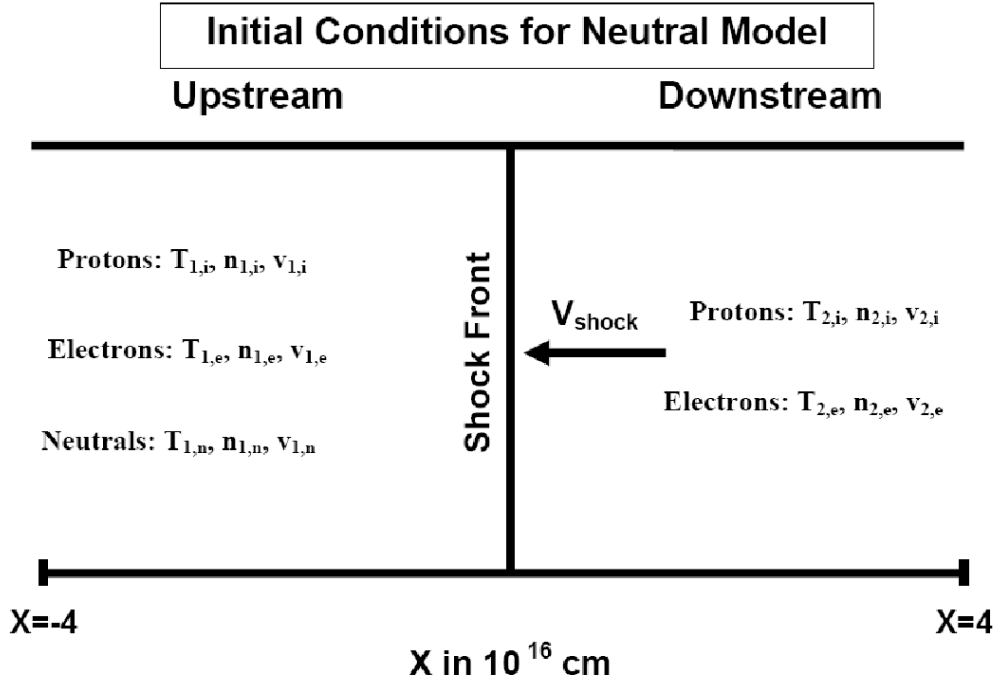,width=14cm}
  \caption{The initial setup of the neutral model.  There are four mean free path of length on either side of the shock.  The \label{shockmodel}}
  }
  \def\baselinestretch{2.0}
\end{figure}

 A neutral particle is chosen randomly from this
distribution and is tracked starting at the shock front shown as
x=0 in Figure \ref{shockmodel}. A random thermal velocity is
chosen from a gaussian distribution and assigned to the neutral
particle. Time and distance are incremented and the probability of
each of the three interactions (ionization, excitation, or charge
transfer) are calculated.

Although the interaction of neutrals with the shock front is the
driving cause of the simulations, many factors contribute to this
interaction.  Detailed in Table \ref{parameterstable}, the shock
speed, neutral fraction, magnetic field and orientation, and the
degree of temperature equilibration between electrons and protons,
$\alpha$, are varied to study the heating with respect to neutrals
at the shock front.

\begin{table}[h]
\def\baselinestretch{1.0}
\centering \caption{Summary of Parameters Studies for Neutral
Simulation \label{parameterstable}}

\smallskip
\begin{tabular}[h]{|c|c|}

\hline
 Parameter & Value \\\hline

 Speed (km s$^{-1}$) & 300, 500, 1000, 1500, 2000, 2500, 3000\\ \hline
 Neutral Fraction& 0.2, 0.3, 0.5, 0.9\\\hline

Magnetic Field (Gauss)& 3 $\times$ 10$^{-6}$\\ \hline

Magnetic Angle (Degrees)& 0, 90, 45 \\\hline

$\frac{T_{e}}{T_{p}}$&0.1,0.5,0.9\\\hline
\end{tabular}
\end{table}
\def\baselinestretch{2.0}

\medskip
\section{Atomic Interactions}

Three atomic interactions or a lack of interaction were possible
for a particle at each time step: ionization, charge transfer,
H$\alpha$ emission (excitation), or simple transport with no
atomic processes taking place. The atomic interaction was then
chosen by a Monte Carlo method based on the calculation of the
probabilities of each of the three atomic interactions.

In general, the probability that a specific atomic process could
take place is defined as
\begin{equation}
P=tn \sigma v
\end{equation}
where\\
n=number density of target particles \\
t=time (time step in the model)\\
$\sigma$=cross section\\
v=velocity of particle.\\

To calculate the atomic rates of these interactions, several
sources were used.  For many of the rates, direct numerical
integration over two Maxwellian distributions was preformed by
\citet{lam96} and \citet{lam90} to find the cross section of the
interaction.

The charge transfer or electron capture reaction is as follows:
\begin{equation}
H^{+} + H \rightarrow H + H^{+}.
\end{equation}
Charge transfer occurs between fast protons and slow neutrals or
slow protons and fast neutrals.  For charge transfers occurring
downstream of the shock and charge transfer between a slow neutral
and a fast proton, data from the \citet{lam96} and \citet{lam90}
papers were used.

For upstream charge transfer, the calculation for the cross
sections were taken from the Redbook Atomic Data Tables
\citep{bar90}. The authors used a Chebyshev fitting method to fit
experimental atomic data for this interaction. The minimum energy
of an incoming proton is E$_{min}$=1.28 $\times$ 10$^{-1}$ eV/amu
with a maximum incoming proton energy of E$_{max}$=6.3 $\times$
10$^{5}$ eV/amu. This encompasses the entire energy range of
interest for this simulation.  This cross section was then
multiplied by the velocity and the density of the target particles
in order to obtain a rate.

Ionization of a hydrogen atom in the simulation can occur by a
interaction with either a proton or an electron, as shown by
Equations \ref{pion} and \ref{eion}.
\begin{equation}\label{pion}
H + H^{+}\rightarrow 2H^{+} + e^{-}
\end{equation}
or by electrons
\begin{equation}\label{eion}
H + e^{-} \rightarrow H^{+} + 2e^{-}
\end{equation}

The reaction rate for ionization by a proton upstream when the
thermal velocity of the ambient protons are small is given by
\citet{fan95}. The rate for ionization by an electron is taken
from \citet{sch}.   The downstream ionization rate was calculated
from the Laming et al. papers.

The excitation of protons to produce the Balmer $\alpha$ line
emission involves the transition of a hydrogen atom from the
ground state to an excited state and the subsequent decay back to
the ground state with the emission of a photon, h$\nu$.
\begin{equation}
H + e^{-}\rightarrow H^{*} + e^{-}\rightarrow H + h\nu
\end{equation}

\begin{equation}
 H + p^{+}  \rightarrow H^{*} +p^{+} \rightarrow H + h\nu
\end{equation}
For the excitation of a slow neutral downstream, the rates were
taken from the Laming et al. papers. Excitation could also occur
of a fast neutral. The probability that this excitation would
occur again used the Laming et al. papers. Upstream, excitation of
a slow neutral can occur. The rate for excitation by an electron
into the 3p state and the combined 3s and 3d states are taken from
\citet{calad}.

Once these probabilities were calculated at a specific location
and time, they were fed into a subroutine that used a Monte Carlo
algorithm to choose the interaction.  Each interaction was then
recorded and the velocity of the particle was modified.  If the
particle was ionized, it was assumed to leave the system and was
no longer tracked because the new proton will quickly thermalize
with the downstream plasma. Downstream of the shock, if the
particle was excited the emission was recorded but the velocity
left unchanged. If charge transfer occurred the proton that was
created is assumed to quickly thermalize with the downstream
plasma and is not tracked. The fast neutral produced by the
downstream charge transfer was the particle that was followed.

Upstream of the shock, H$\alpha$ emission did not affect the
velocity. Ionization or charge exchange with a fast neutral
created a fast proton upstream. This was then tracked to
understand the preheating of the plasma that could occur as
described in the next section.

\medskip
\section{Upstream Precursor Generated By Fast Neutrals}
When a proton undergoes charge exchange downstream, it becomes a
fast neutral.  This fast neutral, if oriented correctly, can pass
through the shock front upstream without interacting with the
magnetic field.  There is a possibility that the fast neutral will
be ionized or charge transfer and therefore become a fast proton
upstream. As this fast proton travels upstream, it heats the
surrounding environment. According to \citet{spi56}, Equation
\ref{energyloss} describes the rate that a fast particle loses
energy as

\begin{equation}\label{energyloss}
\frac{dE}{dt}=-\frac{m \omega^{2}}{t_{s}}
\end{equation}
where\\
$\omega$=velocity of the particle\\
m=mass of particle\\
t$_{s}$=time for energy exchange by Coulomb collisions\\

This is the basis of a neutral precursor. The amount of heat is
calculated based not only on Coulomb collisions, but also on the
generation of Lower Hybrid waves by the velocity component due to
gyration about the magnetic field.  Lower Hybrid waves have been
studied extensively in the heliosphere as a source of heating
\citep{cai02,sha98}. As studied by \citet{sha97} if there are
electrons present they are heated by the lower hybrid waves if the
Alfvenic velocity was less than the gyration speed. The energy
deposited upstream was then recorded. The energy that is deposited
due to wave interaction contributes to the thermal energy of the
electrons. The Coulomb heating is assumed to increase the thermal
energy of the protons.

\section{Results of Simulations}

\begin{figure}
\def\baselinestretch{1.0}
  \centering{ \epsfig{file=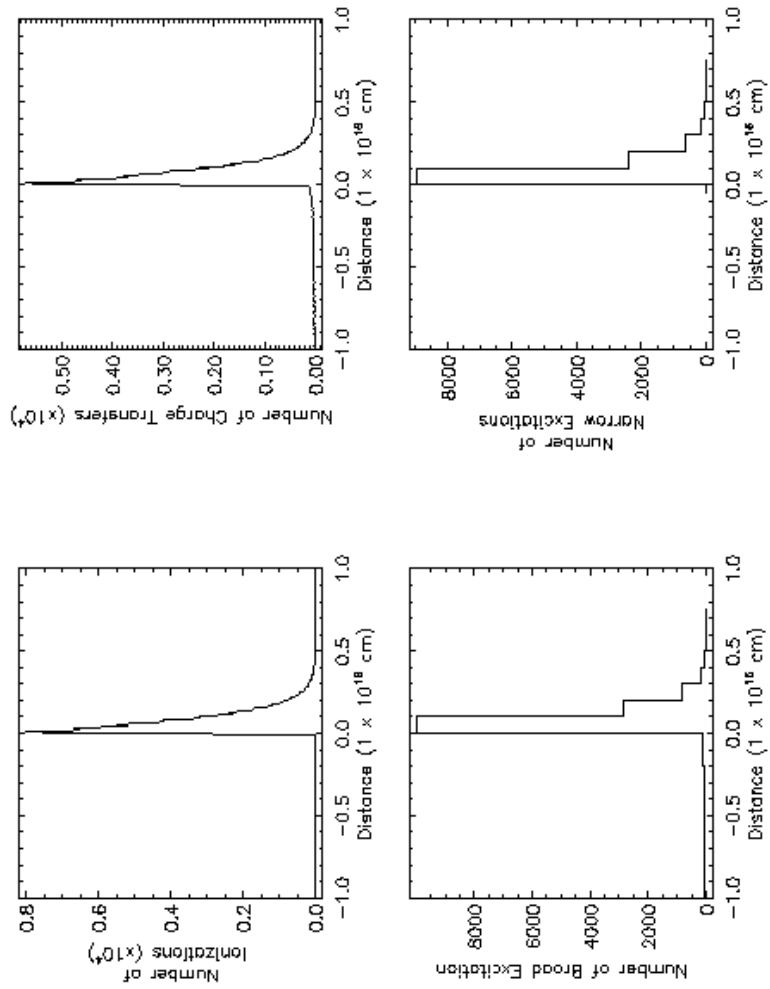,angle=180,width=14cm}
  \caption{Atomic interactions versus distance from the shock
  front for a 3000 km s$^{-1}$ shock with 30 $\%$ neutral fraction
  and $\alpha$=0.1 in a parallel shock.  The shock is at x=0,
  upstream is negative x and downstream is the positive x
  direction.
 \label{modelresults} } }
  \def\baselinestretch{2.0}
\end{figure}

Using a specific example, the atomic reactions were plotted versus
distance from a parallel shock front in Figure \ref{modelresults}.
The initial neutral fraction is 30$\%$, the shock speed was 3000
km s$^{-1}$ and $\alpha$ is 0.1.
The ionization of the neutrals is within one ionization mean free
path of the shock front.  The ionization drops off sharply after
the shock. The number of charge transfers also drops off sharply
downstream of the shock.  The broad excitation is centered on the
shock front.  In this example, there are 2000 interactions
upstream of the shock, which is 2\% of the total neutrals,
(1$\times$10$^{5}$), examined in the code.  The narrow emission
falls off quickly downstream and no emission is seen upstream.

\begin{figure}
\def\baselinestretch{1.0}
  \centering{
  \epsfig{file=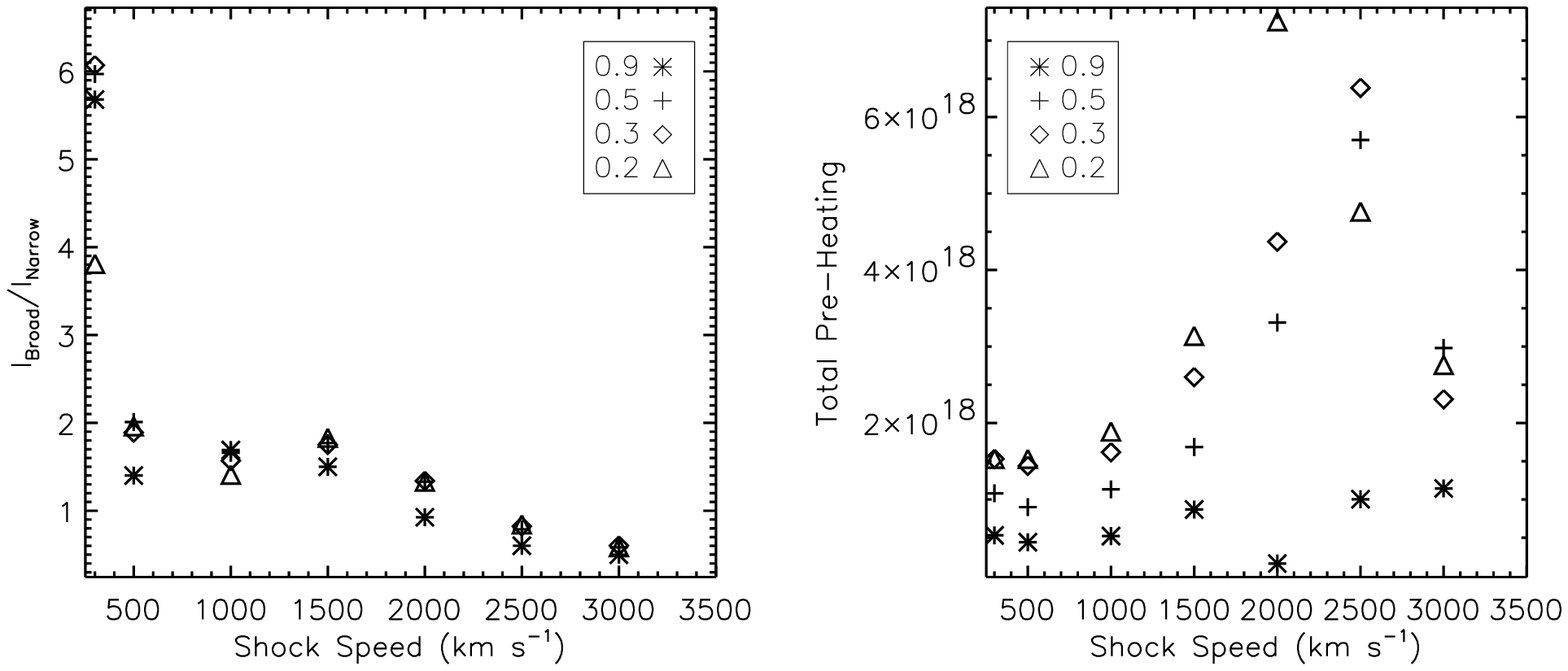,clip=,angle=90,width=6cm}
  \caption{Plot of the broad to narrow intensity ratio versus
  shock speed for a parallel shock and $\alpha$=0.1.  Variations
  in pre-shock neutral density is shown for 20\%, 30\%, 50\% and
  80\% as detailed in the legend. \label{r1par}  } }
  \def\baselinestretch{2.0}
\end{figure}

\begin{figure}
\def\baselinestretch{1.0}
  \centering{
  \epsfig{file=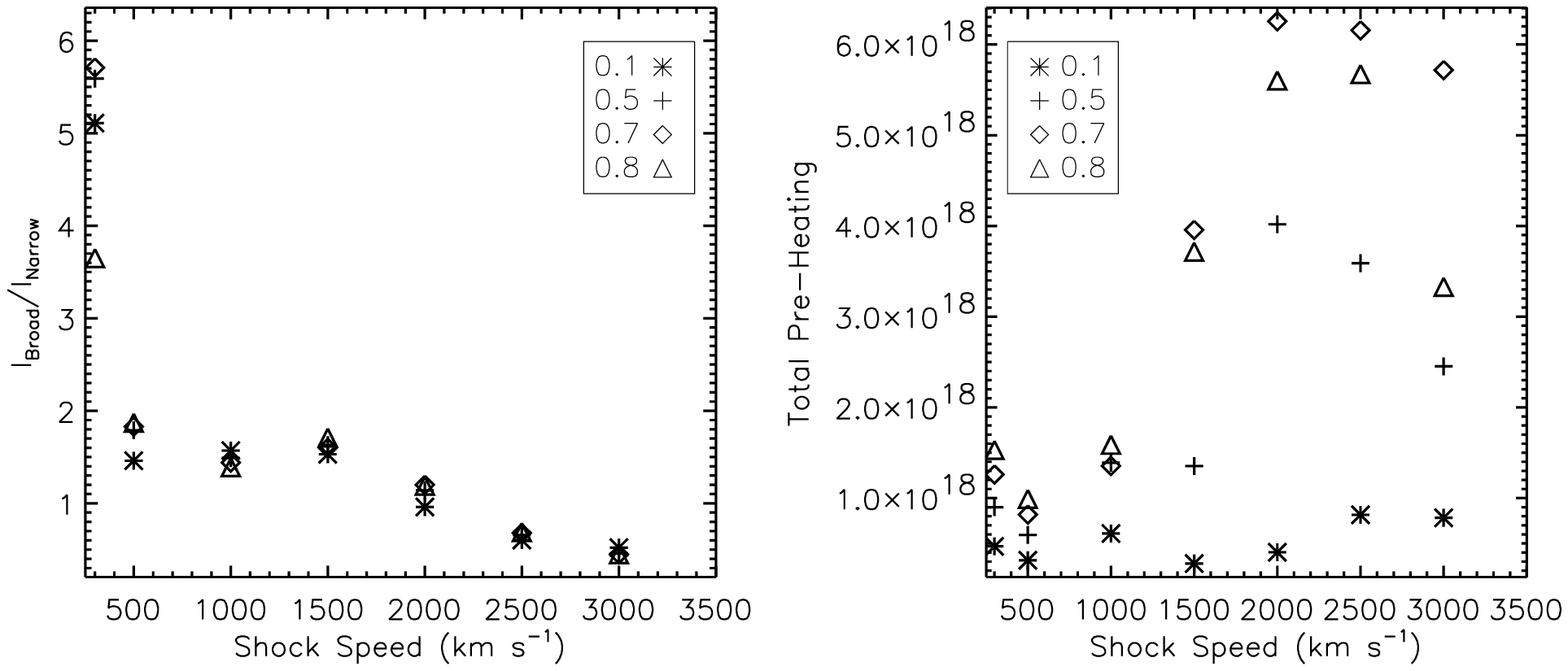,clip=,angle=90,width=6cm}
  \caption{Plot of the broad to narrow intensity ratio versus shock speed for a 45 degree shock and $\alpha$=0.1. Variations
  in pre-shock neutral density is shown for 20\%, 30\%, 50\% and
  80\% as detailed in the legend.\label{r1quasi}}
  }
  \def\baselinestretch{2.0}
\end{figure}

\begin{figure}
\def\baselinestretch{1.0}
  \centering{
  \epsfig{file=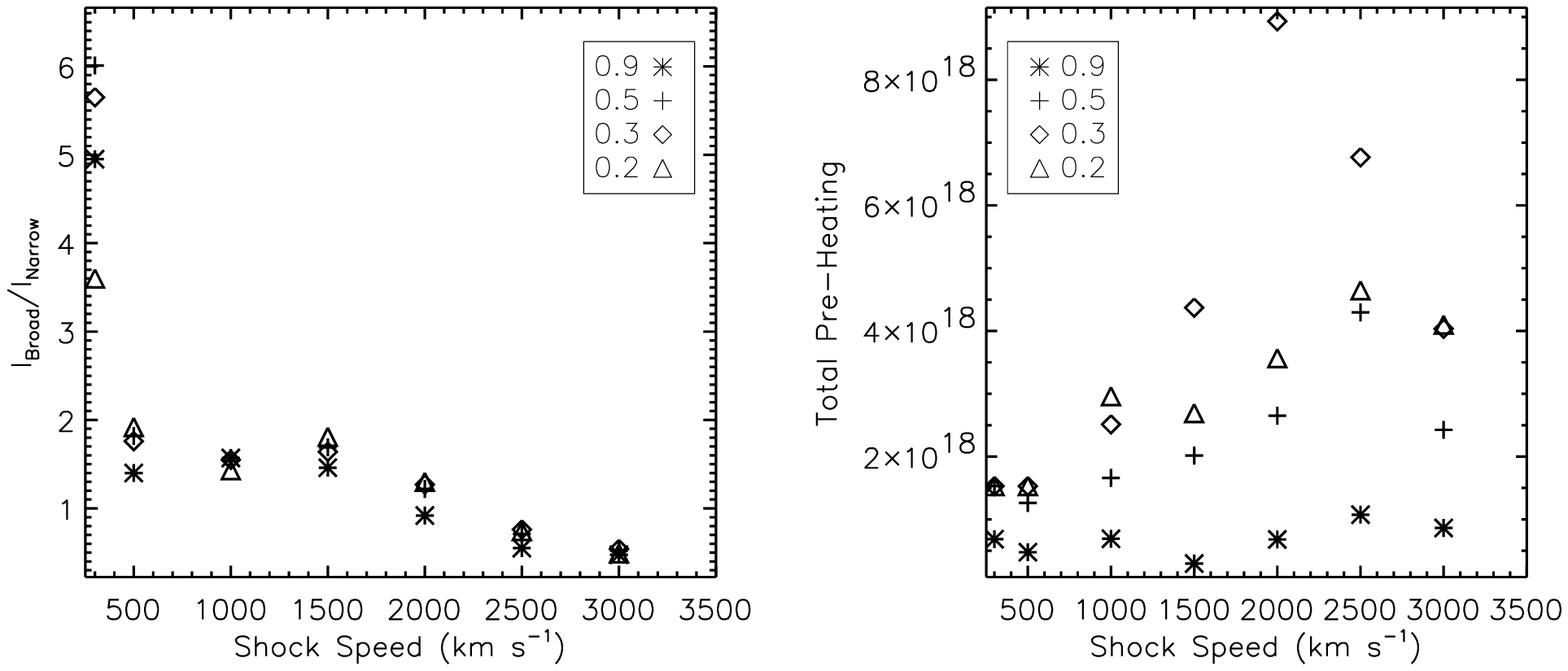,clip=,angle=90,width=6cm}
  \caption{Plot of the broad to narrow intensity ratio versus shock speed for a perpendicular shock and $\alpha$=0.1.Variations
  in pre-shock neutral density is shown for 20\%, 30\%, 50\% and
  80\% as detailed in the legend.\label{r1perp}}
  }
  \def\baselinestretch{2.0}
\end{figure}

In addition to keeping track of the interactions versus distance
from the shock front, the broad to narrow ratios versus shock
speed for each angle were plotted in Figures \ref{r1par} -
\ref{r1perp}. Each plotting symbol corresponds to specific neutral
fraction.

Figure \ref{r1par} plots the broad to narrow intensity ratio for
$\alpha$=0.1 for a parallel shock.  Although overall the intensity
ratios decrease with increasing shock speed, very little
difference is seen between the various neutral fractions. The
slowest shock studied with a velocity of 300 km sec$^{-1}$ was the
most efficient producer of the broad component.  Almost 12\% of
the neutrals came back upstream in this study.  The high amount of
neutrals were then available for creation of a broad component.
Figure \ref{r1quasi} plots the intensity ratio for a 45 degree
shock. The overall trend of the intensity ratio to decrease with
increasing shock speed still holds true. However, at this angle,
the various neutral fractions do affect the intensity ratio.  The
perpendicular plot, Figure \ref{r1perp}, is similar to the
parallel plot as they both indicate that the slowest shocks are
most effective in producing broad emission.  This is due to the
higher charge transfer rate and lower ionization rate at the shock
speed and neutral fraction.

\section{Discussion}
By varying the parameters of the study, the relative importance of
each parameter can be examined as to their contribution to the
formation of a broad and narrow component of the H$\alpha$
emission.

\subsection{Pre-shock neutral fraction}
The pre-shock neutral fraction was varied from 20$\%$ to 80 $\%$.
At high speeds, v$_{shock}$ $\ge$ 2000, the intensity ratio drops
because the charge transfer cross section declines.  The neutral
fraction did not affect the intensity ratios for large shock
speeds, indicating the decrease in charge transfer cross section
has a larger effect than the number of neutrals.  The intensity
ratios were affected by the neutral fraction when the shock speed
was below 500 km sec$^{-1}$.

\citet{lim96} did not find a large variation for a small neutral
fraction. However, they did not account for magnetic interaction
of the ratio of electron to proton temperature which we will now
examine.

\subsection{Electron and Proton Equilibration}
\begin{figure}
\def\baselinestretch{1.0}
  \centering{
  \epsfig{file=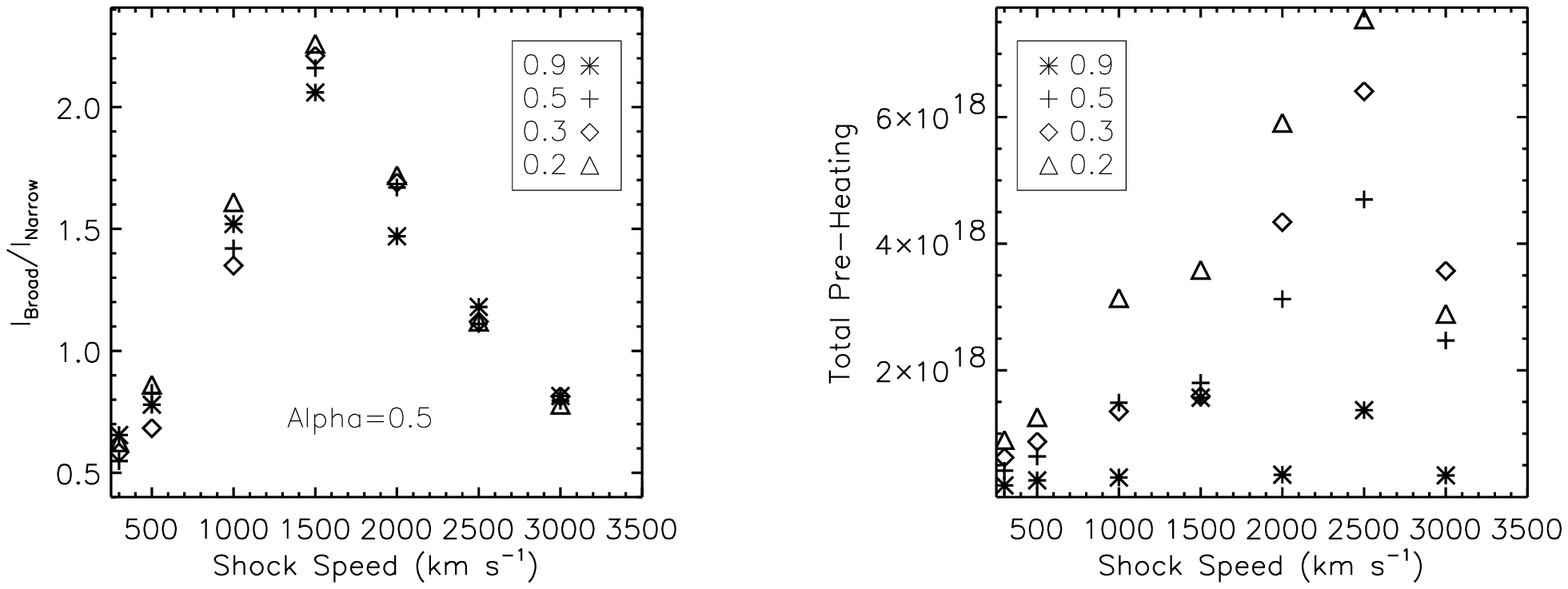,clip=,angle=90,width=6cm}
  \caption{Plot of the broad to narrow intensity ratio versus
  shock speed for a parallel shock and $\alpha$=0.5.\label{r2par}  }
  }
  \def\baselinestretch{2.0}
\end{figure}

\begin{figure}
\def\baselinestretch{1.0}
  \centering{
  \epsfig{file=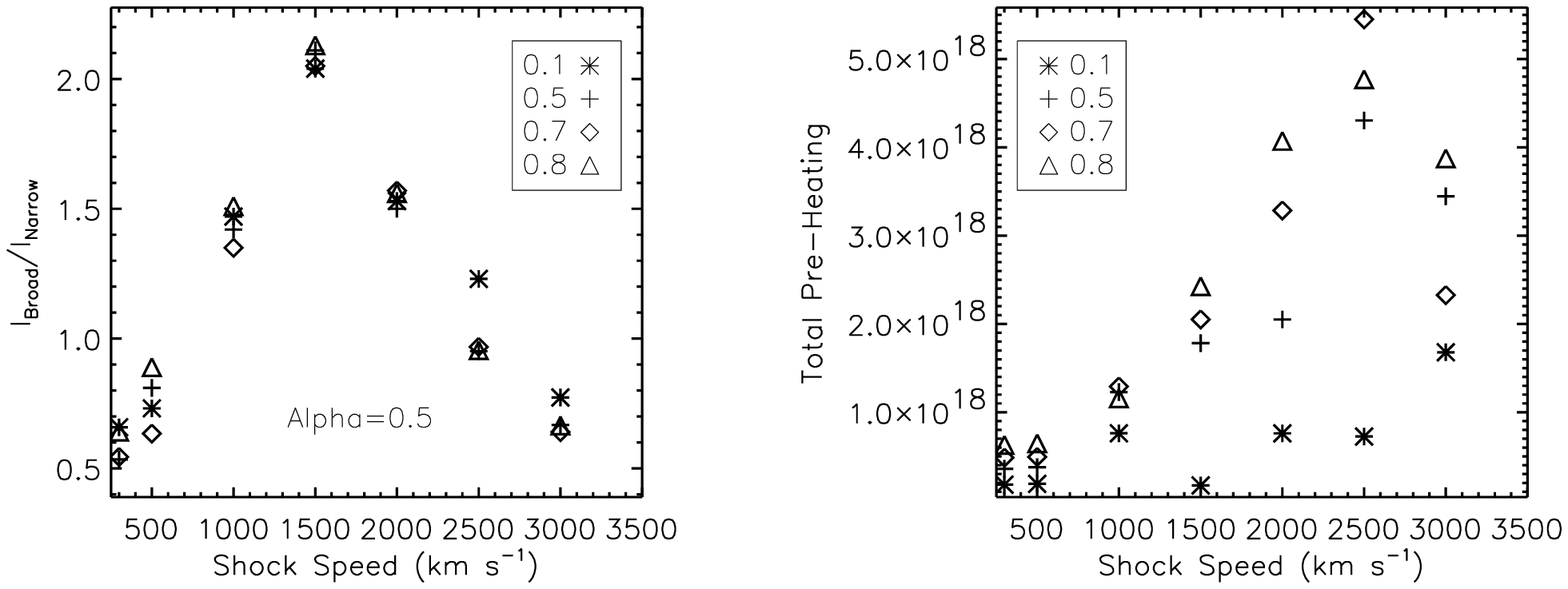,clip=,angle=90,width=6cm}
  \caption{Plot of the broad to narrow intensity ratio versus
  shock speed for a 45 degree shock and $\alpha$=0.5.\label{r2quasi} }
  }
  \def\baselinestretch{2.0}
\end{figure}

\begin{figure}
\def\baselinestretch{1.0}
  \centering{
  \epsfig{file=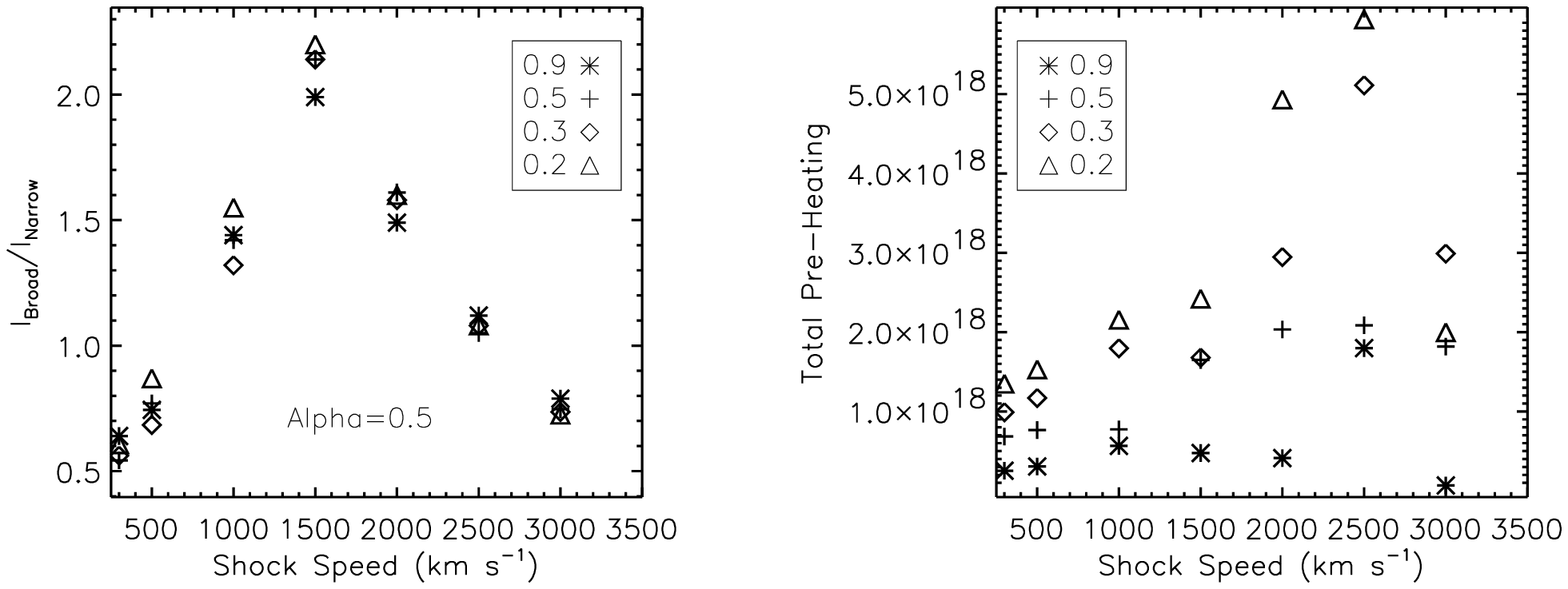,clip=,angle=90,width=6cm}
  \caption{Plot of the broad to narrow intensity ratio versus
  shock speed for a perpendicular shock and $\alpha$=0.5.\label{r2perp}}
  }
  \def\baselinestretch{2.0}
\end{figure}

\begin{figure}
\def\baselinestretch{1.0}
  \centering{
  \epsfig{file=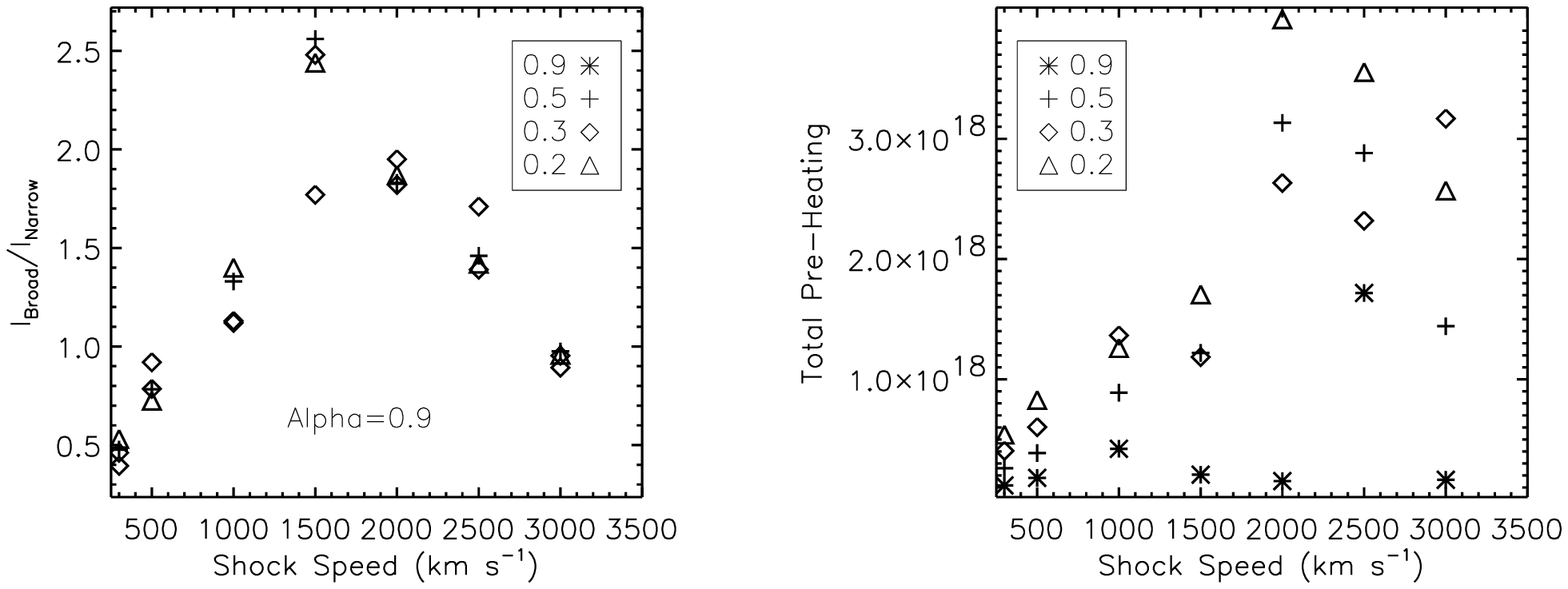,clip=,angle=90,width=6cm}
  \caption{Plot of the broad to narrow intensity ratio versus
  shock speed for a parallel shock and $\alpha$=0.9. \label{r3par} }
  }
  \def\baselinestretch{2.0}
\end{figure}

\begin{figure}
\def\baselinestretch{1.0}
  \centering{
  \epsfig{file=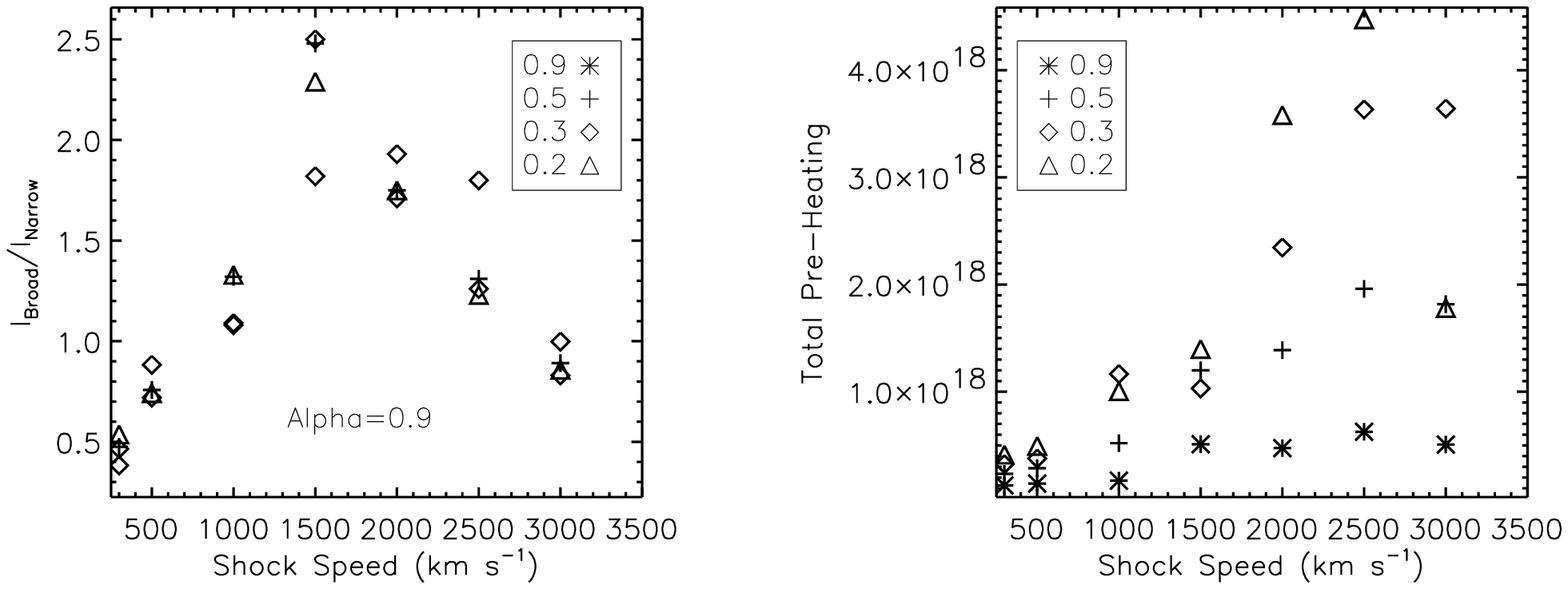,clip=,angle=90,width=6cm}
  \caption{Plot of the broad to narrow intensity ratio versus
  shock speed for a 45 degree shock and $\alpha$=0.9. \label{r3quasi}}
  }
  \def\baselinestretch{2.0}
\end{figure}

\begin{figure}
\def\baselinestretch{1.0}
  \centering{
  \epsfig{file=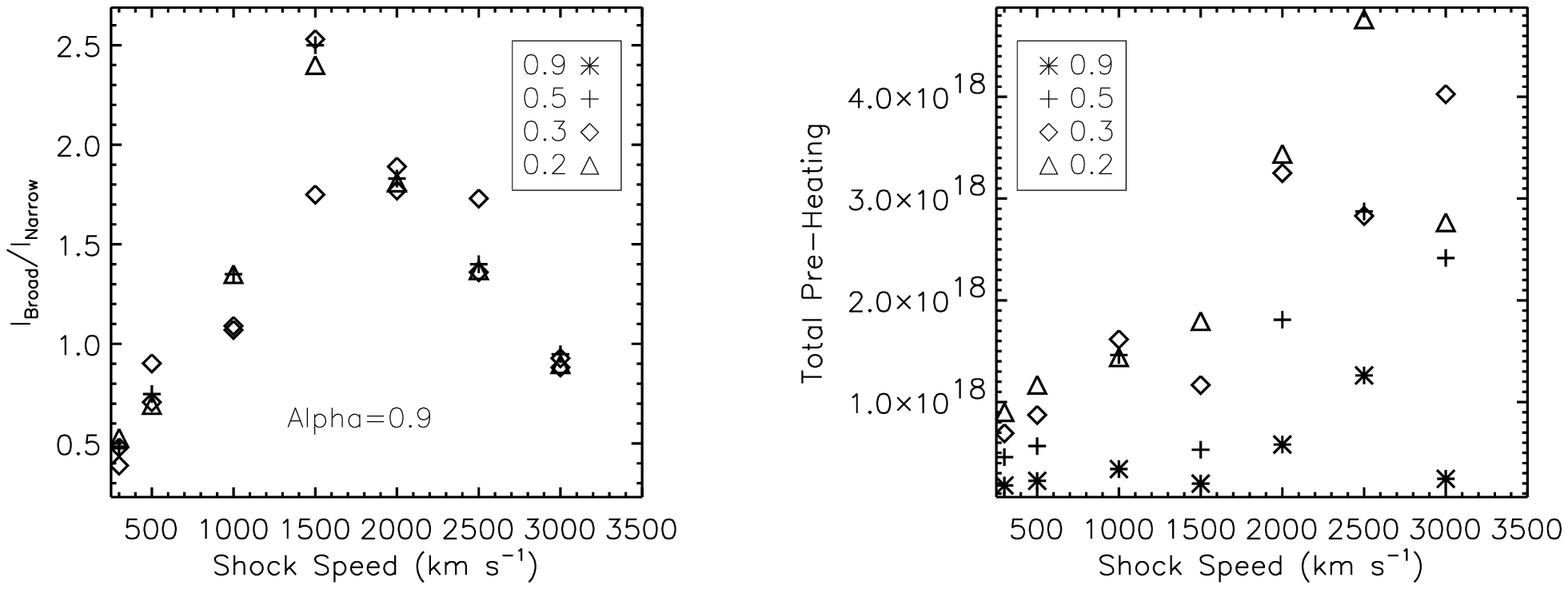,clip=,angle=90,width=6cm}
  \caption{Plot of the broad to narrow intensity ratio versus
  shock speed for a perpendicular shock and $\alpha$=0.9.\label{r3perp}}
  }
  \def\baselinestretch{2.0}
\end{figure}

The ratio of electron to proton temperature, $\alpha$, plays a key
role in the resulting broad to narrow intensity ratio.  From the
previous Figures, Figures \ref{r1par} - \ref{r1perp}, $\alpha$=0.1
shows little variation in the broad to narrow intensity ratio.
However, when the electron temperature is closer to the proton
temperature the broad to narrow ratio versus shock speed varies.
The intensity ratios increase until the shock speed is 1500 km
s$^{-1}$ and then the ratios decreases as seen in Figures
\ref{r2par} - \ref{r3perp}.


As was the case in the previous section, when a shock reaches
velocities higher than 2000 km s$^{-1}$, the cross section for
charge transfer for protons over 2 keV drops dramatically causing
fewer broad emissions to be produced.

\subsection{Magnetic Angle Effects}
The orientation of the magnetic field to the shock normal has been
shown in the previous chapters to play a role in heating of ions
as the shock passes.  Three orientations were examined in the
neutral simulation, $\theta$=0, 45, 90. These angles were chosen
to simulate a parallel, quasi-perpendicular, and perpendicular
shock. The orientation of the magnetic field as in the parallel
shock case can aid in a particle escaping upstream, whereas a
perpendicular shock would be more likely to prohibit the motion of
particles upstream. Each angle did produce different broad to
narrow ratios, although very little difference was seen between
the parallel and perpendicular cases. The parallel shock in
general consistently produced the highest broad to narrow ratios
for a given speed and neutral fraction.  The difference in heating
varied greatly between the perpendicular and parallel shock. This
will be discussed in a following section.

\subsection{Effect of Shock Speed on Neutral Signatures}
From past observations of supernova remnants, Table \ref{ibtable}
was constructed.  Shock speed is the one factor that is readily
available for comparison.  As seen in the previous sections, many
factors influence the intensity ratios.  The shock speed changes
this ratio significantly when the shock speed is greater than 1500
km s$^{-1}$.  Beyond this shock speed, the intensity ratio varies
little yet decreases with increasing shock speed.

The supernova remnants in Table \ref{ibtable} fall within the
range of values for the broad to narrow intensity ratio found
through this simulation. Further study would make this a useful
tool for predicting interactions and possibly even magnetic angle
and neutral fraction.

\begin{table}[h]
\def\baselinestretch{1.0}
\centering \caption{Summary of Broad to Narrow Intensity Ratios
for H$\alpha$\label{ibtable}}
\smallskip
\begin{tabular}[h]{|c|c|c|c|}

\hline

 Remnant & Ratio & v$_{shock}$& Model Ratio\\ \hline
 Tycho  & 1.08$\pm$0.16 & 1800$\pm$100\footnotemark{a,b}&1.04-2.4 \\\hline

SN1006 &0.73$\pm$0.06 &2890$\pm$100\footnotemark{c}& 0.47-1.1\\
\hline

 0519-69.0 &0.8 $\pm$0.2 & 1380$\pm$200\footnotemark{b}&1.0-1.4 \\\hline

 0548-70.4 & 1.1$\pm$0.2& 780$\pm$110 \footnotemark{b}& 1.0-1.4\\\hline
\end{tabular}
\\
\def\baselinestretch{1.0}
\footnotetext(a) \citet{kwc87}(b) \citet{smi91} (c) \citet{gha02}
\end{table}
\def\baselinestretch{2.0}

\subsection{Pre-Heating by Neutrals}
The simulation presented here set out to answer questions about
the pre-heating that could occur with a neutral population present
at a collisionless shock front.  It could easily be assumed that
as the number of neutrals increases the heating would also
increase.  In the perpendicular and parallel cases, little
variance was seen in the heating based on neutral fraction. The
one parameter that effected heating was the degree of thermal
equilibrium between electrons and protons. A magnetic orientation
of 45 degrees displayed a spread of heating versus initial neutral
fraction.

In the case of $\alpha$=0.1, there is a decrease in preheating
with increasing shock velocities.  The lowest pre-shock neutral
fraction, 20$\%$ has the most relative pre-heating.  For the cases
of $\alpha$=0.5 and 0.9, similar trends of increasing heating with
decreasing initial neutral fraction present. If the electrons are
cooler than protons and there are few neutrals, the heating
mechanism is relying on shock speed to increase heating. Whereas
when the electrons are closer in temperature to the protons, the
shock speed is not the dominate mechanism available for heating.
\begin{figure}
\def\baselinestretch{1.0}
  \centering{
  \epsfig{file=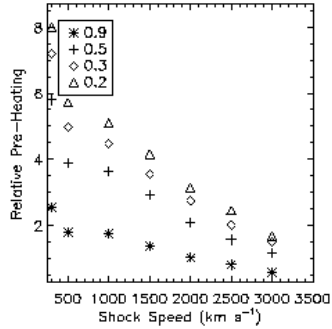,clip=,angle=90,width=5cm}
  \caption{Plot of the upstream heating versus shock speed for a 45 degree shock and $\alpha$=0.1. \label{heat1} }
  }
  \def\baselinestretch{2.0}
\end{figure}
\begin{figure}
\def\baselinestretch{1.0}
  \centering{
  \epsfig{file=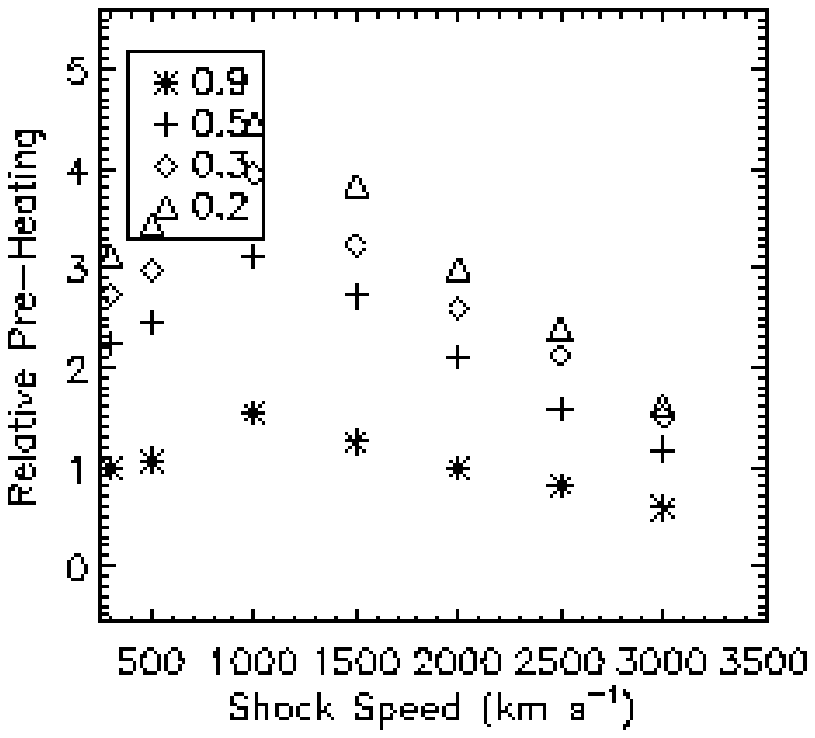,clip=,angle=90,width=5cm}
  \caption{Plot of the upstream heating versus shock speed for a 45 degree shock and $\alpha$=0.5.\label{heat2} }
  }
  \def\baselinestretch{2.0}
\end{figure}
\begin{figure}
\def\baselinestretch{1.0}
  \centering{
  \epsfig{file=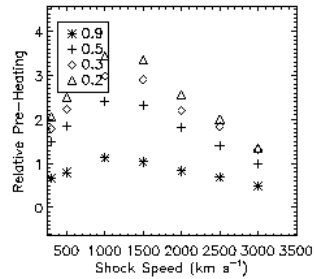,clip=,angle=90,width=5cm}
  \caption{Plot of the upstream heating versus shock speed for a 45 degree shock and $\alpha$=0.9.\label{heat3}}
  }
  \def\baselinestretch{2.0}
\end{figure}
To answer some of the fundamental questions about neutrals at the
shock front, a summary of the finding for heating of all shocks
follows.  The number of particles that made it upstream and were
available to perform pre-heating was about 9$\%$ of the total
particle simulation with as low as 2\% and as high as 14\% of the
particles preheating the shock. The heating varied based on
magnetic orientation of the shock. Per heating interaction, the
particles in the parallel shock gave up the most energy followed
by a quasi perpendicular shock.  The perpendicular shock displayed
the least amount of preheating.

\section{Conclusions}
Neutrals play a key role in the observational spectral from shocks
associated with supernova remnants.  The degree of temperature
equilibrium between electrons and protons played the largest role
in the outcome of the broad to narrow intensity ratios.  However,
once the degree of equilibrium approaches one, the shock speed and
the neutral fraction were more effective in changing and modifying
the intensity ratio.  The magnetic angle did play a role in
heating yet further study is necessary to conclude a trend.

A full 3-D MHD particle simulation would be ideal to understand
the physics of the neutrals at the shock front. However, using
this 2-D model observed data could be simulated.  Future work
could include the shock going through a non-homogenous or clumpy
material.  In addition, the magnetic field plays a key role in
particle transport that the study of magnetic inhomogeneities and
structure would also improve the simulations as well as MHD waves
and instabilities that are available to heat the plasma. Future
work could include heavier ions as they would act as another fluid
in the simulations. The interaction of heavy atoms as neutrals or
ions at low charge states would be an interesting application of
the code.



\def\baselinestretch{1}

\chapter{Conclusions and Future Work}

\def\baselinestretch{1.66}


\goodbreak
This thesis has performed the most comprehensive study to date of
the heating of ions and neutrals in collisionless shocks. The
heavy ion heating was examined in several different environments
as a function of M$_{A}$, $\beta$, and magnetic field orientation.
The heating based on mass fractionation was discussed in both
supernova remnants and the shocks that propagate before a CME. The
neutral population of a pre-shock plasma was also investigated to
examine the atomic processes involved in the electromagnetic
emission from a shock as well as to help interpret and understand
the observed astronomical data from shocks.  The integration of
different measurements allows us to use the data to make
predictions about magnetic field orientation and the magnetic
energy present, which is key to understanding particle heating and
acceleration.

\section{Summary of Work}
\subsection{Supernova Remnant Shock Heating}
In Chapter 2, the ion heating in SN1006 was examined using
observations from the FUSE Satellite.  Due to the faintness of the
current observations only the OVI and the Ly-$\beta$ spectral
lines were available to determine the less than mass proportional
heating.  Using past observations, the less than mass proportional
heating was confirmed.  The plasma $\beta$ of the SNR is $\sim$
0.1 which is similar to that of the heliospheric shocks studied
indicating that the heating process is magnetically dominated.
However, unlike the heliosphere the heavy ions in the supernova
are less than mass proportionally heated.

Supernova remnants present many obstacles that make studying them
difficult: distance from observation, few photons, no clear
indication of magnetic orientation, etc. Although more
observations and multi-wavelength studies could be integral to
understanding the current data, this thesis examines the plasma
physics of the heliosphere shocks and how we can apply the in situ
measurements to understand the astronomical data. This study is
opening new frontiers in understanding the magnetic field of an
astronomical object by using shock heating data.

\subsection{CMEs Heating}
Chapter 3 dealt with in situ data taken from the collisionless
shocks in front of Coronal Mass Ejections.  Each shock was
classified according to magnetic angle to the shock normal.  The
Mach number and the plasma beta were then examined for their
effect on the ion heating.  Quasi-perpendicular shocks were found
to heat ions more than quasi-parallel shocks.  The Mach number was
not found to have a definitive effect on the heating.

Parallel shocks heated the ions less than the perpendicular
shocks.  The heating decreased with increasing $\beta$.  A
correlation of the decrease in magnetic energy that is indicated
by an increasing $\beta$ with the decrease in heating implies that
magnetic effects are dominant in the heating process. A parallel
shock was also compared to a multi-fluid derivation of the Rankine
Hugoniot conditions.  For all but the energy term, the ion data
matched the expected results.  The energy term does include a
potential which could be derived from the difference between the
predicted and actual values.

Future observations in the heliosphere of CME shocks will be
conducted by STEREO.  The Solar TErrestrial RElations Observatory
(STEREO) Mission will explore the 3-D shock structure, which is of
importance to the evolution of the particle distributions. The
study of the deviations of the particle distributions from
Maxwellian is of interest to understand the heating, acceleration,
and relaxation processes in the solar wind.

\begin{figure}
\def\baselinestretch{1.0}
  \centering{ \epsfig{file=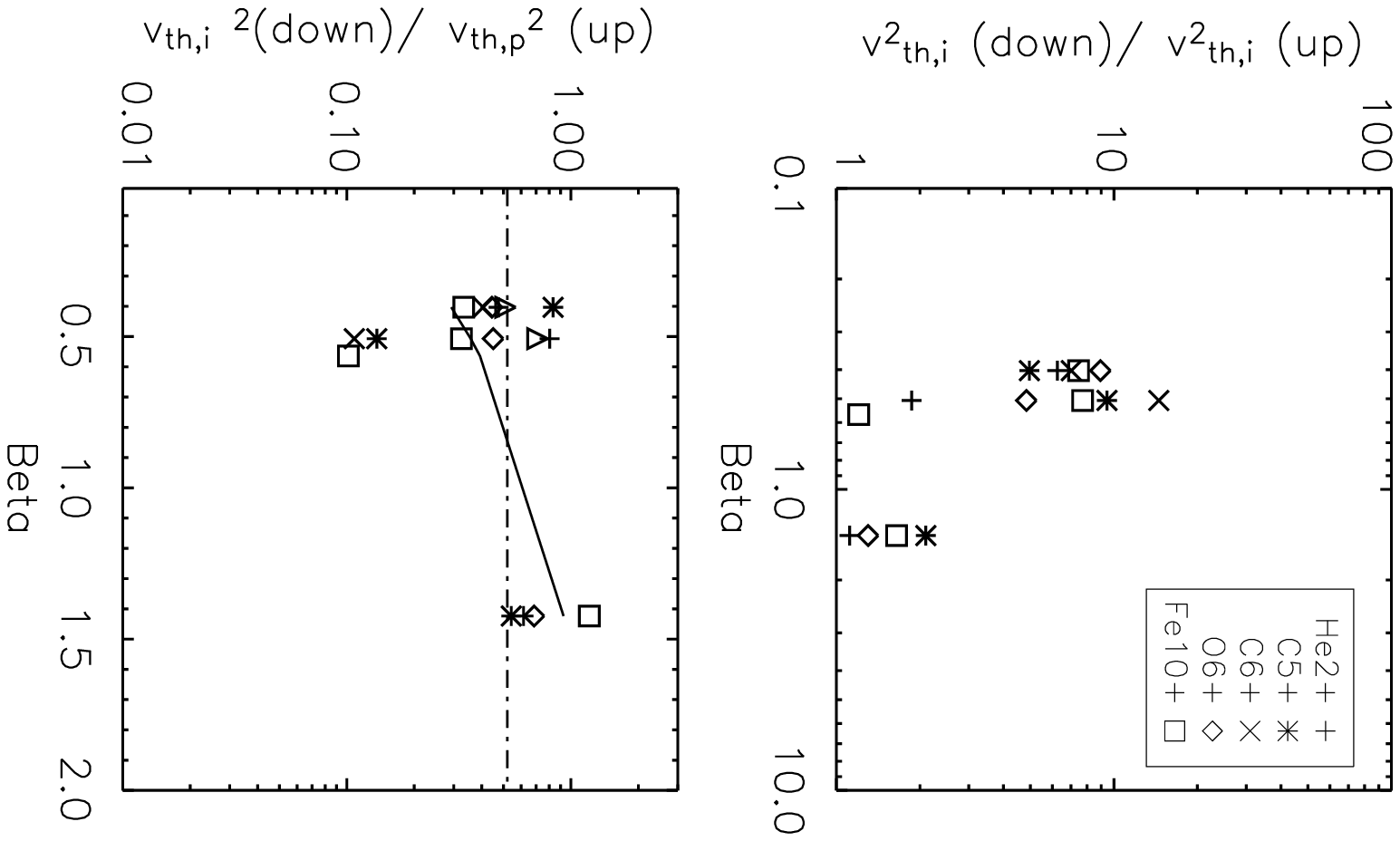,clip=,angle=90,width=11cm}
\caption{Downstream Temperature Ratio versus Plasma $\beta$ for
parallel shocks.  The x-axis is plasma $\beta$.  The y-axis is the
ratio of downstream ion temperature to proton temperature.  The
dashed horizontal line is the value for SN1006.  The solid line is
a fit to the heating of the parallel shocks in the study.  The
intersection of these two lines at $\beta$ =0.77 is the predicted
plasma $\beta$ for the supernova. \label{superpred}} }
  \def\baselinestretch{2.0}
\end{figure}

An interesting consequence of this thesis is the ability to use in
situ heliospheric data to make predictions and further examine
astronomical data. One such case is the prediction of the plasma
$\beta$ for the supernova environment based on the observed
heating.  A strong correlation of decreasing ion heating with
increasing plasma $\beta$ was found.  Using the data from the
Supernova 1006 study and the CME shock study for parallel shocks,
Figure \ref{superpred} was created.  The northwest region of
SN1006 is assumed to have a parallel shock due to the galactic
magnetic field orientation.  The x-axis is the plasma $\beta$. The
y-axis is the ratio of upstream heating of an ion to a proton.  A
linear fit to the CME data is shown as the solid black line.  The
heating seen in the supernova is the dashed horizontal line. Where
the dashed line meets the fit of the data is the predicted plasma
$\beta$ for the supernova.  The prediction for the plasma $\beta$
is 0.77. This varies from the $\beta$ calculated from the
densities obtained from observations. The value is a factor of 7
higher than the calculated $\beta$.  The higher value coincides
with the lack of magnetic energy in the region. In the future,
heliospheric trends can be utilized to obtain a range of probable
$\beta$ values for supernova remnant shocks and other astronomical
collisionless shocks.

\subsection{Neutral at the Shock Front: Source of Heating}
Chapter Four explored the role of neutrals at a collisionless
shock front.  Neutrals are important for producing emission lines
as well as the atomic interactions and our understanding of these
interactions.  The degree of thermal equilibrium between protons
and electrons played a surprising role in the broad to narrow
intensity ratios.

One important factor was found to be the orientation of the
magnetic field for the ratio of broad to narrow components.  Using
the broad to narrow intensity ratio from SN1006, we try to predict
the magnetic orientation of the shocks in the northeast and
northwest.  For SN1006, an $\alpha$ of $\sim$0.1 is
observed\citep{vink03}. Using Figures \ref{r1par} -\ref{r1perp}
and the shock velocity from Table \ref{ibtable}, one could
estimate that the shock is a quasi-perpendicular shock. This is
what the northeast rim of SN1006 would be if the assumptions about
the Interstellar magnetic field orientation is correct.  This also
is in line with the heating seen in the CME work.  The northeast
rim of SN1006 is the region where X-ray emission is bright as well
as the observations of gamma rays were reported \citep{bam03}.

This proves to be an interesting tool.  Magnetic fields and their
orientation are notoriously difficult to measure.  If the
intensity ratio and a measurement of the electron and proton
temperatures can be used to at least create a limit to the
magnetic parameter, much progress can be made in understanding
acceleration methods that rely on magnetic angle.

\goodbreak
\section{Collisionless Shocks in the Interstellar Medium}

As we have seen, collisionless shocks that form from interactions
within the interstellar medium (ISM) heat the ISM in ways that are
not well understood. As this dissertation has shown, shocks can
reveal many characteristics of plasma interactions such as
temperature, density, magnetic field strength, and ionization
state.  Our understanding of the physics of these heating
mechanisms is increased through the comparative study of these
shocks. For future work there are three main prongs of work that
would complement each other: analysis of multi-wavelength
observation shock data, theoretical shock characterization of the
system, and Magnetohydrodynamics (MHD) modeling of the systems
using the BATS-R-US 3-D MHD code. This unique blend of theory,
observation, and modeling gives a comprehensive view of the shock
and how it interacts and subsequently heats the plasma.

\subsection{Analysis of Observational Shock Data}
Supernova remnants, CMEs, and non-relativistic jets of stellar
systems all have collisionless shocks.  By looking at a broad
range of shocks both in the heliosphere and in astrophysical
systems, we study a larger range of parameters, such as velocity
and density, resulting in constraints and improved input for
future theoretical models.  To increase our understanding of shock
heating use of multi-wavelength data is necessary.

In this sense, radio observations of the jets and shocks would
prove useful. At radio wavelengths, synchrotron emission from
electrons will give a rough estimate of non-thermal electron
density.  If an estimate of the energy and mass outflow of the
jets is obtained, the mass infall rate can be estimated.
Supernovas such as Tycho, Kepler, and SN1006 have been observed in
the x-ray and radio.  The radio and X-ray data from the SNR could
be used to study shock trends such as the evolution of temperature
with time, velocity distribution of the particles and the broad to
narrow components of H$\alpha$ and shock emission.

 The first data set is of Cygnus Loop, the middle aged
supernova remnant with a shock speed of $\sim$350 km s$^{-1}$. The
remnant was observed with the MMT long slit spectrograph in the
3300-5000 \AA\  wavelength range.  After an initial review of the
data, several lines are available for study, Ne III, Ne V, OII and
Balmer H-$\beta$. The higher order H lines will more accurately
pinpoint the temperature and ionization state of the plasma.  The
data will afford a mass-to-charge ratio versus temperature
calculation as long as the radiation of the shock is low.  The
mass-to-charge ratio played a key factor in determining the
heating in shocks in the heliosphere in Chapter 3. Using the
varying speed and strength of prior observations of supernova, a
trend in the velocity versus the mass-to-charge ratio can be
calculated.

Another interesting data set is a Chandra observation of the jets
of the symbiotic stars of R Aquarii.  Chandra x-ray data show
magnetic interactions leading to accretion mechanisms and
acceleration of the plasma outward from the star.  More
observations of the system are planned as it reveals an evolution
of the jet with time.  The jets have been observed in the radio
(Stark et al. 1992) as well as the x-ray wavelengths. The current
observations show a collimated flow with a shock forming where the
jet meets the ISM. This system will be of key interest to the MHD
modelling efforts.

Solar wind data containing many shocks with varying parameters
will be obtained from the Advanced Composition Explorer (ACE)
satellite.  ACE is situated 240 R$_{E}$ upstream of the Earth.  We
will examine the data in an effort to improve the understanding of
the energetics of the solar wind. This also allows for an
examination of heavier ions such as Si and Mg to determine the
role of heating minor ions in a shock front. The solar data set
allows for a study of the distribution of thermal speeds of ions
at the shock front which could be non-Maxwellian leading to
information on the number of high energy particles available for
acceleration processes. The root mean square of the magnetic field
can be compared to thermal energy to test for turbulence.

\subsection{Theoretical Shock Characterization}

 The Rankine-Hugoniot(RH) conservation equations describe how the temperature,
pressure, and density pre-shock (upstream) relates to the
post-shock (downstream) temperature, pressure and density.
Magnetic fields can be included in these conservation equations
making them acceptable for use in systems where there is
significant magnetic energy.   Simplifications are made in many
approximations used for shock analysis.  Turbulence and viscosity
are ignored, as well as the temperature difference between
electrons and protons.   In addition, the distribution of
particles is generally assumed to be Maxwellian which can be far
from reality in shocked astrophysical plasmas.  These assumptions
need to be re-explored as the interpretation for more precise data
is needed.  A multi-fluid approach like that of \citet{bur91}, is
needed to successfully interpret shock data.

Because these shocks are expanding into a neutral ISM, the model
previously developed in Chapter 4 to study neutral interaction
will be used to characterize the shock.  The Monte Carlo code will
be updated to include the effect of turbulence, viscosity, and
temperature differences between species in the plasma.  In
addition, cosmic ray acceleration based on these shocks will be
considered by studying the ion heating and the distribution
functions of particles.

Most of the work done on shock heating originates from
heliospheric data.   Advances made in understanding shock heating
in the heliosphere will be applied to observations outside the
heliosphere.  A coherent scheme will be built to convert from the
particle detection methods in the heliosphere to the photon
detection methods in astronomy to further the ability to compare
the shocks and other types of data.  Results from the revised
Monte Carlo code will show possible emission mechanisms and
intensities of lines which will quantify the density, pressure and
temperature upstream and downstream.  The results of the study of
the emission mechanisms and data correlation will determine the
most important observable in a shock front and therefore enhance
modelling including these parameters. This will also allow us to
focus our observations on the characteristics that will lead to
the most fruitful data set. Feedback from the analysis of
observations will improve the theoretical model.

\subsection{MHD Modelling Using BATS-R-US}

BATS-R-US , a block adaptive grid code, is a first principles MHD
model that has been used to study the Earth's magnetosphere,
comets, and the heliosphere.  This code can be used for any system
that satisfies the MHD conservation equations.  Using the MHD
model, simulations of the observed characteristics of the
jet-shock-ISM system will be performed and therefore be able to
better constrain the temperature, density, and magnetic field
present.  The first simulation will be of the R Aquarii binary
system.  Chandra observations reveal two x-ray jets coming from
the system (Kellogg et. al. 2001).  R Aquarii is a symbiotic star
system made up of a Mira-type mass-losing variable with a white
dwarf companion. The system is relatively close to the Earth (~200
pc) facilitating the study of these jets. The system's orbit is
highly inclined (70 degrees) to the line of sight with the
accretion disk of the hot companion edge-on.  The magnetic field
is assumed to be a dipole field that is aligned with the axis of
rotation of the white dwarf, which for simplicity, is taken to be
perpendicular to the plane of the orbit.  The simulations would
start with the simplest assumptions of a three species plasma
(protons, electrons, and neutrals).  The input into the code will
require the density, temperature and magnetic field of the white
dwarf and the accretion disk.  The resulting output would be the
jet density, temperature and magnetic field.  The output will also
contain a time evolution of the system.  This model will place
tight limits on the temperatures and densities originally needed
to produce the jet therefore constraining the temperature and
density of the binary system. BATS-R-US has been used to simulate
other jet systems.  A jet sheet that is formed at the intersection
of the heliosphere and the ISM has been found by modeling done by
Opher et al. 2004.  These high resolution models simulate the
pressure, temperature, speed, and magnetic field of the system. In
addition, supernova shocks of SN1006 and Cygnus loop will be
simulated to determine the debated magnetic field orientation and
density.  Through this similarity study, BATS-R-US can be used as
a predictive tool for shocks and jets.

Solar Physics has a plethora of in situ data and physical models
backing up the data analysis.  Astronomical observational
capabilities are beginning to be able to resolve objects with
sizes similar to that of the heliosphere.  However, it is still a
great challenge to understand the physics based on the emission
mechanisms for the few photons observed.  In summary, this
research would use multi-wavelength observations to study shocks,
use these observations to better model the system and then use an
advanced 3-D MHD code, BATS-R-US to simulate the time evolution of
the system.  Eventually this research will use the combination of
a shock physics model and the MHD code to predict and analyze
observed data.  Future studies will use the MHD code as a
predictive and analytic tool for the jets associated with HH
objects, protoplanetary disks, comets, coronal mass ejections, and
supernova shocks.  The completion of this project will give me a
solid foundation to work to bridge astrophysics and space physics.

\goodbreak
\smallskip

\goodbreak
\section{Future of Collisionless Shock Research}
The cross disciplinary mix of physics, astronomy, and space
science can only help to further each of the disciplines.  This
thesis used a unique combination of systems from each of the
disciplines in order to understand the heating of heavy ions and
neutrals in collisionless shocks.  This facilitates the
advancement of plasma astrophysics and the understanding of the
cosmic ray puzzle.

Future work in collisionless shock research lies in a
multi-wavelength, multi-system approach. By using the models
already in existence and theories from different subfields of
astrophysics, the understanding of collisionless shock physics
will increase dramatically.

\bibliographystyle{astron}
\bibliography{kekbib}

\begin{thebibliography}{}

\bibitem[\protect\astroncite{{Alfv{\' e}n}}{1945}]{alf45}
{Alfv{\' e}n}, H.: 1945,
\newblock {\em \mnras} {\bf 105}, 3

\bibitem[\protect\astroncite{Anderson et~al.}{2003}]{fuse03}
Anderson, B., Sankrit, R., and Dupuis, J.: 2003,
\newblock {\em FUSE Observer's Guide}

\bibitem[\protect\astroncite{{Bale} et~al.}{2002}]{bal02}
{Bale}, S.~D., {Hull}, A., {Larson}, D.~E., {Lin}, R.~P., {Muschietti}, L.,
  {Kellogg}, P.~J., {Goetz}, K., and {Monson}, S.~J.: 2002,
\newblock {\em \apjl} {\bf 575}, L25

\bibitem[\protect\astroncite{Bamba et~al.}{2003}]{bam03}
Bamba, A., Yamazaki, R., Ueno, M., and Koyama, K.: 2003,
\newblock {\em \apj} {\bf 589}, 827

\bibitem[\protect\astroncite{{Barnett} et~al.}{1990}]{bar90}
{Barnett}, C.~F., {Hunter}, H.~T., {Fitzpatrick}, M.~I., {Alvarez}, I.,
  {Cisneros}, C., and {Phaneuf}, R.~A.: 1990,
\newblock {\em NASA STI/Recon Technical Report N} {\bf 91}, 13238

\bibitem[\protect\astroncite{{Baumjohann} and {Treumann}}{1997}]{bau97}
{Baumjohann}, W. and {Treumann}, R.~A.: 1997,
\newblock {\em {Basic space plasma physics}},
\newblock London: Imperial College Press, |c1997

\bibitem[\protect\astroncite{{Belcher} et~al.}{1993}]{bel93}
{Belcher}, J.~W., {Lazarus}, A.~J., {McNutt}, R.~L., and {Gordon}, G.~S.: 1993,
\newblock {\em \jgr} {\bf 98(17)}, 15177

\bibitem[\protect\astroncite{Berdichevsky et~al.}{1997}]{ber97}
Berdichevsky, D., Geiss, J., Gloeckler, G., and Mall, U.: 1997,
\newblock {\em \jgr} {\bf 102}, 263

\bibitem[\protect\astroncite{{Burgi}}{1991}]{bur91}
{Burgi}, A.: 1991,
\newblock {\em \jgr} {\bf 96(15)}, 17689

\bibitem[\protect\astroncite{{Cairns} and {Zank}}{2002}]{cai02}
{Cairns}, I.~H. and {Zank}, G.~P.: 2002,
\newblock {\em \grl} {\bf 29}, 47

\bibitem[\protect\astroncite{{Callaway}}{1994}]{calad}
{Callaway}, J.: 1994,
\newblock {\em Atomic Data and Nuclear Data Tables} {\bf 57}, 9

\bibitem[\protect\astroncite{Cane and Richardson}{2003}]{cra03}
Cane, H. and Richardson, I.: 2003,
\newblock {\em \jgr} {\bf 108}, 1156

\bibitem[\protect\astroncite{Carroll and Ostlie}{1996}]{car96}
Carroll, B. and Ostlie, D.: 1996,
\newblock {\em An Introduction to Modern Astrophysics},
\newblock Addison Wesley

\bibitem[\protect\astroncite{{Chandrasekhar}}{1984}]{cha84}
{Chandrasekhar}, S.: 1984,
\newblock {\em Science} {\bf 226}, 497

\bibitem[\protect\astroncite{Charles and Seward}{1995}]{cha95}
Charles, P. and Seward, F.: 1995,
\newblock {\em Exploring the X-ray Universe},
\newblock Cambridge University Press

\bibitem[\protect\astroncite{Chen}{1984}]{che84}
Chen, F.: 1984,
\newblock {\em Introduction to Plasma Physics and Controlled Fusion},
\newblock Plenum Press

\bibitem[\protect\astroncite{Chevalier et~al.}{1980}]{kcr}
Chevalier, R., Kirshner, R., and Raymond, J.: 1980,
\newblock {\em \apj} {\bf 235}, 186

\bibitem[\protect\astroncite{Chevalier and Raymond}{1978}]{cr78}
Chevalier, R. and Raymond, J.: 1978,
\newblock {\em \apj} {\bf 225}, L27

\bibitem[\protect\astroncite{{Chevalier}}{1982}]{che82}
{Chevalier}, R.~A.: 1982,
\newblock {\em \apjl} {\bf 259}, L85

\bibitem[\protect\astroncite{Draine}{1980}]{dra80}
Draine, B.: 1980,
\newblock {\em \apj} {\bf 241}, 1021

\bibitem[\protect\astroncite{Ellison and Reynolds}{1991}]{ell92}
Ellison, D. and Reynolds, S.: 1991,
\newblock {\em \apj} {\bf 382}, 242

\bibitem[\protect\astroncite{{Fang} et~al.}{1995}]{fan95}
{Fang}, C., {Feautrier}, N., and {Henoux}, J.-C.: 1995,
\newblock {\em \aap} {\bf 297}, 854

\bibitem[\protect\astroncite{Finland}{1998}]{spg98}
Finland, S. P. G.~O.: 1998,
\newblock {\em Space Physics Textbook},
\newblock spaceweb@oulu.fi

\bibitem[\protect\astroncite{Fisk et~al.}{1974}]{fis74}
Fisk, L., Kozlovsky, B., and Ramaty, R.: 1974,
\newblock {\em \aplett} {\bf 190}, L35

\bibitem[\protect\astroncite{{Fisk}}{2003}]{fis03}
{Fisk}, L.~A.: 2003,
\newblock {\em Journal of Geophysical Research (Space Physics)} {\bf 108(A4)},
  7

\bibitem[\protect\astroncite{Flower et~al.}{1985}]{flo85}
Flower, D., des Forets, G.~P., and Hartquist, T.: 1985,
\newblock {\em \mnras} {\bf 216}, 775

\bibitem[\protect\astroncite{Forbes}{2000}]{for00}
Forbes, T.: 2000,
\newblock {\em \jgr} {\bf 105(A10)}, 23153

\bibitem[\protect\astroncite{{Fuselier} and {Schmidt}}{1997}]{fus97}
{Fuselier}, S.~A. and {Schmidt}, W.~K.~H.: 1997,
\newblock {\em \jgr} {\bf 102(11)}, 11273

\bibitem[\protect\astroncite{{Garrard} et~al.}{1998}]{gar98}
{Garrard}, T.~L., {Davis}, A.~J., {Hammond}, J.~S., and {Sears}, S.~R.: 1998,
\newblock {\em Space Science Reviews} {\bf 86}, 649

\bibitem[\protect\astroncite{Ghavamian et~al.}{2002}]{gha02}
Ghavamian, P., Winkler, P., Raymond, J., and Long, K.: 2002,
\newblock {\em \apj} {\bf 572}, 888

\bibitem[\protect\astroncite{{Gloeckler} et~al.}{1998}]{glo98}
{Gloeckler}, G., {Cain}, J., {Ipavich}, F.~M., {Tums}, E.~O., {Bedini}, P.,
  {Fisk}, L.~A., {Zurbuchen}, T.~H., {Bochsler}, P., {Fischer}, J.,
  {Wimmer-Schweingruber}, R.~F., {Geiss}, J., and {Kallenbach}, R.: 1998,
\newblock {\em Space Science Reviews} {\bf 86}, 497

\bibitem[\protect\astroncite{{Gombosi}}{1999}]{gom99}
{Gombosi}, T.~I.: 1999,
\newblock {\em {Physics of the Space Environment}},
\newblock Physics of the Space Environment, ISBN 052159264X, Cambridge
  University Press, 1999.

\bibitem[\protect\astroncite{Goodman}{1992}]{goo92}
Goodman, R.: 1992,
\newblock {\em LAKOTA STAR KNOWLEDGE: STUDIES IN LAKOTA STELLAR THEOLOGY 2nd
  Edition},
\newblock Sinte Gleshka University Press,
\newblock information at P.O. Box 490 Rosebud SD 57570-0490 605-856-2368

\bibitem[\protect\astroncite{{Gosling} et~al.}{1997}]{gos97}
{Gosling}, J.~T., {Bame}, S.~J., {Feldman}, W.~C., {McComas}, D.~J., {Riley},
  P., {Goldstein}, B.~E., and {Neugebauer}, M.: 1997,
\newblock {\em \grl} {\bf 24}, 309

\bibitem[\protect\astroncite{{Gosling} et~al.}{1974}]{gos74}
{Gosling}, J.~T., {Hildner}, E., {MacQueen}, R.~M., {Munro}, R.~H., {Poland},
  A.~I., and {Ross}, C.~L.: 1974,
\newblock {\em \jgr} {\bf 79(18)}, 4581

\bibitem[\protect\astroncite{{Habbal} et~al.}{1997}]{hab97}
{Habbal}, S.~R., {Woo}, R., {Fineschi}, S., {O'Neal}, R., {Kohl}, J., {Noci},
  G., and {Korendyke}, C.: 1997,
\newblock {\em \apjl} {\bf 489}, L103+

\bibitem[\protect\astroncite{{Hefti}}{1998}]{hef98}
{Hefti}, S.: 1998,
\newblock {\em SWICS/ACE Technical Note}

\bibitem[\protect\astroncite{{Hester} et~al.}{1994}]{hes94}
{Hester}, J.~J., {Raymond}, J.~C., and {Blair}, W.~P.: 1994,
\newblock {\em \apj} {\bf 420}, 721

\bibitem[\protect\astroncite{Hetherington}{1996}]{cha00}
Hetherington, B.: 1996,
\newblock {\em A chronicle of pre-telescopic astronomy},
\newblock John Wiley and Sons.,
\newblock Paul Charbonneau,Great Moments in the History of Solar Physics
  copyright2000

\bibitem[\protect\astroncite{{Horowitz} and {Li}}{1999}]{hor99}
{Horowitz}, C.~J. and {Li}, G.: 1999,
\newblock {\em Physical Review Letters} {\bf 82}, 5198

\bibitem[\protect\astroncite{Jones and Pye}{1988}]{jp88}
Jones, L. and Pye, J.: 1988,
\newblock {\em \mnras} {\bf 238}, 567

\bibitem[\protect\astroncite{{Kahler}}{1992}]{kah92}
{Kahler}, S.~W.: 1992,
\newblock {\em \araa} {\bf 30}, 113

\bibitem[\protect\astroncite{{Kaufmann}}{1991}]{kau91}
{Kaufmann}, W.: 1991,
\newblock {\em Universe 3rd. Ed.},
\newblock W.H. Freeman and Company, New York

\bibitem[\protect\astroncite{{Kennel} et~al.}{1985}]{ken85}
{Kennel}, C.~F., {Edmiston}, J.~P., and {Hada}, T.: 1985,
\newblock {\em Washington DC American Geophysical Union Geophysical Monograph
  Series} {\bf 34}, 1

\bibitem[\protect\astroncite{Kirshner et~al.}{1987}]{kwc87}
Kirshner, R., Winkler, P., and Chevalier, R.: 1987,
\newblock {\em \apj} {\bf 315}, L135

\bibitem[\protect\astroncite{{Korreck} et~al.}{2004}]{kor04}
{Korreck}, K.~E., {Raymond}, J.~C., {Zurbuchen}, T.~H., and {Ghavamian}, P.:
  2004,
\newblock {\em \apj} {\bf 615}, 280

\bibitem[\protect\astroncite{{Laming}}{1990}]{lam90}
{Laming}, J.~M.: 1990,
\newblock {\em \apj} {\bf 362}, 219

\bibitem[\protect\astroncite{{Laming}}{2004}]{lam04}
{Laming}, J.~M.: 2004,
\newblock {\em \apj} {\bf 604}, 874

\bibitem[\protect\astroncite{{Laming} et~al.}{1996}]{lam96}
{Laming}, J.~M., {Raymond}, J.~C., {McLaughlin}, B.~M., and {Blair}, W.~P.:
  1996,
\newblock {\em \apj} {\bf 472}, 267

\bibitem[\protect\astroncite{Lee and Wu}{2000}]{lee00}
Lee, L. and Wu, B.: 2000,
\newblock {\em \apj} {\bf 535}, 1014

\bibitem[\protect\astroncite{{Lee} et~al.}{1987}]{l87}
{Lee}, L.~C., {Mandt}, M.~E., and {Wu}, C.~S.: 1987,
\newblock {\em \jgr} {\bf 92(11)}, 13438

\bibitem[\protect\astroncite{{Lee} et~al.}{1986}]{lee86}
{Lee}, L.~C., {Wu}, C.~S., and {Hu}, X.~W.: 1986,
\newblock {\em \grl} {\bf 13}, 209

\bibitem[\protect\astroncite{{Lembege} et~al.}{2004}]{lem04}
{Lembege}, B., {Giacalone}, J., {Scholer}, M., {Hada}, T., {Hoshino}, M.,
  {Krasnoselskikh}, V., {Kucharek}, H., {Savoini}, P., and {Terasawa}, T.:
  2004,
\newblock {\em Space Science Reviews} {\bf 110}, 161

\bibitem[\protect\astroncite{{Leroy} et~al.}{1982}]{ler82}
{Leroy}, M.~M., {Winske}, D., {Goodrich}, C.~C., {Wu}, C.~S., and
  {Papadopoulos}, K.: 1982,
\newblock {\em \jgr} {\bf 87(16)}, 5081

\bibitem[\protect\astroncite{Lim and Raga}{1995}]{lim95}
Lim, A. and Raga, A.: 1995,
\newblock {\em \mnras}

\bibitem[\protect\astroncite{Lim and Raga}{1996}]{lim96}
Lim, A. and Raga, A.: 1996,
\newblock {\em \mnras} {\bf 280}, 103

\bibitem[\protect\astroncite{Long et~al.}{2003}]{lon03}
Long, K., Reynolds, S., and Raymond, J.: 2003,
\newblock {\em \apj} {\bf 586}, 1162

\bibitem[\protect\astroncite{{MacQueen}}{1980}]{mac80}
{MacQueen}, R.~M.: 1980,
\newblock {\em Royal Society of London Philosophical Transactions Series A}
  {\bf 297}, 605

\bibitem[\protect\astroncite{{McComas} et~al.}{1998}]{mcc98}
{McComas}, D.~J., {Bame}, S.~J., {Barker}, P., {Feldman}, W.~C., {Phillips},
  J.~L., {Riley}, P., and {Griffee}, J.~W.: 1998,
\newblock {\em Space Science Reviews} {\bf 86}, 563

\bibitem[\protect\astroncite{{McKee} and {Ostriker}}{1977}]{mck77}
{McKee}, C.~F. and {Ostriker}, J.~P.: 1977,
\newblock {\em \apj} {\bf 218}, 148

\bibitem[\protect\astroncite{{Moore} et~al.}{1999}]{moo99}
{Moore}, R.~L., {Falconer}, D.~A., {Porter}, J.~G., and {Suess}, S.~T.: 1999,
\newblock {\em \apj} {\bf 526}, 505

\bibitem[\protect\astroncite{{Moos} et~al.}{2000}]{moo00}
{Moos}, H.~W., {Cash}, W.~C., {Cowie}, L.~L., {Davidsen}, A.~F., {Dupree},
  A.~K., {Feldman}, P.~D., {Friedman}, S.~D., {Green}, J.~C., {Green}, R.~F.,
  {Gry}, C., {Hutchings}, J.~B., {Jenkins}, E.~B., {Linsky}, J.~L., {Malina},
  R.~F., {Michalitsianos}, A.~G., {Savage}, B.~D., {Shull}, J.~M., {Siegmund},
  O.~H.~W., {Snow}, T.~P., {Sonneborn}, G., {Vidal-Madjar}, A., {Willis},
  A.~J., {Woodgate}, B.~E., {York}, D.~G., {Ake}, T.~B., {Andersson}, B.-G.,
  {Andrews}, J.~P., {Barkhouser}, R.~H., {Bianchi}, L., {Blair}, W.~P.,
  {Brownsberger}, K.~R., {Cha}, A.~N., {Chayer}, P., {Conard}, S.~J.,
  {Fullerton}, A.~W., {Gaines}, G.~A., {Grange}, R., {Gummin}, M.~A.,
  {Hebrard}, G., {Kriss}, G.~A., {Kruk}, J.~W., {Mark}, D., {McCarthy}, D.~K.,
  {Morbey}, C.~L., {Murowinski}, R., {Murphy}, E.~M., {Oegerle}, W.~R., {Ohl},
  R.~G., {Oliveira}, C., {Osterman}, S.~N., {Sahnow}, D.~J., {Saisse}, M.,
  {Sembach}, K.~R., {Weaver}, H.~A., {Welsh}, B.~Y., {Wilkinson}, E., and
  {Zheng}, W.: 2000,
\newblock {\em \apjl} {\bf 538}, L1

\bibitem[\protect\astroncite{Murthy and Henry}{1995}]{mur95}
Murthy, J. and Henry, R.: 1995,
\newblock {\em \apj} {\bf 448}, 848

\bibitem[\protect\astroncite{{Ogilvie} et~al.}{1980}]{ogi80}
{Ogilvie}, K.~W., {Bochsler}, P., {Geiss}, J., and {Coplan}, M.~A.: 1980,
\newblock {\em \jgr} {\bf 85(14)}, 6069

\bibitem[\protect\astroncite{Otte et~al.}{2003}]{ott03}
Otte, B., Dixon, W., and Sankrit, R.: 2003,
\newblock {\em \apj} {\bf 586}, L53

\bibitem[\protect\astroncite{{Otte} et~al.}{2004}]{ott04}
{Otte}, B., {Dixon}, W.~V.~D., and {Sankrit}, R.: 2004,
\newblock {\em \aplett} {\bf 606}, L143

\bibitem[\protect\astroncite{{Papadopoulos}}{1985}]{pap85}
{Papadopoulos}, K.: 1985,
\newblock in {\em Advances in Space Plasma Physics}, pp 289--+

\bibitem[\protect\astroncite{{Pye} et~al.}{1981}]{pye81}
{Pye}, J.~P., {Pounds}, K.~A., {Rolf}, D.~P., {Smith}, A., {Willingale}, R.,
  and {Seward}, F.~D.: 1981,
\newblock {\em \mnras} {\bf 194}, 569

\bibitem[\protect\astroncite{Ray}{1958}]{ray58}
Ray, D.~J.: 1958,
\newblock {\em The ALASKA SPORTSMAN},
\newblock reprinted in AURORA BOREALIS The Amazing Northern Lights by S.I.
  Akasofu Alaska Geographic Volume 6 Number 2 1979

\bibitem[\protect\astroncite{{Raymond} et~al.}{1995}]{ray95}
{Raymond}, J.~C., {Blair}, W.~P., and {Long}, K.~S.: 1995,
\newblock {\em \aplett} {\bf 454}, L31+

\bibitem[\protect\astroncite{{Reynolds}}{2004}]{rey04}
{Reynolds}, S.~P.: 2004,
\newblock {\em Advances in Space Research} {\bf 33}, 461

\bibitem[\protect\astroncite{{Reynolds} and {Gilmore}}{1986}]{rey86}
{Reynolds}, S.~P. and {Gilmore}, D.~M.: 1986,
\newblock {\em \aj} {\bf 92}, 1138

\bibitem[\protect\astroncite{{Scholz} and {Walters}}{1991}]{sch}
{Scholz}, T.~T. and {Walters}, H.~R.~J.: 1991,
\newblock {\em \apj} {\bf 380}, 302

\bibitem[\protect\astroncite{{Schweizer} and {Middleditch}}{1980}]{sch80}
{Schweizer}, F. and {Middleditch}, J.: 1980,
\newblock {\em \apj} {\bf 241}, 1039

\bibitem[\protect\astroncite{{Sedov}}{1959}]{sed59}
{Sedov}, L.~I.: 1959,
\newblock {\em {Similarity and Dimensional Methods in Mechanics}},
\newblock Similarity and Dimensional Methods in Mechanics, New York: Academic
  Press, 1959

\bibitem[\protect\astroncite{{Shapiro} et~al.}{1998a}]{sha98}
{Shapiro}, V.~D., {Bingham}, R., {Dawson}, J.~M., {Dobe}, Z., {Kellett}, B.~J.,
  and {Mendis}, D.~A.: 1998a,
\newblock {\em Physica Scripta Volume T} {\bf 75}, 39

\bibitem[\protect\astroncite{{Shapiro} et~al.}{1998b}]{sha97}
{Shapiro}, V.~D., {Bingham}, R., {Dawson}, J.~M., {Dobe}, Z., {Kellett}, B.~J.,
  and {Mendis}, D.~A.: 1998b,
\newblock {\em Physica Scripta Volume T} {\bf 75}, 39

\bibitem[\protect\astroncite{{Smith} et~al.}{1998}]{smi98}
{Smith}, C.~W., {L'Heureux}, J., {Ness}, N.~F., {Acu{\~ n}a}, M.~H., {Burlaga},
  L.~F., and {Scheifele}, J.: 1998,
\newblock {\em Space Science Reviews} {\bf 86}, 613

\bibitem[\protect\astroncite{{Smith} et~al.}{1991}]{smi91}
{Smith}, R.~C., {Kirshner}, R.~P., {Blair}, W.~P., and {Winkler}, P.~F.: 1991,
\newblock {\em \apj} {\bf 375}, 652

\bibitem[\protect\astroncite{{Smith} et~al.}{1994}]{smi94}
{Smith}, R.~C., {Raymond}, J.~C., and {Laming}, J.~M.: 1994,
\newblock {\em \apj} {\bf 420}, 286

\bibitem[\protect\astroncite{{Sonett} and {Abrams}}{1963}]{son63}
{Sonett}, C.~P. and {Abrams}, I.~J.: 1963,
\newblock {\em \jgr} {\bf 68(17)}, 1233

\bibitem[\protect\astroncite{{Spitzer}}{1956}]{spi56}
{Spitzer}, L., J.: 1956,
\newblock {\em Physics of Fully Ionized Gases},
\newblock Interscience Publishers, Inc.

\bibitem[\protect\astroncite{{Stepanova} and {Kosovichev}}{2000}]{ste00}
{Stepanova}, T.~V. and {Kosovichev}, A.~G.: 2000,
\newblock {\em Advances in Space Research} {\bf 25}, 1855

\bibitem[\protect\astroncite{{Stone} et~al.}{1998}]{sto98}
{Stone}, E.~C., {Frandsen}, A.~M., {Mewaldt}, R.~A., {Christian}, E.~R.,
  {Margolies}, D., {Ormes}, J.~F., and {Snow}, F.: 1998,
\newblock {\em Space Science Reviews} {\bf 86}, 1

\bibitem[\protect\astroncite{{Strom}}{1994}]{str94}
{Strom}, R.~G.: 1994,
\newblock {\em \aap} {\bf 288}, L1

\bibitem[\protect\astroncite{{Tanimori} et~al.}{1998}]{tan98}
{Tanimori}, T., {Hayami}, Y., {Kamei}, S., {Dazeley}, S.~A., {Edwards}, P.~G.,
  {Gunji}, S., {Hara}, S., {Hara}, T., {Holder}, J., {Kawachi}, A., {Kifune},
  T., {Kita}, R., {Konishi}, T., {Masaike}, A., {Matsubara}, Y., {Matsuoka},
  T., {Mizumoto}, Y., {Mori}, M., {Moriya}, M., {Muraishi}, H., {Muraki}, Y.,
  {Naito}, T., {Nishijima}, K., {Oda}, S., {Ogio}, S., {Patterson}, J.~R.,
  {Roberts}, M.~D., {Rowell}, G.~P., {Sakurazawa}, K., {Sako}, T., {Sato}, Y.,
  {Susukita}, R., {Suzuki}, A., {Suzuki}, R., {Tamura}, T., {Thornton}, G.~J.,
  {Yanagita}, S., {Yoshida}, T., and {Yoshikoshi}, T.: 1998,
\newblock {\em \aplett} {\bf 497}, L25+

\bibitem[\protect\astroncite{Tennekes and Lumley}{1972}]{ten57}
Tennekes, H. and Lumley, J.: 1972,
\newblock {\em A First Course in Turbulence},
\newblock The MIT Press

\bibitem[\protect\astroncite{{Tidman}}{1969}]{tid69}
{Tidman}, D.~A.: 1969,
\newblock in {\em Plasma Instabilities in Astrophysics}, pp 229--+

\bibitem[\protect\astroncite{Universe}{2000}]{mto04}
Universe: 2000,
\newblock {\em Myths: Tonatiuh},
\newblock http://www.windows.ucar.edu Accessed December 4 2004

\bibitem[\protect\astroncite{{van Ballegooijen} and {Martens}}{1989}]{van89}
{van Ballegooijen}, A.~A. and {Martens}, P.~C.~H.: 1989,
\newblock {\em \apj} {\bf 343}, 971

\bibitem[\protect\astroncite{{Vasyliunas} and {Siscoe}}{1976}]{val76}
{Vasyliunas}, V.~M. and {Siscoe}, G.~L.: 1976,
\newblock {\em \jgr} {\bf 81(10)}, 1247

\bibitem[\protect\astroncite{{Vink} et~al.}{2003}]{vink03}
{Vink}, J., {Laming}, J.~M., {Gu}, M.~F., {Rasmussen}, A., and {Kaastra},
  J.~S.: 2003,
\newblock {\em \aplett} {\bf 587}, L31

\bibitem[\protect\astroncite{{Wang} and {Sheeley}}{1990}]{wan90}
{Wang}, Y.-M. and {Sheeley}, N.~R.: 1990,
\newblock {\em \apj} {\bf 355}, 726

\bibitem[\protect\astroncite{{Willingale} et~al.}{1996}]{wil96}
{Willingale}, R., {West}, R.~G., {Pye}, J.~P., and {Stewart}, G.~C.: 1996,
\newblock {\em \mnras} {\bf 278}, 749

\bibitem[\protect\astroncite{{Winkler} et~al.}{2003}]{wink03}
{Winkler}, P.~F., {Gupta}, G., and {Long}, K.~S.: 2003,
\newblock {\em \apj} {\bf 585}, 324

\bibitem[\protect\astroncite{{Woo} and {Habbal}}{1997}]{woo97}
{Woo}, R. and {Habbal}, S.~R.: 1997,
\newblock {\em \grl} {\bf 24}, 1159

\bibitem[\protect\astroncite{{Zertsalov} et~al.}{1976}]{zer76}
{Zertsalov}, A.~A., {Vaisberg}, O.~L., and {Temnyi}, V.~V.: 1976,
\newblock {\em Cosmological Research} {\bf 14}, 257

\bibitem[\protect\astroncite{{Zhao} et~al.}{1991}]{zha91}
{Zhao}, X., {Ogilvie}, K.~W., and {Whang}, Y.~C.: 1991,
\newblock {\em \jgr} {\bf 96}, 5437

\end{thebibliography}
\startabstractpage{ION HEATING IN COLLISIONLESS SHOCKS IN
SUPERNOVAE AND THE HELIOSPHERE}{Kelly Elizabeth
Korreck}{Co-Chairpersons: Thomas H. Zurbuchen and John C. Raymond}
\pagestyle{empty}

Collisionless shocks play a role in many astrophysical phenomena,
from coronal mass ejections (CMEs) in the heliosphere to supernova
remnants.  Their role in heating and accelerating particles is
well accepted yet the exact mechanism for ion heating is not well
understood.  Two systems, CMEs and supernova remnants, were
examined to determine the heating of heavy ions as they pass
through collisionless shocks thus providing a seed population for
cosmic ray acceleration processes.  Three parameters are examined,
the plasma beta, the Mach number of the shock and the magnetic
angle of the shock.  CMEs heat heavy ions preferentially.  This is
in contrast to the supernova data which shows less than mass
proportional heating.  In addition to these studies, heating in
astrophysical systems involves neutral atoms. A Monte Carlo model
simulated neutral particles as they pass through the shock.
Neutrals can create a precursor to the shock additionally heating
the plasma.  This work uses in situ data from the heliosphere to
study astronomical systems because of common shock properties is a
unique way to study magnetic components of shocks remotely.

\end{document}